\documentclass[11pt,a4paper]{article}  

\usepackage[margin=1in]{geometry}     
\usepackage{newtxmath}                

\usepackage{amsmath}
\usepackage{amssymb}

\usepackage{setspace}                 
\usepackage{ragged2e}

\usepackage{titlesec}
\titleformat{\section}
  {\normalfont\Large\bfseries}{\thesection}{1em}{}
\titleformat{\subsection}
  {\normalfont\large\bfseries}{\thesubsection}{1em}{}

\usepackage{float}
\usepackage{caption}
\usepackage{booktabs}
\usepackage{threeparttable}
\usepackage{makecell}
\usepackage{siunitx}
\renewcommand{\thetable}{\arabic{table}}
\renewcommand{\thefigure}{\arabic{figure}}

\usepackage{graphicx}
\usepackage[export]{adjustbox}
\usepackage{subfig}
\usepackage{pdflscape}
\usepackage{rotating}

\usepackage{enumitem}
\setlist[itemize]{noitemsep,topsep=0pt,left=1em}

\usepackage{hyperref}
\hypersetup{
    colorlinks=true,
    linkcolor=blue,
    citecolor=blue,
    urlcolor=blue
}

\usepackage[american]{babel}
\usepackage{csquotes}
\usepackage[authordate,backend=biber]{biblatex-chicago}
\usepackage{cleveref}

\title{Racial bias, colorism, and overcorrection\thanks{An earlier version of this paper was presented at conferences and seminars hosted by the University of Zurich (2024), the Erasmus School of Economics in Rotterdam (2024), the University of Reading (2025, online), and the University of Tübingen (2025). We thank participants for their helpful comments and suggestions. We appreciate individual reviews and comments from Wallace Hendricks, Sem Manna, Francesco Maria Esposito, Ignacio Palacios-Huerta, Thomas Peeters, Carl Singleton, and Stefan Szymanski. We are especially grateful to Teodora Szasz for her guidance as well as to Anjali Adukia, Alex Eble, Emileigh Harrison, and Hakizumwami Birali Runesha for sharing their code. Ethical approval for this study was received by the Institutional Review Board (IRB) at the Faculty of Economics and Social Sciences of the University of Tübingen under the registry number A2.5.4-323\textunderscore hb. The authors have no financial interest in the topic of this paper, and there are no conflicts of interest. All errors are our own.}}

\author{
\begin{tabular}{ccc}
\makecell{Kenneth Colombe\footnote{Corresponding author. University of Bonn, Department of Economics. Email: s72kcolo@uni-bonn.de} \\ \small \textit{University of Bonn}} & \makecell{Alex Krumer\footnote{Molde University College, Faculty of Business Administration and Social Sciences. Email: alex.krumer@himolde.no} \\ \small \textit{Molde University College}} \\
\\
\makecell{Rosa Lavelle-Hill\footnote{University of Basel, Faculty of Humanities and Social Sciences. Email: rosa.lavelle-hill@unibas.ch} \\ \small \textit{University of Basel}} & \makecell{Tim Pawlowski\footnote{University of Tübingen, Institute of Sports Science, LEAD Graduate School and Research Network, Interfaculty Research Institute for Sports and Physical Activity. Email: tim.pawlowski@uni-tuebingen.de} \\ \small \textit{University of Tübingen}} 
\end{tabular}
}

\date{}

\begin{document}

{\setstretch{1.0}
\maketitle
\thispagestyle{empty}
\vspace{0.5em}
\center{\date{\today}}
\vspace{2em}
\begin{abstract}
This paper examines whether increased awareness can affect racial bias and colorism. For this purpose, we utilize a natural experiment arising from the widespread publicity of \textcite{price2010}, which served as an external shock, intensifying scrutiny of racial bias in men’s basketball officiating. We investigate refereeing decisions in a similar setting, the Women’s National Basketball Association (WNBA), which is known as a progressive institution with a longstanding commitment to diversity, equity, and inclusion (DEI) policy. We apply state-of-the-art artificial intelligence and machine learning techniques to systematically predict race and objectively measure skin tone. Our empirical strategy exploits the quasi-random assignment of referees to games, combined with high-dimensional fixed effects, to estimate the relationship between the racial and skin tone compositions of referees and players, as well as foul-calling behavior. Our results show no significant racial bias before the intense media coverage. However, afterward, we find evidence of overcorrection, according to which a player earns fewer fouls when facing more referees from the opposite race and skin tone. Even though this overcorrection seems to wear off over time, we highlight the need to consider baseline levels of bias before applying any prescription with direct relevance to policymakers and organizations, given the recent discourse on DEI.
\noindent
\\
\vspace{2em}
\end{abstract}
}

\clearpage
\setcounter{page}{1}
\pagenumbering{arabic}

\section{Introduction} \label{introduction}
\begin{quote}
    \textit{“A person is aware that they are biased in favor of X; or else they are concerned about the possibility that they might be biased in favor of X; or at least, they are concerned to avoid the appearance of being biased in favor of X. Because of this, they bend over backward in order not to exhibit bias in favor of X. But they overshoot and end up treating X less favorably than they would if they were unbiased.”}
    \flushright{— \parencite{kelly2022}}
\end{quote}
Since the seminal work of Gary Becker on the economics of discrimination \parencite{becker1957}, it has become one of the most researched topics in economics \parencite{abrevaya2012}. Discrimination has been found in a variety of settings that include but are not limited to law enforcement \parencite{antonovics2009, knowles2001}, justice system \parencite{shayo2011}, housing market \parencite{ahmed2008}, labor market \parencite{bertrand2004}, and sports \parencite{price2010, szymanski2000}.\footnote{See \textcite{lang2020} for a comprehensive review of empirical literature on the economics of discrimination in the labor market and criminal justice system.}

At the time of Becker’s publication in 1957, racial discrimination in the USA was still legal. Today, racial discrimination of any kind is illegal. However, this does not mean that racial discrimination has fully disappeared, and therefore, there is room for increasing awareness of the problem in organizations that have experienced discrimination. Intuitively, such organizations are expected to benefit from increased awareness, which is likely to reduce bias. For example, \textcite{price2010}, one of the two studies mostly related to our paper, investigated the racial bias of referees in the National Basketball Association (NBA). The authors found that players earned significantly more fouls from the crew that had more referees of a different race. That paper provoked extensive media attention, including front-page coverage in the \textit{New York Times} \parencite{schwarz2007} and other major media outlets. The widespread coverage was used in \textcite{ppw2018}, the second study that guides our work, as an exogenous shock to show that NBA referee racial bias disappeared following increased awareness.\footnote{It is also worth mentioning a recent study by \textcite{alesina2024} that showed that an increased awareness among Italian teachers reduced the gap between the grades of native versus immigrant students.}

However, it is also plausible to assume that many (hopefully most) organizations have few, if any, cases of existing systematic discrimination. Then the question is how high-profile discriminatory exposures, à-la the above-mentioned NBA case, would affect such organizations. Ideally, organizations lacking systematic discrimination would not change their behavior during relevant and intensified public debates. However, \textit{unintended} overcorrection can occur when individuals or institutions seek to avoid or compensate for past biases, whether perceived or actual. This behavior may lead to an unintentional preference for a particular group. According to \textcite{mendes2013}, such overcorrection may happen because people may be concerned about appearing racially prejudiced.

Even though overcorrection can be closely linked to the widely researched topic of discrimination, the phenomenon of overcorrection has received relatively little attention from economists. To the best of our knowledge, the only field study to document overcorrection in an empirical setting is \textcite{chowdhury2024}, who used home advantage in professional cricket as the main driver of bias. They investigated international cricket games where the local umpires were known to be biased in favor of their home nation teams and, therefore, were replaced by neutral umpires. However, because of the COVID-19 pandemic, the local umpires resumed officiating games of their home nation teams. During these games, the local umpires favored the away teams in a way that is in line with overcorrection. 

We are not aware of an economic paper on overcorrection in the context of racial bias. Most of the evidence on overcorrection in the racial context comes from lab studies in experimental psychology. For example, \textcite{harber1998} found that when evaluating a poorly written essay, White raters rated the content more positively if they thought that the author was Black than when they believed the author was White. Similarly, \textcite{mendes2013} found that the majority group members engaged in more positive behavior toward stigmatized or minority group members compared to non-stigmatized or ingroup members. In a more recent study, \textcite{howe2022} showed that White patients made more effort to socially engage with Black providers of medicine than with White or Asian.\footnote{There was no difference in engagement between White and Asian providers}

In this paper, we fill a gap in the literature by investigating whether an exogenous increase in awareness of racial discrimination caused overcorrection within an established institution with a longstanding commitment to diversity, equity, and inclusion (DEI), where explicit discrimination is unlikely. The institutional setting and its policies resemble those adopted later across the public and private sectors following the focus on racial injustice in 2020 \parencite{biden2021,wef2022}. More specifically, we build on the efforts of \textcite{price2010} and \textcite{ppw2018}, but expand them in several ways. We investigate possible racial bias in a different, though relevant setting, the Women’s National Basketball Association (WNBA). Since its founding, the WNBA has been synonymous with social justice, actively breaking barriers and advocating for equality \parencite{mathewson2020}. If WNBA’s referees indeed have no systematic racial and skin tone bias (we provide evidence supporting this claim), it would make the WNBA a perfect case to investigate the effect of widespread media coverage on racial discrimination in a non-discriminatory organization.

Our paper also contributes to a growing body of research using artificial intelligence and machine learning to uncover discrimination in settings such as children’s books \parencite{adukia2023, szasz2023}, sports \parencite{kamel2023}, online job networking \parencite{evsyukova2025}, and algorithmic fairness research \parencite{thong2023}. First, for predictions of race, we use the FairFace algorithm, which is trained on a racially balanced dataset of more than 100,000 images and designed to mitigate racial biases in facial recognition \parencite{karkkainenfairface}. Second, we also extend the analysis to skin tone, recognizing that race is subjective \parencite{smedley2005} and analyzing only race risks obscuring a key driver of discrimination. Substantial evidence regarding inter- and intra-race skin tone preferences \parencite{dixon2017,malema2011, telles2004} confirms that reducing the analysis to race alone oversimplifies the task. Discrimination based on skin tone persists even within racial and ethnic groups \parencite{monk2021}, including African Americans \parencite{hill2002,keith1991}, Latin Americans \parencite{camposvazquez2021,golashboza2008}, and Asians \parencite{malema2011,wagatsuma1967}. While the examination of colorism is not necessarily novel \parencite{goldsmith2006,hersch2006,hersch2008}, existing studies commonly relied on binary skin color classification, presenting a similar binary classification issue as past racial studies \parencite{branigan2013}. Moreover, earlier efforts often lacked objective, quantitative, and scalable measures of skin tone measurement. Specifically, previous papers typically relied on surveys such as the National Survey of Black Americans (NSBA) or the Multi-City Study of Urban Inequality (MCSUI), which are limited by factors including lack of representativeness, small sample sizes, and imprecise measurement. We address these critical measurement limitations in past research following the approach proposed by \textcite{adukia2023}. More precisely, we measure skin tone continuously on a 0 to 100 scale based on the CIELAB color space, as it offers an objective and standardized method for communicating color information, especially skin color \parencite{connolly1997,weatherall1992}.\footnote{Color spaces are mathematical models that include three or four components to represent color information \parencite{kolkur2017}.} This system aligns with how humans see color, enabling accurate color measurement and comparison \parencite{moreland2009}. Thus, we can test not only the effect of the difference in race between the players and the referee crew on the personal foul calls based on a binary definition (Black vs non-Black), but also the effect of the difference in skin tone in a continuous spectrum.

As expected, following the methodology presented in \textcite{ppw2018} and based on close to 13,000 individual fouls, we do not find significant evidence that WNBA referees exhibited systematic racial bias in their foul calls in the years between 2004 and 2006 (before the widespread media coverage of the \textcite{price2010} results). However, based on more than 17,000 individual fouls in the years between 2007 and 2010, we find evidence of overcorrection. When using a binary variable, our results suggest that players receive fewer fouls when officiated by a different race crew. When using the more informative continuous skin tone variable, our estimates substantially gain in precision, showing that referees call fewer fouls the greater the distance is between the referee's and the players' skin tone. We interpret this finding with increased awareness of racial discrimination, creating an unintended consequence in an organization whose referees were already free of racial bias. Remarkably, in the years between 2011 and 2014, which is another extension to the analyses also presented in \textcite{ppw2018}, we find a return to the ``pre-awareness'' levels.

The rest of this paper is organized as follows: Section \ref{wnba} provides an overview of the WNBA context and the officiating framework, Section \ref{method} outlines the methodology and data used in the analysis, Section \ref{results} presents the empirical results, Section \ref{discussion} engages in discussion exploring potential mechanisms driving the observed behavior and acknowledges some limitations, and finally, Section \ref{conclusion} concludes.

\section{The Women's National Basketball Association} \label{wnba}
Like the National Basketball Association (NBA) for men, the WNBA is the premier league for women’s basketball. The similar names underscore the NBA’s direct involvement in creating the WNBA in 1997 as its sister league, initially owning all the teams \parencite{voepel2014}. Starting in 2002, the NBA sold the franchises, some to their NBA operator counterparts and others to third parties \parencite{hart2023}. Despite this interconnectedness, both leagues differ in two relevant ways. First, the players of both leagues have always kept their interests separate through independent player unions that collectively bargain with their respective leagues on issues such as salary and benefits. Second, referees of the two leagues remained separate until 2017, when the National Basketball Referees Association (NBRA) was brought in to collectively negotiate and represent both leagues’ referees from then on (Women’s National Basketball Association, 2017).\footnote{Other notable competition- and gameplay-related differences between the leagues are the number of teams and games, quarter length, ball size, and three-point line distance.} 

\subsection{Officiating} \label{officiating}
Every season, each of approximately 30 referees in the WNBA officiates about 20 regular-season games.\footnote{Insider knowledge about officiating practices discussed in this section was gained through interviews with long-time WNBA referee Amy Bonner \parencite{lahaye2020} and direct discussions with another veteran referee who wished to remain anonymous.} A WNBA season lasts between May and September/October. Referees see each team one to five times per season and rarely work games in the same city without at least a week between assignments. These practical constraints prevent referee assignments from being completely random. However, we follow the procedure presented in Price and Wolfers (2010) to show that for each year in our sample, the number of non-Black referees is unrelated to the number of Black starters (five players who start the game). The results of these tests are presented in Table~\ref{table:blackstartersperteamAI} (as well as  Table~\ref{table:blackstartersperteamhr} for human raters and Table~\ref{table:tonestartersperteam} for skin tone), suggesting that the assignment decision is unrelated to the racial characteristics of the teams.

\begin{table}[H]
\centering
\captionsetup{
  justification=centering,
  singlelinecheck=false,
  labelsep=none,
  font=normalsize
}
\caption{\\Black starters per team and the distribution of refereeing crews by race}\vspace{-0.8em}
\label{table:blackstartersperteamAI}
\begin{threeparttable}
  \small
  \setlength\tabcolsep{4pt}      
  \renewcommand{\arraystretch}{0.8} 
  \begin{tabular}{
    @{}l
    S[table-format=4.2]
    S[table-format=4.2]
    S[table-format=4.2]
    S[table-format=4.2]
    S[table-format=1.3]
    @{}}
    \toprule\toprule
& \multicolumn{4}{c}{Black starters per team} \\
\cmidrule(lr){2-5}
Season & {0 non-Black refs} & {1 non-Black ref} & {2 non-Black refs} & {3 non-Black refs} & {\makecell[cb]{\(\chi^2\) test of \\ independence\(^{\alpha}\) \\ (p-value)}} \\
\midrule
    2004 & 4.00 & 3.43 & 3.48 & 3.62 & .443 \\
    2005 & 3.50 & 3.14 & 3.33 & 3.35 & .862 \\
    2006 & 3.92 & 3.43 & 3.47 & 3.29 & .261 \\
    2007 & 3.25 & 3.36 & 3.23 & 3.27 & .418 \\
    2008 & 3.93 & 3.64 & 3.81 & 3.79 & .456 \\
    2009 & 3.77 & 3.81 & 3.81 & 3.73 & .934 \\
    2010 & 3.83 & 3.88 & 3.81 & 3.68 & .060 \\
    2011 & 4.00 & 4.01 & 3.82 & 3.82 & .650 \\
    2012 & 4.28 & 4.13 & 3.96 & 4.06 & .333 \\
    2013 & 4.32 & 4.28 & 4.30 & 4.30 & .808 \\
    2014 & 4.00 & 3.94 & 3.85 & 3.93 & .541 \\
\\
\makecell[l]{Sample size (\% of \\ all player-games)} & {\makecell{296 \\ (6.27)}} & {\makecell{1,306 \\ (27.68)}} & {\makecell{2,002 \\ (42.43)}} & {\makecell{1,114 \\ (23.61)}} & {n = 4,718} \\
\bottomrule\bottomrule
  \end{tabular}
  \begin{tablenotes}[para,flushleft]
    \setstretch{0.5}\tiny
    \setlength{\parskip}{0pt}
    \setlength{\parindent}{0pt}
    \setlength{\itemsep}{-1pt}
    \scriptsize
    \item \textit{Notes:} Testing if the number of non-Black referees is unrelated to the number of black starters for each year in our sample. Racial classifications determined by FairFace. Human rater classifications are included in the appendix. Each observation is a team×game observation.
    \(^{\alpha}\)Final column tests:  
    \(H_0\): \# non-Black referees is independent of \# Black starters.
  \end{tablenotes}
\end{threeparttable}
\end{table}

Referees arrive in the host city the night before a game. On game day, they meet off-site to discuss the matchup, focusing on specific concerns such as recent rule changes. They arrive at the arena 75-90 minutes before tip-off. After the game, they gather in the locker room to review pivotal plays, including any game-deciding calls, technical fouls, or instances of unsportsmanlike conduct. They then submit a detailed report to the league office and receive a copy of the game video for individual evaluation. It is important to note that with referee training heavily emphasizing and incentivizing the correct adjudication of fouls and considering the league’s inherent culture of inclusivity (see Section \ref{social}), it is the league’s position that specific bias training concerning race and/or skin color is deemed unnecessary.

\subsection{Social justice} \label{social}
\begin{quote}
    \textit{``[Basketball can] help us understand that if we say we care about values like equality, diversity, and inclusion then we must reduce implicit biases not just in the workplace, but at home; not just in our laws, but in our hearts and minds—because that’s what’s expected on the court. And while all of us—athletes and fans alike—face moral choices professionally and personally that can cause discomfort and require risk, today there can no longer be a moral separation between what we do and who we are.''}
    \parencite{borders2018}
\end{quote}
\begin{flushright}
    — Lisa Borders (WNBA Commissioner, 2016 to 2018)
\end{flushright}

The quote above emphasizes the WNBA’s commitment to social justice, and its advocacy for equality \parencite{mathewson2020}. The WNBA is notable among professional sports leagues because of its strong emphasis on DEI.\footnote{It is evident in who they accept in the league or the treatment of those who do not conform. For example, a former US Senator and team owner, Kelly Loeffler, was pressured to sell her stake in the Atlanta Dream following her letter to league commissioner Cathy Engelbert opposing the Black Lives Matter movement \parencite{graham2021}. Players even played a decisive role in campaigning for her political opponent, who eventually won \parencite{gregory2021}.} Our interviewed former referee confirms the league’s position that its established commitment to these values obviates the need for bias training.

The WNBA boasts numerous firsts, including the first women to secure a Nike signature shoe deal, the first professional league to launch a pride campaign and dedicate a season to social justice \parencite{booker2020}, the first professional team to wear ``Black Lives Matter'' shirts, and the first professional league to feature married teammates \parencite{teng2021}.

Moreover, WNBA players consistently speak out on critical issues such as sexuality and equal pay \parencite{weiner2020}. Mistie Bass, a ten-year veteran, made the point clear, saying, ``\textit{We are a walking protest at all times as a W.N.B.A. athlete, if you think about it, we have so many different stigmas. We’re just constantly in the fight. I don’t think we have ever not been in a fight for equality, for justice}'' \parencite{weiner2020}. The players have now institutionalized their fight by forming their own Justice, Equity, Diversity, and Inclusion (JEDI) Committee. Moreover, in solidarity, the league office promotes several initiatives related to civic engagement, voting, and educational resources.\footnote{WNBPA Handbook https://wnbpa.com/wp-content/uploads/2024/08/Digital-WNBPA-Union-Handbook.pdf and WNBA Social Justice page https://www.wnba.com/social-justice.} 

\section{Methodology} \label{method}
\subsection{Data and measures} \label{data}

We leverage publicly available data spanning 2004 through 2014.\footnote{Box score statistics are from Her Hoop Stats https://herhoopstats.com/. Images are collected from the WNBA League Office https://stats.wnba.com/, gettyimages https://www.gettyimages.com/, or the National Basketball Referee Association https://www.nbra.net/nba-officials/. Relevant biographical information is taken from Basketball Reference https://www.basketball-reference.com/wnba/.} As in \textcite{price2010} and \textcite{ppw2018}, we use regular season games for our analyses. For players, our data includes detailed game-level statistics, career measures, and individual characteristics such as birthdate, height, weight, and nationality. We also have career statistics and individual characteristics for coaches and referees. Moreover, we systematically collect one front-facing photo per person, prioritizing images from official pre-season media day photoshoots. We conduct a simple Google Images search if no official photo meeting our criteria is available.\footnote{Images obtained by Google search only account for 18\% of our sample size and are mostly of high quality.} Our key measures, race and skin tone, are derived from the pictures as described below.

\begin{figure} [h]
\centering
\captionsetup{width=0.85\linewidth}
\captionsetup{labelfont=normalfont,labelsep=space,justification=centering}
\caption{\\FairFace Racial Predictions}\vspace{-0.8em}
\captionsetup{justification=raggedright, singlelinecheck=false}
\includegraphics[width=0.85\linewidth]{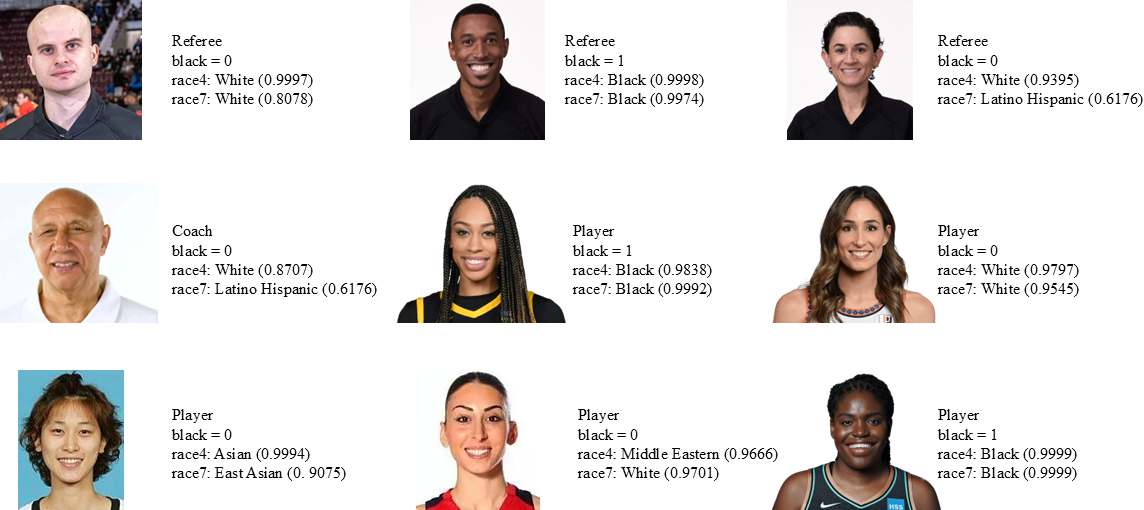}
\caption*{\justifying\scriptsize \textit{Notes:} An individual is coded as Black if the FairFace prediction algorithm categorizes them as Black under either the ``race4'' classification (White, Black, Asian, Indian) or ``race7'' classification (White, Black, Latino Hispanic, East Asian, Southeast Asian, Indian, Middle Eastern). Otherwise, they are coded as non-Black (0). The pictures in this Figure provide different examples across gender, racial profiles, image quality, and types.}
\label{fig:figff}
\end{figure}

Our first approach to measuring race is using the FairFace algorithm \parencite{karkkainenfairface}. FairFace is a machine learning model trained on a racially balanced dataset of over 108,000 images. As such, and in contrast to other algorithms using face datasets that overrepresent Caucasian faces, it is explicitly designed to mitigate racial biases in facial recognition.\footnote{See \textcite{karkkainenfairface} for algorithm details and performance metrics: https://github.com/dchen236/FairFace.} FairFace provides two race classifications, ``race4'' (White, Black, Asian, Indian) and ``race7'' (White, Black, Latino Hispanic, East Asian, Southeast Asian, Indian, Middle Eastern), and assigns probability scores to each category. We used both classifications to create our race variable. An individual is coded as Black (1) if FairFace classifies them as Black in either ``race4'' or ``race7''; otherwise, they are coded as non-Black (0) (see Figure~\ref{fig:figff}).

To validate the FairFace predictions of race, we implemented a second approach closely following \textcite{price2010}, i.e., we used human raters to assign race as Black or non-Black, aligning with the US Census definition of Black or African American.\footnote{https://www.census.gov/topics/population/race/about.html.} One male and one female rater independently rated each player, coach, and referee based on their photos. For each photo, raters indicated ``Black'' or ``non-Black'' based on their first impression and then rated their confidence level on a scale of 1 (unsure) to 4 (very confident). We reviewed the ratings and assigned a third rater to resolve discrepancies. We calculate Kappa scores to measure the overall agreement between the FairFace classifications and human raters (see Appendix Table~\ref{table:kappa}). There was 94.8\% agreement, which is ``almost perfect agreement'' \parencite{mchugh2012}.

Our third approach follows the ``Image-to-Data Pipeline'' in \textcite{adukia2023}, dividing the process into two main parts, i.e., face segmentation and skin tone classification. Face segmentation starts with employing a deep-learning approach called a Fully Connected Convolutional Neural Network (FC-CNN) along with the 2D Face Alignment library to detect 68 facial landmarks.\footnote{For more details, see: https://github.com/1adrianb/2D-and-3D-face-alignment?tab=readme-ov-file} These landmarks are refined using a convex hull method to generate a precise binary skin mask that isolates the facial region. This process predicts periphery landmarks like the jawline, eyebrows, and lips. From these landmarks, we create a binary skin mask to isolate the facial region before applying this mask to the original image to extract the face and segment the skin for color classification. A Hue Saturation Value (HSV) color space-based filter is applied to ensure that only reliable skin pixels are retained, thus minimizing any lighting artifacts.\footnote{The HSV color space represents colors using intuitive values based on tint, saturation, and tone. Importantly, HSV decouples color information (hue and saturation) from brightness (value), making it particularly effective at identifying skin pixels even under uneven lighting conditions \parencite{vezhnevets2003, bora2015, shaik2015}. This conversion robustly isolates skin pixels by reducing the influence of shadows and variations in illumination.} 

\begin{figure}
\centering
\captionsetup{labelfont=normalfont,labelsep=none,justification=centering}
\caption{\\Sample Distribution of L*, a*, and b* values}\vspace{-0.8em}
\captionsetup{justification=raggedright, singlelinecheck=false}
{\label{a}\includegraphics[width=.50\linewidth]{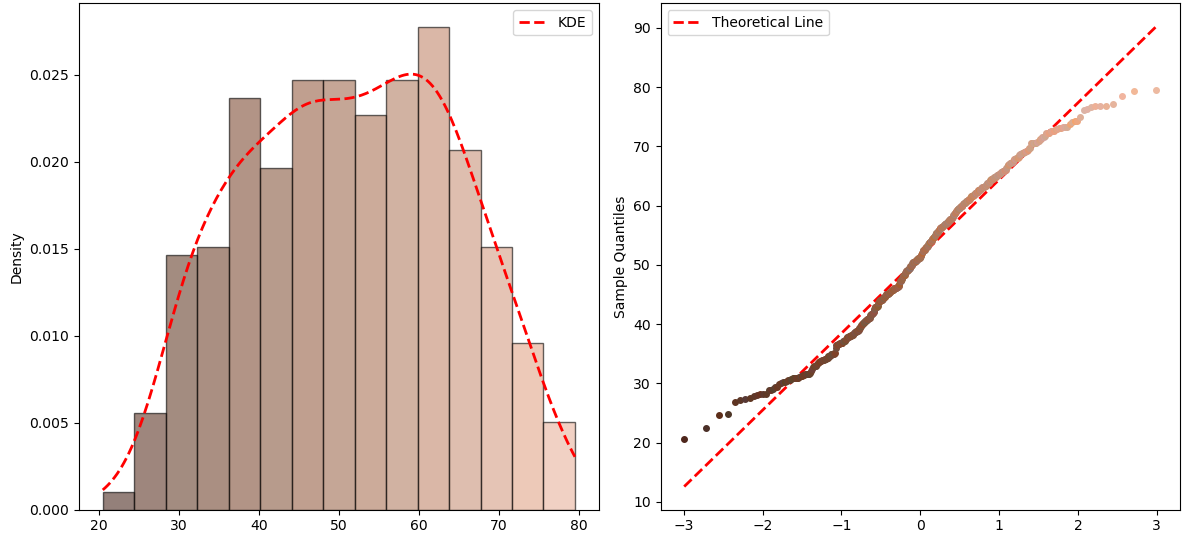}}\hfill
{\label{b}\includegraphics[width=.50\linewidth]{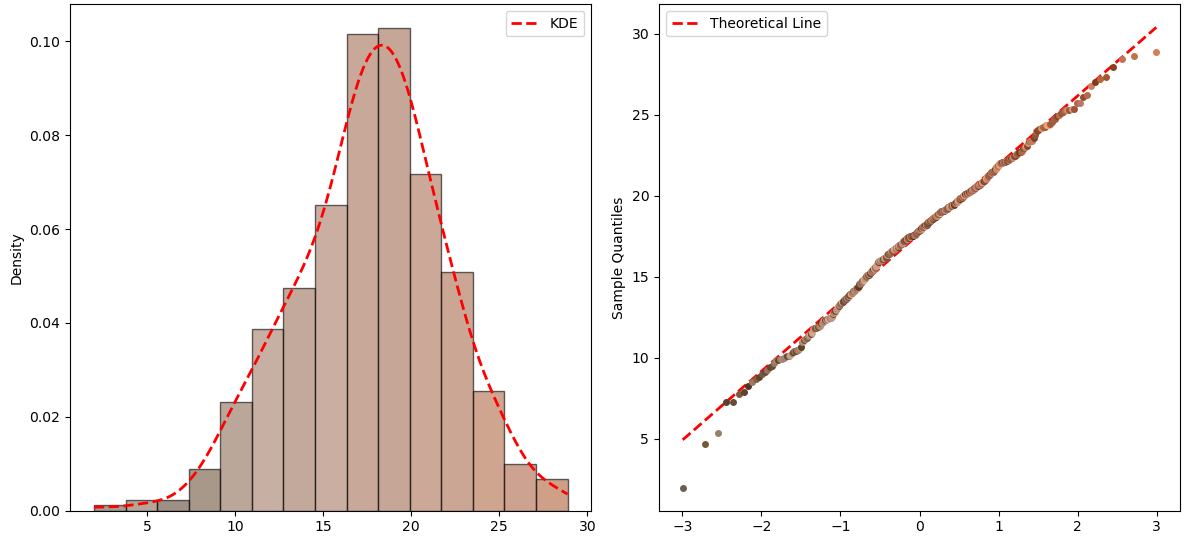}}\par 
{\label{c}\includegraphics[width=.50\linewidth]{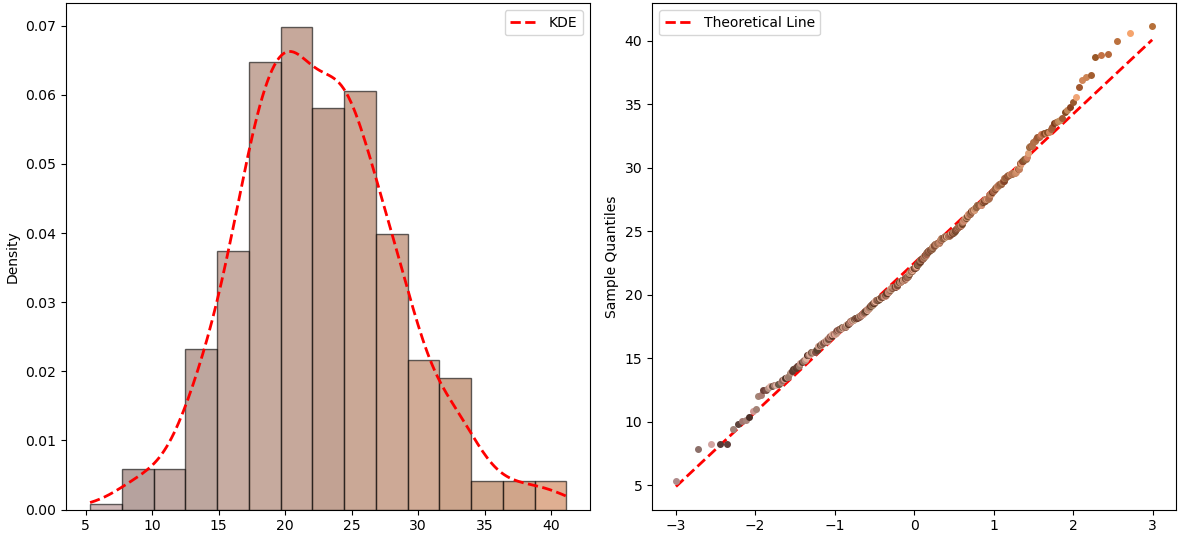}}
\captionsetup{justification=centering} 
\caption*{\justifying\scriptsize \textit{Notes:} This figure shows clockwise the normal distributions of the L*, a*, and b* values, respectively, obtained as described in \textcite{adukia2023}. It confirms that L* values go, from left to right, dark to light and that a* and b* do not necessarily correspond to skin tint as no obvious light to dark pattern arises.}
\label{fig:figlab}
\end{figure}

\begin{figure}
\centering
\captionsetup{width=1\linewidth}
\captionsetup{labelfont=normalfont,labelsep=space,justification=centering}
\caption{\\Image-to-data pipeline}\vspace{-0.8em}
\captionsetup{justification=raggedright, singlelinecheck=false}
\includegraphics[width=1\linewidth]{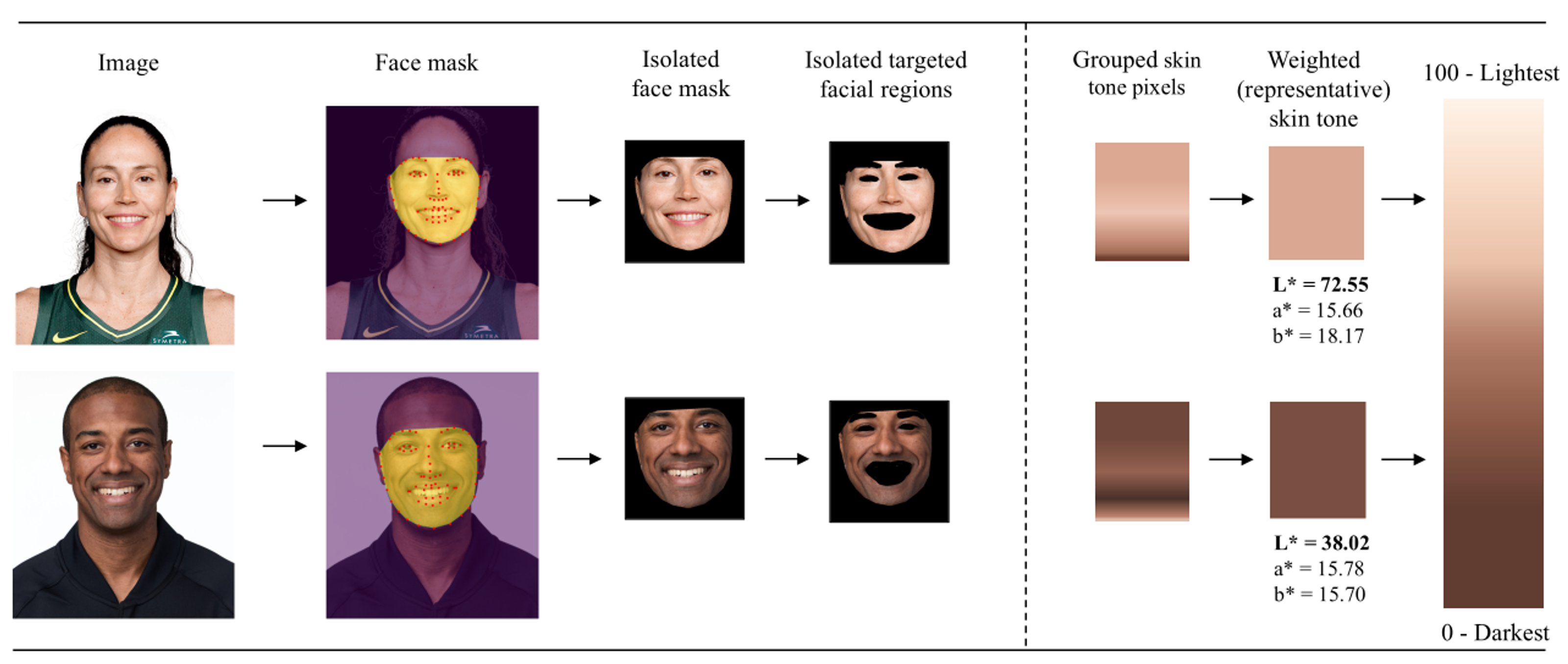}
\caption*{\justifying\scriptsize \textit{Notes:} This figure illustrates the steps of the skin tone analysis for a player (Sue Bird) and a referee (John Butler), divided into two main parts. 1) Face segmentation (\textit{left-hand side}): The face segmentation process begins by converting headshot images to the red-green-blue (RGB) format, before extracting a face mask using a Fully Connected Convolutional Neural Network (FC-CNN). After this, the FaceAlignment library is used to detect 68 specific facial landmarks, further refined using a Convex Hull method to isolate the targeted facial region. Finally, a Hue Saturation Value (HSV) color space-based filter is applied to ensure that only reliable skin pixels are retained, thus minimizing any lighting artifacts. 2) Skin tone classification (\textit{right-hand side}): We use k-means clustering to identify the 5 most dominant colors of the skin tone pixels and calculate the representative skin tone in the CIELAB (L*a*b*) color space using a weighted average of cluster sizes. For our further analysis, we focus on ``perceptual skin tint'' \parencite{adukia2023}, which is the L* value measured on a scale of 0–100, with 0 representing the darkest and 100 the lightest. This process is automated with Python, ensuring reproducibility and efficiency, with complete details provided in Appendix Section~\ref{amethods}.}
\label{fig:mlpipe}
\end{figure}

\begin{figure} [t]
\centering
\captionsetup{width=0.9\linewidth}
\captionsetup{labelfont=normalfont,labelsep=space,justification=centering}
\caption{\\Full sample WNBA faces sorted by L*}\vspace{-0.8em}
\captionsetup{justification=raggedright, singlelinecheck=false}
\includegraphics[width=0.9\linewidth]{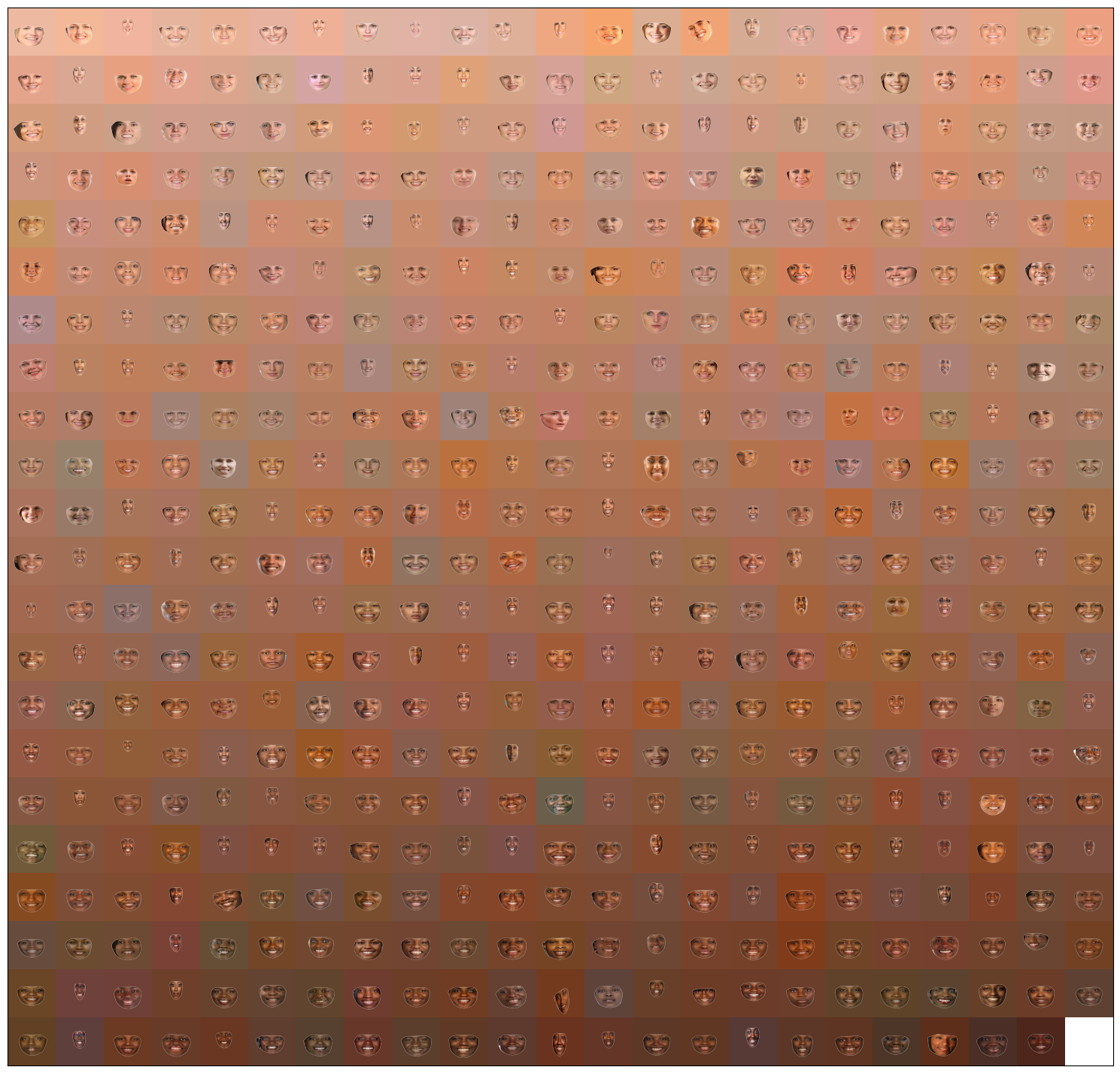}
\caption*{\justifying\scriptsize \textit{Notes:} The Image-to-Data pipeline was applied to all pictures in our sample and sorted by the skin tint (L*) from lightest (top left) to darkest (bottom right). Skin tint is on a scale from 0 being the darkest to 100 being the lightest. Faces are the face segmentation output described in the pipeline, and the background color of each box is based on the representative skin tone.}
\label{fig:facematrix}
\end{figure}

Next, we use k-means clustering to identify the 5 most dominant colors of the skin tone pixels and calculate the representative skin tone in the CIELAB (L*a*b*) color space using a weighted average of cluster sizes. In the CIELAB space, L* represents lightness or perceived brightness of a color, while a* and b* represent the red-green and the yellow-blue axes, respectively. All together create a three-dimensional space where each color occupies a unique position based on its L*, a*, and b* coordinates.\footnote{Note, this is a departure from the common way to measure skin tone using RGB values \parencite{robst2011, agha2024}, because the RGB color space does not reflect human perception when assessing skin color variation \parencite{thong2023}. Rather, RGB images combine color and intensity information for each pixel, making it challenging to create a robust method of detecting skin tones due to the impact of lighting and shadows on pixel intensity \parencite{chandrappa2011,shaik2015}.} Figure~\ref{fig:figlab} displays the distribution of these L*a*b* values across our dataset. For our analysis, we focus on ``perceptual skin tint'' \parencite{adukia2023}, which is the L* value measured on a scale of 0–100, with 0 representing the darkest and 100 the lightest. Figures~\ref{fig:mlpipe} and~\ref{fig:facematrix} illustrate the pipeline’s predictions on sample images and show the dataset sorted by L*. 

In total, our sample includes 505 players and 69 referees who participated in 2,359 games over 11 seasons. Table~\ref{table:samplesize} presents the sample size and the distribution of players by color according to the FairFace algorithm (Panel A) and human raters (Panel B) for the entire period (2004-2014). We provide separate tables of sample size for each investigated period (2004-2006, 2007-2010, and 2011-2014) in Appendix \Cref{table:samplesize0406,table:samplesize0710,table:samplesize1114}.

According to the FairFace algorithm, there are 335 Black and 170 non-Black players, as well as 23 Black and 46 non-Black referees. According to human raters, the numbers are similar, but not identical (335 Black and 170 non-Black players, as well as 27 Black and 42 non-Black referees). By looking at the distribution of perceptual skin tint values for Black and non-Black individuals, we see upper and lower bounds (min-max) as well as a considerable overlap between Black and non-Black individuals. Finally, we created a variable that represents the average absolute difference in skin tone (L*) between a player and each member of the three-person referee crew. As described in Panel C of Table~\ref{table:samplesize}, there were 1,876 unique crews with the mean absolute distance of 8.39 (sd = 4.19).

Table~\ref{table:plcharsff} presents the summary statistics for the different players, referees, and game characteristics along with a comparison of the mean values between Black and non-Black players. These summary statistics reveal significant differences between Black and non-Black players in minutes played, number of fouls, points, and different adjusted per 40 minutes characteristics that relate to players’ performance. We will return to these differences when discussing our empirical strategy. We also see differences in the referee crew composition for Black and non-Black players. In Appendix \Cref{table:plcharsff0406,table:plcharsff0710,table:plcharsff1114,table:plcharshr,table:plcharshr0406,table:plcharshr0710,table:plcharshr1114}, we present summary statistics for each investigated period (2004-2006, 2007-2010, and 2011-2014) per each race classification (FairFace and human raters). In general, the results presented in the tables for each investigated period show no significant pattern of associations between the race of players and the referee crew.

\begin{table}[H]
\centering
\captionsetup{
  justification=centering,
  labelsep=none,
  font=normalsize
}
\caption{\\Sample size}\vspace{-0.8em}
\label{table:samplesize}
\begin{threeparttable}
  \renewcommand{\arraystretch}{0.8}
  \small
  \begin{tabular}{lccc}
    \toprule
    & Black & Non-Black & Total \\
    \midrule
    \multicolumn{4}{c}{\textit{A. FairFace}} \\
    No.\ of players            & 335       & 170        & 505     \\
    \quad Mean L* value        & 45.00     & 64.06      & 51.42    \\
    \quad \quad (SD)           & (10.39)   & (7.14)     & (13.03)  \\
    \quad \quad Min–max        & 20.52–71.26 & 41.85–79.48 & 20.52–79.48 \\
    \addlinespace
    No.\ of referees           & 23        & 46         & 69      \\
    \quad Mean L* value        & 42.49     & 59.77      & 54.01    \\
    \quad \quad (SD)           & (10.62)   & (9.09)     & (12.59)  \\
    \quad \quad Min–max        & 20.56–60.75 & 41.77–76.32 & 20.56–76.32 \\
    \addlinespace
    No.\ of games              & 2,359     & 2,356      & 2,359    \\
    No.\ of player–games       & 33,606    & 12,272     & 45,878   \\
    No.\ of player–minutes     & 707,374   & 245,560    & 952,934  \\
    \addlinespace
    \multicolumn{4}{c}{\textit{B. Human raters}} \\
    No.\ of players            & 335       & 170        & 505      \\
    \quad Mean L* value        & 45.04     & 64.00      & 51.42    \\
    \quad \quad (SD)           & (10.43)   & (7.21)     & (13.03)  \\
    \quad \quad Min–max        & 20.52–71.26 & 41.85–79.48 & 20.52–79.48 \\
    \addlinespace
    No.\ of referees           & 27        & 42         & 69      \\
    \quad Mean L* value        & 43.94     & 60.48      & 54.01    \\
    \quad \quad (SD)           & (10.83)   & (8.89)     & (12.59)  \\
    \quad \quad Min–max        & 20.56–63.32 & 43.21–76.32 & 20.56–76.32 \\
    \addlinespace
    No.\ of games              & 2,359     & 2,354      & 2,359    \\
    No.\ of player–games       & 33,527    & 12,351     & 45,878   \\
    No.\ of player–minutes     & 704,010   & 248,924    & 952,934  \\
    \addlinespace
    \multicolumn{4}{c}{\textit{C. Referee crew compositions}} \\
    No.\ of unique crews       &           &            & 1,876    \\
    \quad Mean abs.\ L* diff.  &           &            & 8.39     \\
    \quad \quad (SD)           &           &            & (4.19)   \\
    \quad \quad Min–max        &           &            & 0.414–33.46 \\
    \bottomrule
  \end{tabular}
  \begin{tablenotes}[para,flushleft]
    \setstretch{0.5}\tiny
    \setlength{\parskip}{0pt}
    \setlength{\parindent}{0pt}
    \setlength{\itemsep}{-1pt}
    \item \scriptsize \textit{Notes:} L* values represent skin tone measurements using the CIELAB color space. The conditional descriptives are based on the racial classifications made by FairFace predictions and human raters as discussed in Section \ref{data}.
  \end{tablenotes}
\end{threeparttable}
\end{table}

\begin{table}[H]
\centering
\captionsetup{
  justification=centering,
  labelsep=none,
  font=normalsize
}
\caption{\\Summary statistics}\vspace{-0.8em}
\label{table:plcharsff}
\begin{threeparttable}
\renewcommand{\arraystretch}{0.8}
\small
\begin{tabular}{lccc}
\toprule\toprule
& \multicolumn{1}{c}{Black players} & \multicolumn{1}{c}{Non-Black players} & \multicolumn{1}{c}{} \\
\cmidrule(lr){2-2} \cmidrule(lr){3-3}
& \multicolumn{1}{c}{\makecell{Mean \\ (SD)}} & \multicolumn{1}{c}{\makecell{Mean \\ (SD)}} & \multicolumn{1}{c}{Difference} \\
\midrule
\multicolumn{4}{c}{Raw player statistics} \\
Minutes played                    & 26.08  & 25.67  & 0.41\tnote{***} \\
                                  & (8.66) & (8.96) & \\
Fouls                             & 2.34   & 2.12   & 0.22\tnote{***} \\
                                  & (1.49) & (1.50) & \\
Points                            & 10.38  & 9.72   & 0.65\tnote{***} \\
                                  & (6.87) & (7.21) & \\
\multicolumn{4}{c}{Player productivity: stats × 40/minutes played} \\
Fouls                             & 3.87   & 3.56   & 0.31\tnote{***} \\
                                  & (2.98) & (3.03) & \\
Points                            & 15.15  & 14.21  & 0.93\tnote{***} \\
                                  & (8.45) & (8.92) & \\
Free throws made                  & 3.08   & 2.69   & 0.39\tnote{***} \\
                                  & (3.35) & (3.30) & \\
Free throws missed                & 1.01   & 0.67   & 0.34\tnote{***} \\
                                  & (1.68) & (1.35) & \\
2 point goals made                & 4.72   & 3.62   & 1.11\tnote{***} \\
                                  & (3.32) & (2.97) & \\
2 point goals missed              & 5.71   & 4.28   & 1.43\tnote{***} \\
                                  & (3.65) & (3.20) & \\                                                                      
3 point goals made                & 0.88   & 1.43   & -0.56\tnote{***} \\
                                  & (1.50) & (1.91) & \\
3 point goals missed              & 1.75   & 2.56   & -0.81\tnote{***} \\
                                  & (2.29) & (2.70) & \\
Offensive rebounds                & 2.11   & 1.52   & 0.59\tnote{***} \\
                                  & (2.38) & (2.05) & \\
Defensive rebounds                & 4.73   & 4.31   & 0.43\tnote{***} \\
                                  & (3.43) & (3.36) & \\
Assists                           & 3.04   & 3.77   & -0.74\tnote{***} \\
                                  & (2.79) & (3.21) & \\
Steals                            & 1.61   & 1.47   & 0.14\tnote{***} \\
                                  & (1.82) & (1.75) & \\
Blocks                            & 0.76   & 0.72   & 0.04\tnote{**} \\
                                  & (1.40) & (1.45) & \\
Turnovers                         & 2.90   & 2.83   & 0.06\tnote{***} \\
                                  & (2.50) & (2.55) & \\
\multicolumn{4}{c}{Game information} \\
Attendance (1,000s)               & 7.83   & 7.80   & 0.03 \\
                                  & (2.47) & (2.41) & \\
Out of contention                 & 0.03   & 0.03   & 0.01\tnote{***} \\
                                  & (0.18) & (0.16) & \\
\bottomrule
\end{tabular}
\end{threeparttable}
\end{table}

\begin{table}[H]
\ContinuedFloat
\centering
\captionsetup{
labelformat=empty,
labelsep=none,
justification=centering,
font=normalsize
}
\caption{\\(Continued)}\vspace{-0.8em}
\label{table:plcharsff}
\begin{threeparttable}
\renewcommand{\arraystretch}{0.8}
\small
\begin{tabular}{lccc}
\toprule\toprule
& \multicolumn{1}{c}{Black players} & \multicolumn{1}{c}{Non-Black players} & \multicolumn{1}{c}{} \\
\cmidrule(lr){2-2} \cmidrule(lr){3-3}
& \multicolumn{1}{c}{\makecell{Mean \\ (SD)}} & \multicolumn{1}{c}{\makecell{Mean \\ (SD)}} & \multicolumn{1}{c}{Difference} \\
\midrule
\multicolumn{4}{c}{Player characteristics} \\
Age                               & 27.37   & 27.75  & -0.38 \\
                                  & (3.98)  & (3.98) & \\
WNBA experience (yrs)             & 5.44    & 5.93   & -0.49\tnote{**} \\
                                  & (3.28)  & (3.62) & \\
All-WNBA this year                & 0.11    & 0.15   & -0.05\tnote{*} \\
                                  & (0.31)  & (0.36) & \\
Center                            & 0.15    & 0.11   & 0.03 \\
                                  & (0.35)  & (0.31) & \\
Forward                           & 0.43    & 0.32   & 0.11 \\
                                  & (0.50)  & (0.47) & \\
Guard                             & 0.42    & 0.57   & -0.15\tnote{*} \\
                                  & (0.49)  & (0.50) & \\
Starter                           & 0.70    & 0.69   & 0.01\tnote{*} \\
                                  & (0.46)  & (0.46) & \\
Height (cm.)                      & 183.40  & 183.40 & -0.00 \\
                                  & (8.42)  & (9.52) & \\
Weight (kg.)                      & 77.20   & 76.26  & 0.95 \\
                                  & (10.44) & (10.62)& \\
\multicolumn{4}{c}{Referees} \\
0 non-Black referees                  & 0.07    & 0.05   & 0.01\tnote{***} \\
                                  & (0.25)  & (0.22) & \\
1 non-Black referees                  & 0.28    & 0.27   & 0.01\tnote{**} \\
                                  & (0.45)  & (0.44) & \\
2 non-Black referees                  & 0.42    & 0.43   & -0.01 \\
                                  & (0.49)  & (0.50) & \\
3 non-Black referees                  & 0.23    & 0.25   & -0.02\tnote{***} \\
                                  & (0.42)  & (0.43) & \\
\text{\#} non-Black referees          & 1.82    & 1.87   & -0.05 \\
                                  & (0.86)  & (0.84) & \\
\bottomrule
\bottomrule
\end{tabular}
\begin{tablenotes}[para,flushleft]
\scriptsize
\item Racial classifications determined by FairFace. 
Human rater classifications are included in the appendix. 
All observations are weighted by minutes played. \textsuperscript{***}p \textless 0.01; \textsuperscript{**}p \textless 0.05; \textsuperscript{*}p \textless 0.1.
\end{tablenotes}
\end{threeparttable}
\end{table}

\subsection{Empirical strategy} \label{empirical}
Studying the effect of the difference between a player and a referee crew's race or color combination on the number of fouls is challenging. A naïve approach of correlating the number of fouls and the difference between a player and a referee crew's race or color would yield biased and inconsistent estimates. This is because players’ characteristics (e.g., defensive and offensive abilities, position, superstar status, experience, etc.) vary between Black and non-Black players (see Table~\ref{table:plcharsff}). Moreover, the individual ability and the status of players within a team may vary over time.\footnote{See \textcite{gschwend2021} for the discussion on change in ability of professional athletes throughout their careers and the use of fixed effects over long versus short periods.} As such, for identifying the effects of interest, different sources of unobserved heterogeneity need to be taken into account.

Our data follows the same players over time, which allows us to use different fixed effects. More precisely, we use player-year fixed effects that allow us to control for any time-invariant player characteristics that are constant within a year (season). Moreover, our data allows us to control for all the game characteristics that are constant within the game, such as stadium attendance or a certain match-up between the teams.

In addition, it is important to refer to the referee selection process. The advantages of our setting lie in (i) the process of assigning referees to games, which does not consider player race or skin tone (see Table~\ref{table:blackstartersperteamAI}), thereby ensuring that our findings are not confounded by subjects sorting to preferred evaluators or vice versa, and (ii) repeated interactions bypassing selection issues. Accordingly, the most basic specification of our model takes the following form:
\setcounter{equation}{0} 
\begin{equation} \label{eq:1}
    \text{Foul rate}_{itgy} = \alpha \text{Black player}_{i} \times \% \text{non-Black referees}_{g} + \theta_{iy} + \gamma_{g} + 
    \textbf{$\rho^{\prime}$}\textbf{X}_{igt} + \epsilon_{itgy},
\end{equation}
\noindent where the dependent variable is the number of fouls earned per 40 minutes (40×fouls/minutes played) by player $i$ from team $t$, in a specific game $g$, in year $y$. The coefficient of interest, $\alpha$, is for the interaction variable between an indicator of whether player $i$ is Black and the fraction of non-Black referees in a specific game $g$. We interpret it as the differential impact of the racial composition of the refereeing crew on Black players relative to non-Black players. The coefficient is identified from the variation in referee crew composition between games. 

$\theta_{iy}$ denotes player-year fixed effects, controlling for time-invariant player characteristics that are constant within a year (season), and $\gamma_{g}$ denotes game fixed effects, controlling for game-invariant characteristics (e.g., referee crew, attendance, and teams' performance up until the specific game). $\textbf{X}_{igt}$ is our vector of controls, which includes a dummy variable indicating whether a player is a starter, whether her team played the game at home, and the race of her coach.\footnote{While \textcite{zhang2017} showed a positive relationship between a player’s playing time and the similarity between the race of the player and the coach in the NBA, \textcite{agha2024} found no such relationship in the WNBA for the 2018-2022 seasons.} All estimates are weighted by the number of minutes played, and standard errors are clustered at the player and game level.\footnote{In contrast to \textcite{price2010}, we opt for two-way clustering at the player and game levels because of possible dependency between the behavior of players and referee crews within a game \parencite{abadie2023}.}

To explore potential colorism in officiating, we mimic the approach described above. Now our variable of interest is \textit{CrewDistance}, a measure calculated as the average absolute difference in skin tone (L* value) between a player and each member of the three-person referee crew. Thus, $\beta$ represents the effect of the average skin tone difference between referees and players on the player’s foul rate:
\begin{equation} \label{eq:2}
    \text{Foul rate}_{itgy} = \beta \; \text{CrewDistance}_{ig} + \theta_{iy} + \gamma_{g} + 
    \textbf{$\rho^{\prime}$}\textbf{X}_{igt} + \epsilon_{itgy},
\end{equation}
\noindent Based on \textcite{ppw2018}, we utilize a natural experiment arising from the widespread publicity of \textcite{price2010}, which served as an external shock intensifying scrutiny around racial bias in men’s basketball officiating. This event potentially disrupted established norms of referee discrimination in the NBA, prompting an investigation into whether similar changes occurred in the women’s context. We divide our analysis into three periods: a baseline period (2004-2006) before media exposure, a treated period (2007-2010) immediately following increased public awareness, and an extended period (2011–2014) to examine the persistence or evolution of any observed effects. We combine the relevant periods to run a pooled regression, extending equations (\ref{eq:1}) and (\ref{eq:2}) by adding an interaction term between a relevant subsequent period (i.e., 2004-2006 along with 2007-2010 or 2007-2010 along with 2011-2014) and our variables of interest $\text{Black player}_{i} \times \% \text{non-Black referees}_{g}$ and $\text{CrewDistance}_{ig}$, respectively.

\section{Results} \label{results}

\subsection{Binary race measure} \label{binary}
We start our analysis using the binary race classification of Black or non-Black as determined by the FairFace algorithm. Recall, a positive (negative) sign of $\hat{\alpha}$ indicates that players are called for more (fewer) fouls, the higher the share of referees with unlike race classifications. As such, and in line with \textcite{price2010} a positive (negative) estimate would be suggestive of racial discrimination (favoritism) in foul calling. In their main specification, \textcite{price2010} report a significant positive estimate of 0.181 (p $<$ 0.05) (see Table 4 in their paper). 

As can be seen in Column 1 of Table~\ref{table:mlrace}, we do not find any significant evidence of racial discrimination in WNBA foul calling for the period before media exposure (pre-awareness period between 2004 and 2006). Compared with the NBA estimate for the same period, our WNBA estimate is small and not significant ($\hat{\alpha}$ = 0.077, p = 0.713). Interestingly, however, in Column 2, we see a negative estimate of -0.183 (p = 0.205) for the immediate period following media exposure (2007-2010). Notably, even though not significant at conventional levels, this estimate is much larger in absolute value and more precise compared to the pre-awareness period. Content-wise, this is in line with overcorrection as discussed before. Finally, in Column 3, we see again a positive but not significant estimate for the extended period following media exposure (2011-2014) ($\hat{\alpha}$ = 0.212, p = 0.199). While not significant at conventional levels, this result suggests an overreaction in the opposite direction. We will come back to this result in the next sub-section when estimating the continuous skin tone measure to see whether this result is robust or not.

\begin{table}[H]
\centering
\captionsetup{
    justification=centering, 
    labelsep=none,           
    font=normalsize          
}
\caption{\\Racial bias among WNBA referees using FairFace}\vspace{-0.8em}
\label{table:mlrace}
\begin{threeparttable}
\renewcommand{\arraystretch}{0.8} 
\footnotesize 
\begin{tabular}{lccccc} 
\toprule
\toprule
 & Pre-awareness & \multicolumn{2}{c}{Post-awareness} & \multicolumn{2}{c}{Change in coefficient} \\
\cmidrule(r){2-2} \cmidrule(r){3-4} \cmidrule(r){5-6}
 & 2004--2006 & 2007--2010 & 2011--2014 & \makecell{From 2004--2006\\to 2007--2010} & \makecell{From 2007--2010\\to 2011--2014} \\
\midrule
Black \(\times\) & 0.077 & -0.183 & 0.212 & 0.076 & -0.182 \\
fraction non-Black referees & \small{(0.206)} & \small{(0.144)} & \small{(0.165)} & \small{(0.205)} & \small{(0.144)} \\
\addlinespace
Post Black \(\times\) &  &  &  & -0.258 & 0.393\textsuperscript{*} \\
fraction non-Black referees &  &  &  & \small{(0.262)} & \small{(0.206)} \\
\addlinespace
N & 12,893 & 17,266 & 15,705 & 30,159 & 32,971 \\
Sample mean & 4.38 & 4.56 & 4.09 & 4.48 & 4.33 \\
\bottomrule
\bottomrule
\end{tabular}
\begin{tablenotes}[para,flushleft]
\setstretch{0.5} 
\tiny
\setlength{\parskip}{0pt} 
\setlength{\parindent}{0pt} 
\setlength{\itemsep}{-1pt} 
\item \scriptsize \textit{Notes:} Dependent variable is defined as 40×fouls/minutes. Each regression includes player-year fixed effects, game fixed effects, whether a player is a starter, whether a player is playing in her home arena, and the skin color of her coach. The last two columns report the results of a pooled regression, including both a relevant primary period and a subsequent period to measure the change in our variable of interest between the two periods. Each observation is weighted by the number of minutes played. Standard errors are clustered at the player and game levels and appear in parentheses. \textsuperscript{***}p \textless 0.01; \textsuperscript{**}p \textless 0.05; \textsuperscript{*}p \textless 0.1. 
\end{tablenotes}
\end{threeparttable}
\end{table}

Overall, when analyzing the pooled data combining the periods between 2004 and 2010 as well as between 2007 and 2014, these temporal patterns remain: $\hat{\alpha}$ is lower in the 2007-2010 period than in the 2004-2006 period (see Column 4) and in the 2011-2014 period (see Column 5). The latter finding is significant at the 10\%-level.

Using race categorizations from human raters instead of FairFace, we confirm the general patterns from before (see Table~\ref{table:hrrace}). However, while $\hat{\alpha}$ is still neither significant in the pre-awareness period ($\hat{\alpha}$ = 0.106, p = 0.541, see Column 1 ) nor in the extended period following media exposure (i.e., 2011-2014 with $\hat{\alpha}$ = 0.218, p = 0.136, see Column 3), the estimate in the 2007–2010 post-awareness period is precise enough to become significant at conventional levels ($\hat{\alpha}$ = -0.258, p = 0.046, see Column 2). Again, when analyzing the pooled data combining the periods between 2004 and 2010 as well as between 2007 and 2014, these temporal patterns remain: $\hat{\alpha}$ is lower in the 2007-2010 period than in the 2004-2006 period (see Column 4) and in the 2011-2014 period (see Column 5). These findings are significant at 10\% and 5\% levels, respectively.

\begin{table}[H]
\centering
\captionsetup{
    justification=centering, 
    labelsep=none,           
    font=normalsize          
}
\caption{\\Racial bias among WNBA referees using human raters}\vspace{-0.8em}
\label{table:hrrace}
\begin{threeparttable}
\renewcommand{\arraystretch}{0.8} 
\footnotesize 
\begin{tabular}{lccccc} 
\toprule
\toprule
 & Pre-awareness & \multicolumn{2}{c}{Post-awareness} & \multicolumn{2}{c}{Change in coefficient} \\
\cmidrule(r){2-2} \cmidrule(r){3-4} \cmidrule(r){5-6}
 & 2004--2006 & 2007--2010 & 2011--2014 & \makecell{From 2004--2006\\to 2007--2010} & \makecell{From 2007--2010\\to 2011--2014} \\
\midrule
Black \(\times\) & 0.106 & -0.258\textsuperscript{**} & 0.218 & 0.108 & -0.257\textsuperscript{**} \\
fraction non-Black referees & \small{(0.172)} & \small{(0.129)} & \small{(0.146)} & \small{(0.172)} & \small{(0.129)} \\
\addlinespace
Post Black \(\times\) &  &  &  & -0.365\textsuperscript{*} & 0.474\textsuperscript{**} \\
fraction non-Black referees &  &  &  & \small{(0.220)} & \small{(0.186)} \\
\addlinespace
N & 12,893 & 17,266 & 15,705 & 30,159 & 32,971 \\
Sample mean & 4.38 & 4.56 & 4.09 & 4.48 & 4.33 \\
\bottomrule
\bottomrule
\end{tabular}
\begin{tablenotes}[para,flushleft]
\setstretch{0.5} 
\tiny
\setlength{\parskip}{0pt} 
\setlength{\parindent}{0pt} 
\setlength{\itemsep}{-1pt} 
\item \scriptsize \textit{Notes:} Dependent variable is defined as 40×fouls/minutes. Each regression includes player-year fixed effects, game fixed effects, whether a player is a starter, whether a player is playing in her home arena, and the skin color of her coach. The last two columns report the results of a pooled regression, including both a relevant primary period and a subsequent period to measure the change in our variable of interest between the two periods. Each observation is weighted by the number of minutes played. Standard errors are clustered at the player and game levels and appear in parentheses.\\
\textsuperscript{***}p \textless 0.01; \textsuperscript{**}p \textless 0.05; \textsuperscript{*}p \textless 0.1.
\end{tablenotes}
\end{threeparttable}
\end{table}

Using a reduced dataset with a sample that only includes observations where both FairFace and human raters agreed on classifications, we find the same temporal patterns as before but (slightly) increased effect sizes (see Table~\ref{table:mlhragree} for more details). Moreover, we test whether referee decisions depend on collective appearance, i.e., a player’s race relative to the racial mix of both competing teams. Our findings hold even after controlling for the racial gap between competing teams (see Appendix Tables~\ref{table:app_race_subsamplesff} and \ref{table:app_race_subsampleshr}). Another possible issue could be with the definitions of the periods. For example, \textcite{ppw2018} used a three-year pre-awareness period and a four-year post-awareness period. To ensure this uneven period length did not drive our results, we re-evaluated foul call trends by dividing the years following awareness into consecutive three-year segments: 2004-2006, 2007-2009, and 2010-2012. Our analysis, presented in Appendix Tables~\ref{table:mlrace333} and \ref{table:hurace333}, demonstrates that, overall, our results hold consistently regardless of whether a 3-year or 4-year post-awareness evaluation period is used.

Summing up, even though several estimates are not precise enough to become significant at conventional levels, the temporal patterns and general trends are consistent throughout: while there is no statistically significant evidence of racial discrimination in WNBA foul calling for the period before media exposure, we do observe a tendency for overcorrection in the period following media exposure, as well as a rebound effect in foul-calling patterns for the extended period following media exposure.

\subsection{Continuous skin tone measure} \label{tone}
In Table~\ref{table:colorsub}, we present the results for equation (\ref{eq:2}) using a continuous skin tone measure. As for the binary race measures, our results using skin tone for the pre-awareness period indicate that no discrimination exists ($\hat{\beta}$ = -0.000, p = 0.963; see column 1). However, in Column 2, we observe a statistically significant negative coefficient that emerges in the immediate post-awareness period ($\hat{\beta}$ = -0.020, p = 0.003), suggesting that there is evidence of overcorrection immediately after widespread media attention, also when using a continuous skin tone measure. 

\begin{table}[H]
\centering
\captionsetup{
    justification=centering, 
    labelsep=none, 
    font=normalsize 
}
\caption{\\Colorism among WNBA referees}\vspace{-0.8em}
\label{table:colorsub}
\begin{threeparttable}
\renewcommand{\arraystretch}{0.8} 
\small 
\begin{tabular}{lccccc}        
\toprule
\toprule
 & Pre-awareness & \multicolumn{2}{c}{Post-awareness} & \multicolumn{2}{c}{Change in coefficient} \\
\cmidrule(r){2-2} \cmidrule(r){3-4} \cmidrule(r){5-6}
 & 2004--2006 & 2007--2010 & 2011--2014 & \makecell{From 2004--2006\\to 2007--2010} & \makecell{From 2007--2010\\to 2011--2014} \\
\midrule
Crew distance & -0.000 & -0.020\textsuperscript{***} & 0.006 & -0.000 & -0.020\textsuperscript{***} \\
 & \small{(0.006)} & \small{(0.007)} & \small{(0.007)} & \small{(0.006)} & \small{(0.007)} \\
\addlinespace
Post crew distance &  &  &  & -0.019\textsuperscript{**} & 0.026\textsuperscript{***} \\
 &  &  &  & \small{(0.009)} & \small{(0.010)} \\
\addlinespace
N & 12,893 & 17,266 & 15,705 & 30,159 & 32,971 \\
Sample mean & 4.38 & 4.56 & 4.09 & 4.48 & 4.33 \\
\bottomrule
\bottomrule
\end{tabular}
\begin{tablenotes}[para,flushleft]
\setstretch{0.5} 
\tiny
\setlength{\parskip}{0pt} 
\setlength{\parindent}{0pt} 
\setlength{\itemsep}{-1pt} 
\item \scriptsize \textit{Notes:} Dependent variable is defined as 40×fouls/minutes. Crew distance is the mean of the absolute distance in skin tone values between the player and each referee in the officiating crew. Each regression includes player-year fixed effects, game fixed effects, whether a player is a starter, whether a player is playing in her home arena, and her coach's skin tone. The last two columns report the results of a pooled regression, including both a primary period and a subsequent period to measure the changes throughout our main sample years. Each observation is weighted by the number of minutes played. Standard errors are clustered at the player and game levels and appear in parentheses.\\
\textsuperscript{***}p \textless 0.01; \textsuperscript{**}p \textless 0.05; \textsuperscript{*}p \textless 0.1.
\end{tablenotes}
\end{threeparttable}
\end{table}

To put this result into perspective, a one standard deviation change in the difference between referees and players L* (sd = 4.27) corresponds to 0.085 fewer fouls per 40 minutes played. In Column 4, we also see that there is a significant difference in $\hat{\beta}$ in the 2007-2010 period compared to the 2004-2006 period (\textit{Post crew distance} = -0.019, p = 0.030). Finally, the extension to 2014, presented in Column 5, shows a significant increase in $\hat{\beta}$ in the 2011-2014 period compared to the 2007-2010 period (\textit{Post crew distance} = 0.026, p = 0.009) back to near pre-awareness levels, as seen in Column 3. Overall, these findings mimic our findings using the binary race categories, though the estimates using the continuous skin tone measure are comparably more precise to become significant even at the 1-percent level.\footnote{In addition, we explicitly test for potential team-level bias in racial composition by adding team-year fixed effects into our estimations. This accounts for the varying share of black players on rosters and minutes played for a specific team in a given year. Including team-year fixed effects did not alter our results. The results of these estimations are available upon request.} This holds even when accounting for the skin tone composition of the teams (Appendix Table~\ref{table:app_tone_subsamples}, Panel B). Finally, it is important to note that the pattern of overcorrection in the 2007-2010 period remains the same in estimations with binary race and continuous skin tone measures. However, the null effect in the 2011-2014 period (Column 3 in Table~\ref{table:colorsub}) suggests that the large overcorrection in the opposite direction, as was observed with the binary race measure, is not replicable with the continuous skin tone measure.

\subsection{In-group discrimination versus out-group favoritism} \label{group}
An often-discussed issue in empirical studies of preferences and bias is the inability to disentangle in-group versus out-group bias \parencite{feld2016, sandberg2018}. The resultant net effect may be the same, but the mechanism is different. For example, \textcite{feld2016} distinguish between ``exophilic'' case, where agents prefer members of other groups, and ``endophobic'' case, where agents discriminate against people like themselves. In our setting, we are able, at least to some extent, to differentiate between in- and out-group bias by adding the quadratic transformation of the CrewDistance variable in equation (\ref{eq:2}). The intuition of this approach is as follows: by exploring any non-linearity between foul rate and \textit{CrewDistance}, we can identify whether the effect occurs at small(er) distances, involving in-group members (i.e., crews and players of similar skin tone), or at large(r) distances, involving out-group members (i.e., crews and players of different skin tone), or both.

\begin{figure}[h]
\centering
\captionsetup{labelfont=normalfont,labelsep=none,justification=centering}
\caption{\\Locally weighted smoothing and marginal effects of crew distance on foul rate}\vspace{-0.8em}
\captionsetup{justification=raggedright, singlelinecheck=false}
\subfloat{\includegraphics[width=.33\linewidth]{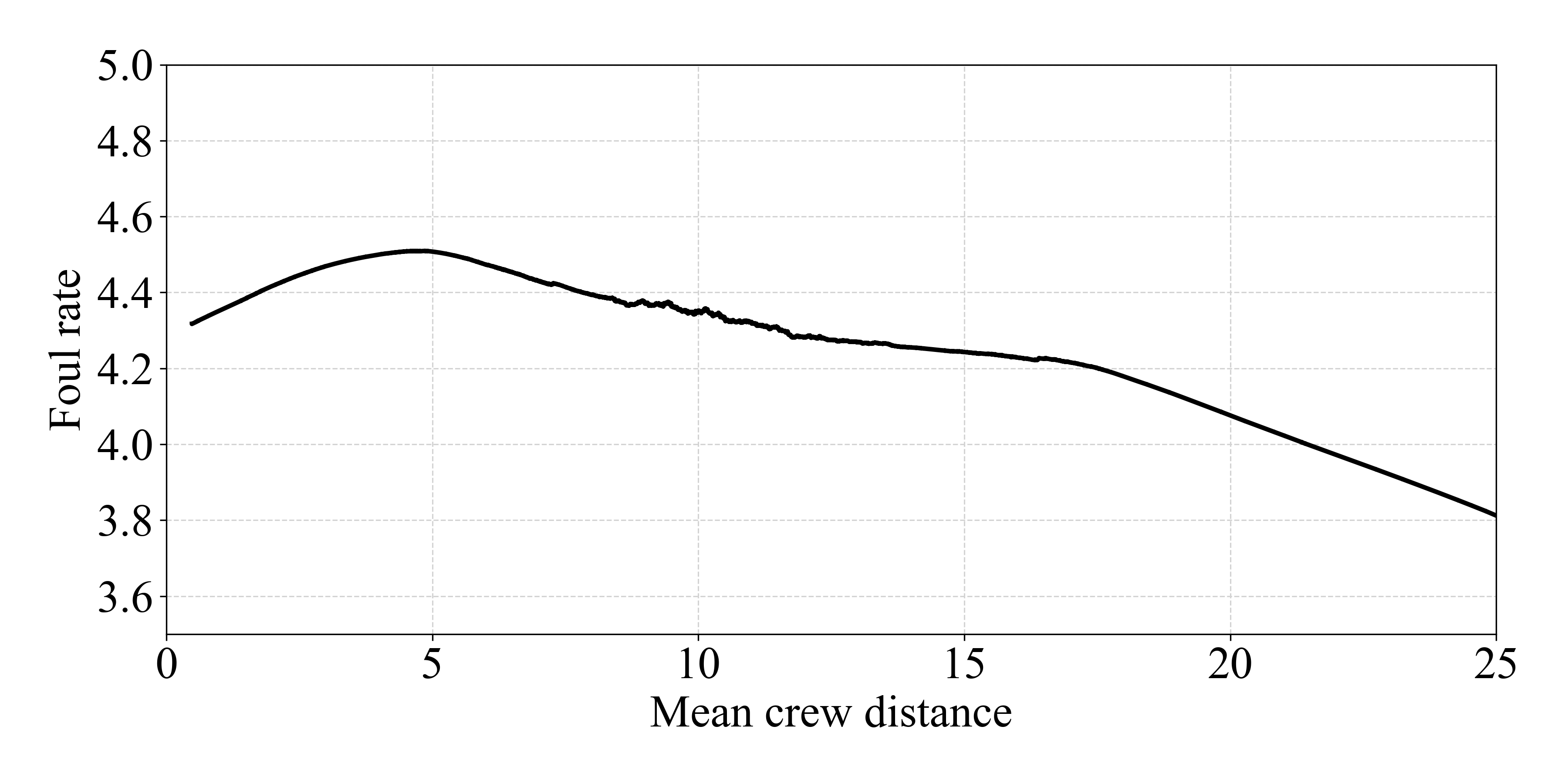}}\hfill
\subfloat{\includegraphics[width=.33\linewidth]{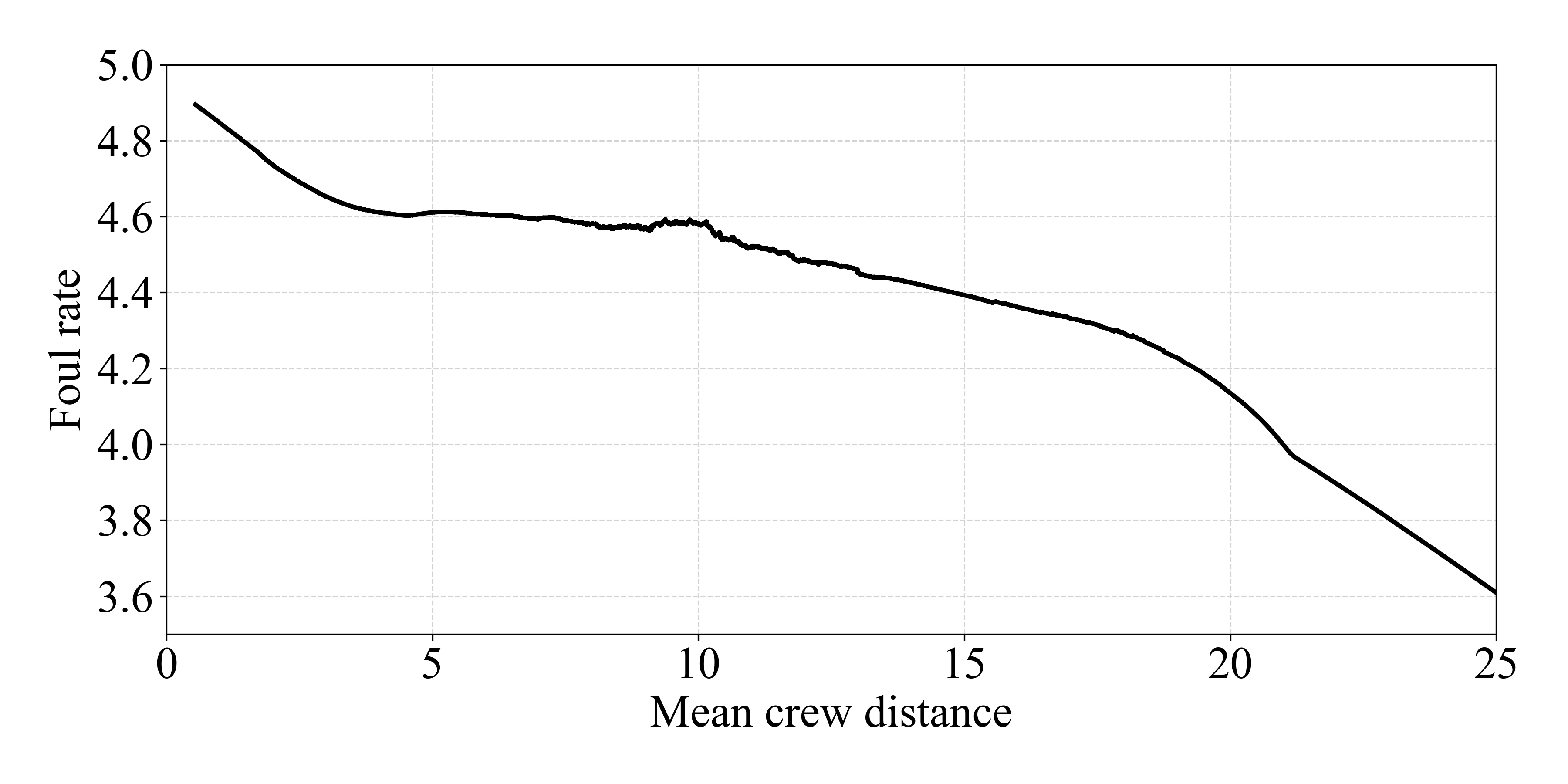}}\hfill
\subfloat{\includegraphics[width=.33\linewidth]{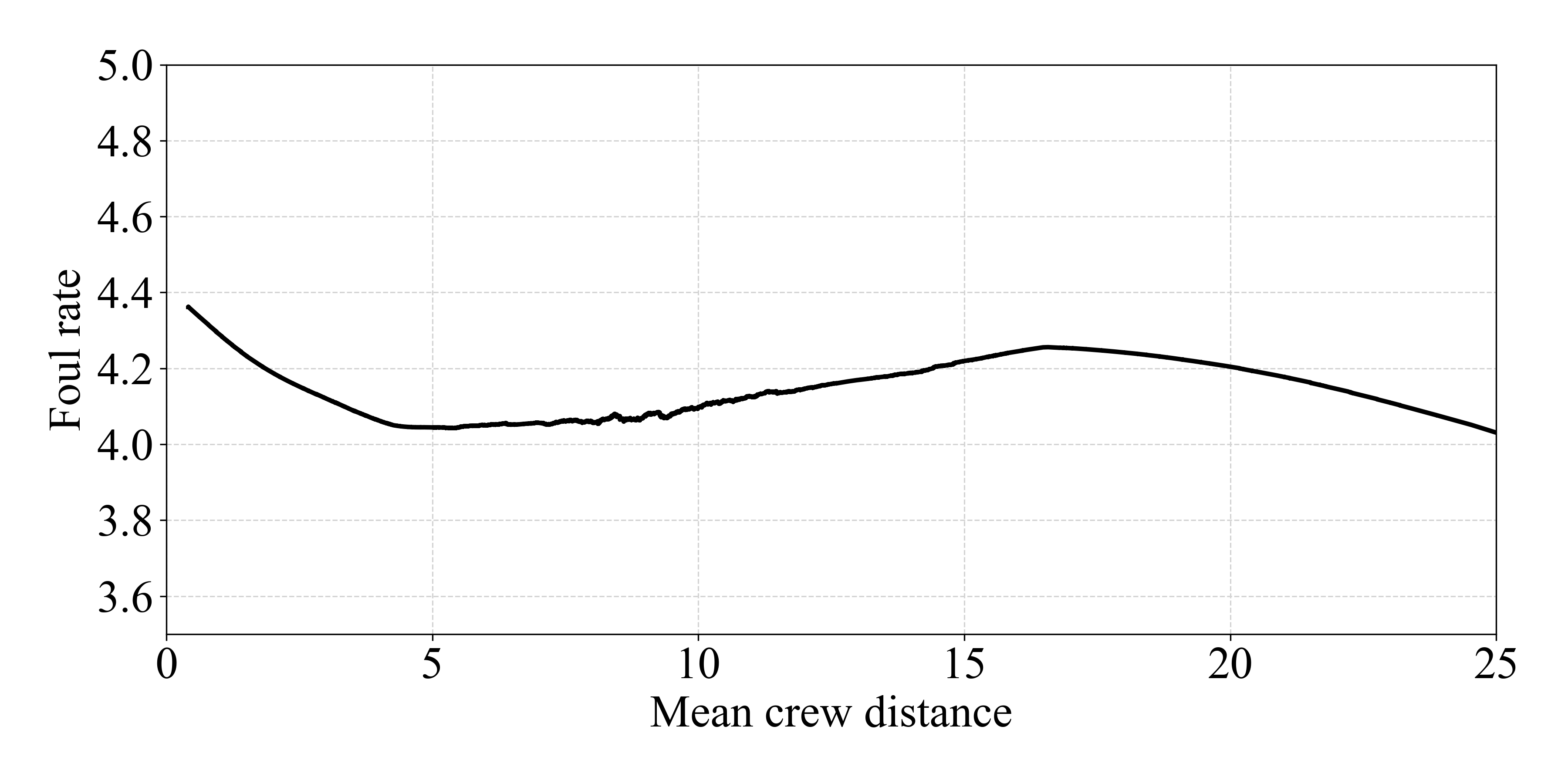}}\\[-1em]
\subfloat{\includegraphics[width=.33\linewidth]{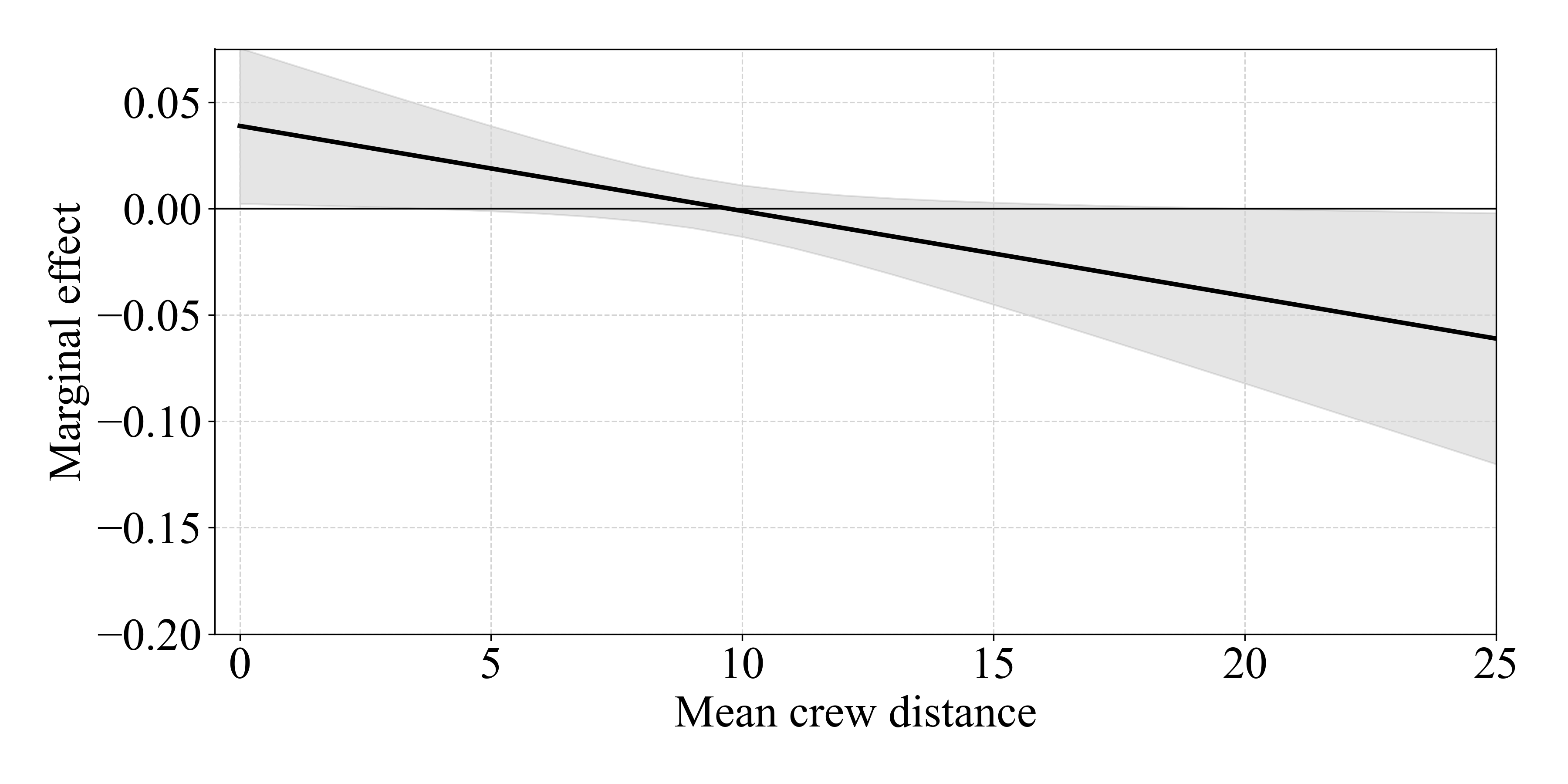}}\hfill
\subfloat{\includegraphics[width=.33\linewidth]{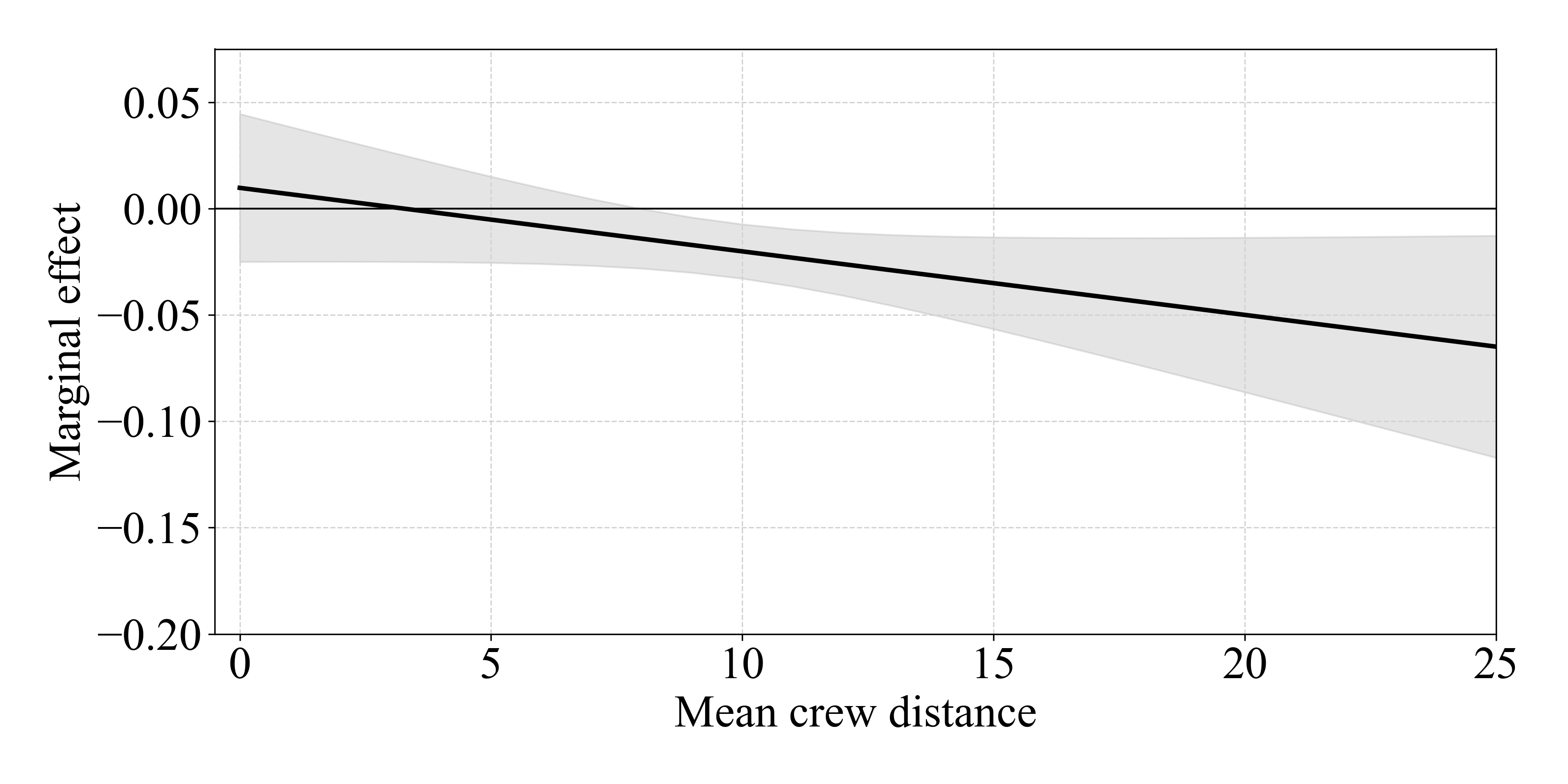}}\hfill
\subfloat{\includegraphics[width=.33\linewidth]{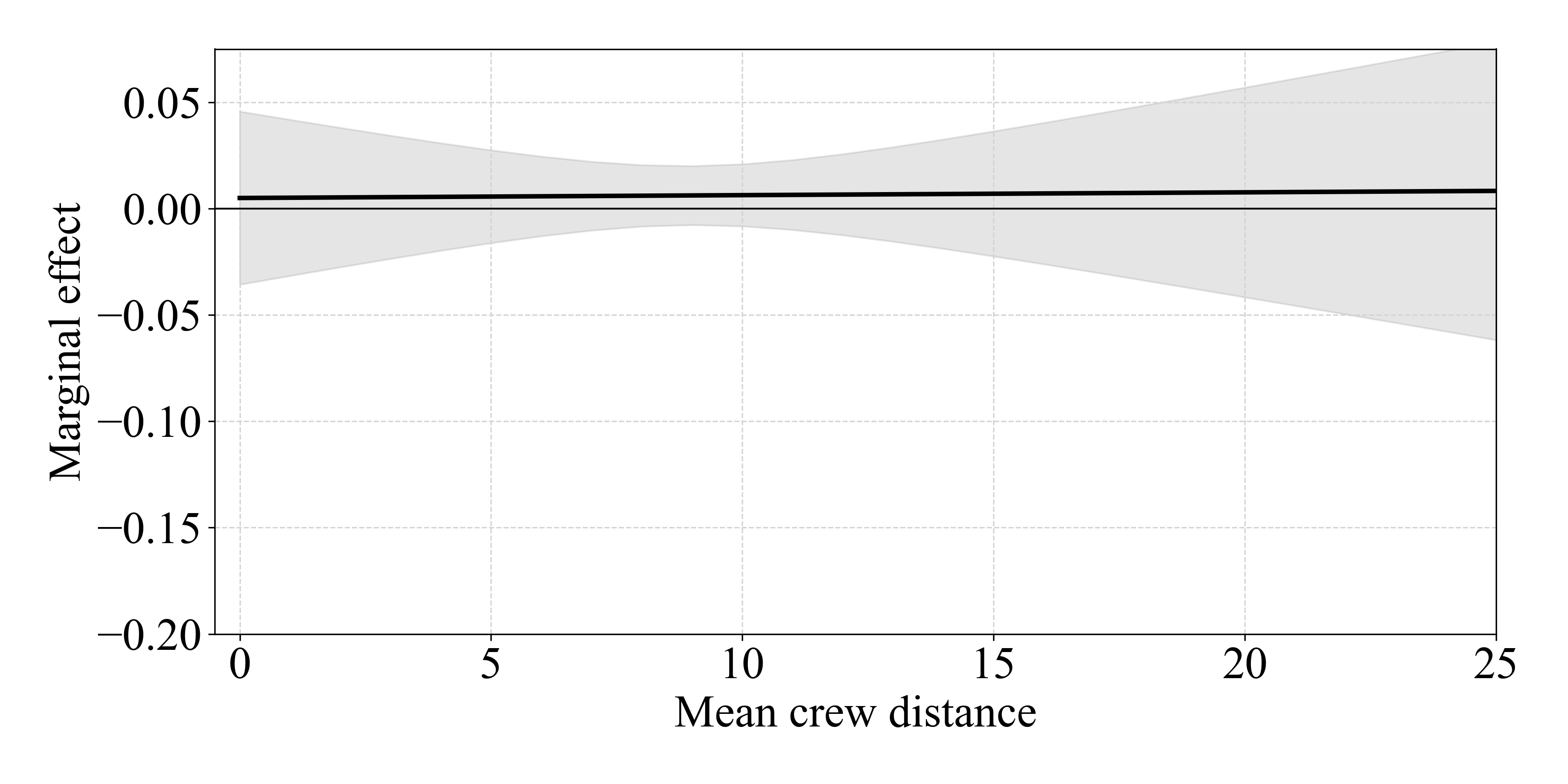}}\\
\caption*{\justifying\scriptsize \textit{Notes:} Each column represents a sampling period: the pre-awareness 2004-2006 period (left), the \textit{immediate} post-awareness 2007-2010 period (middle), and the \textit{later} post-awareness period of 2011-2014 seasons (right). The \emph{top row} shows a locally weighted smoothing (LOWESS) of the relationship between foul rate (y-axis) and the mean of absolute distance in skin tone values between the player and each referee in the officiating crew (x-axis) for each sample, illustrating the raw nonparametric pattern. The \emph{bottom row} depicts the marginal effect of distance derived from the quadratic regression model, which follows the specification from equation (\ref{eq:2}) along with 95\% confidence intervals.}
\label{fig:nonlineareffects}
\end{figure}

Figure~\ref{fig:nonlineareffects} illustrates the non-linearity between foul rate and absolute crew mean distance for the pre-awareness 2004-2006 period (left column), the \textit{immediate} post-awareness 2007-2010 period (middle column), and the rest of the sample period covering 2011-2014 seasons (right column). The top row shows a locally weighted smoothing (LOWESS) of the relationship between foul rate (y-axis) and the mean of absolute distance in skin tone values between the player and each referee in the officiating crew (x-axis) for each sample, illustrating the raw nonparametric pattern. The bottom row depicts the marginal effect of distance derived from the quadratic model, which follows the specification from equation (\ref{eq:2}), along with 95\% confidence intervals. Overall, these plots reveal how the average skin tint difference between each referee and the player influences changes in foul rates across the spectrum. While defining a clear-cut regarding ``group affiliation'' seems arbitrary, in fact, one would expect that such a cut-off is highly idiosyncratic and varies between individuals, it seems reasonable to assume that any crew-athlete constellation closer to zero represents a skin-tint-related in-group while a crew-athlete constellation at larger distances represents a skin-tint-related out-group.

The first two panels exhibit a downward slope as distance increases, indicating that distances (beyond certain values) lead to a negative marginal effect, i.e., an \textit{increase} in the difference between referees' and players’ tone color correlates with a \textit{decrease} in predicted foul rate. Near smaller distances (0–5), the marginal effect hovers around or slightly above zero, suggesting that when distance is minimal, slight differences in skin tone between players and referees do not systematically affect foul calls. If any, there is a slight tendency to \textit{punish} players with \textit{similar} skin tones in the pre-awareness period, though the results are not significant. As distance increases (particularly past 10 or 15, depending on the panel), the marginal effect drops below zero, indicating a negative impact on the foul rate per unit of distance, which is also statistically significant in the 2007-2010 period, suggesting an out-group favoritism rather than in-group discrimination. We see no clear and significant pattern in the third panel that represents the 2011-2014 period.

\subsection{Individual referees} \label{individual}
Since perceptions and consequences of colorism may differ by gender and skin tone of the referees \parencite{abramitzky2023,arceo2014,dixon2017, hill2002, thompson2001}, it is interesting to explore individual, gender, or skin-tone-related referee bias or favoritism. Since we cannot precisely link individual referees to the exact fouls they call, we use the approach of \textcite{sandberg2018} by replacing the $\text{CrewDistance}_{ig}$ variable in equation (\ref{eq:2}) with an interaction term between a dummy for a specific referee and the absolute skin tone distance of each referee to a specific player, replacing $\gamma_{g}$ to control for attendance and team performance up until the specific game. We estimate these models separately for each referee to obtain coefficients indicating how much, on average, a referee deviates from the other referees. Overall, we do not find a clear pattern between the referees' tone color and the players’ tone color and the foul calls. Likewise, no clear pattern emerges when looking at foul calling between female (triangles) and male (circles) referees (see Appendix \Cref{fig:indvref0406,fig:indvref0710,fig:indvref1114}).\footnote{Likewise, including a set of interaction terms regarding referee crew gender when estimating equation (\ref{eq:2}), we do not find evidence in our setting to suggest that referee gender influences the expression of colorism (results are available upon the request).}

\section{Discussion} \label{discussion}
In contrast to the results for the NBA by \parencite{price2010}, we do not find evidence of significant and systematic racial discrimination in the WNBA before the widespread media coverage that was provoked by the publication of the \parencite{price2010} results. Remarkably, however, we observe that \textit{immediately} following the increased public awareness (2007-2010), a player earns fewer fouls when facing more referees from the opposite race. While the estimates are not significant at conventional levels when using the FairFace algorithm, they are precise enough to become significant in the specifications using race categorizations from human raters. Moreover, across specifications, we find a clear and consistent pattern supporting this finding when using our continuous variable of skin tone. 

Overall, this highlights a critical distinction between race and skin tone. In the binary classification by human raters and FairFace, race and skin tone are inherently entangled, and each evaluation relies on skin tone but also cues such as names and/or facial features to make racial assessments. In contrast, the continuous measure isolates skin tone. When doing so, significant results emerge, supporting the notion that skin tone is an important part of the mechanism and may warrant a reexamination of past empirical studies on racial discrimination.

The results in the NBA for the period immediately after the extensive media coverage are the same as in the WNBA, that is, players receive significantly fewer foul calls from the crew that has more opposite race referees. However, while the direction of a potential ``media attention''-effect in the NBA and the WNBA is the same, the crucial difference between both settings is that before media exposure, there was evidence of racial discrimination in the NBA, while there was no such evidence in the WNBA. These ex-ante differences translate into zero racial bias in the NBA in the immediate post-awareness period, but overcorrection in the WNBA.\footnote{The difference in results between the FairFace algorithm and human raters raises important questions about the broader implications of using state-of-the-art algorithms to make judgments traditionally reserved for humans. While we believe the analytical benefits, particularly in terms of scalability, replicability, and objectivity, far outweigh the potential drawbacks in this context, we acknowledge that these technologies introduce new ethical and practical considerations. Although a full discussion is beyond the scope of this paper, we believe there is a pressing need for ongoing dialogue about how best to integrate such advancements responsibly and effectively into empirical research.}

\subsection{Potential mechanisms} \label{mechs}
Our analyses allow us to probe deeper into the mechanisms of overcorrection. First, using the continuous variable of the skin tone, we find evidence that this overcorrection is driven by out-group favoritism rather than by in-group discrimination. This is in line with \textcite{feld2016}, who investigate how student identity influences grader favoritism or discrimination and find a general pattern of favoritism rather than discrimination. Second, it is important to understand whether this out-group favoritism is intentional or not. For that, we dive into literature on the economics of discrimination, whose three main models include taste-based discrimination \parencite{becker1957}, statistical discrimination \parencite{arrow1998,phelps1972}, and implicit discrimination \parencite{bertrand2005}.

The first type of model, taste-based discrimination, assumes that individuals have certain preferences against a certain group of individuals. However, this type is highly unlikely in a competitive market of sports referees \parencite{dohmen2008}. \textcite{price2010} also cast doubts about this mechanism in the NBA. In addition, our results regarding individual referees in the overcorrection period show that there are only very few referees with a significant bias, supporting the notion that taste-based discrimination is the unlikely candidate to explain our results. The second type of model, statistical discrimination, assumes asymmetric information where race is taken as a signal. As such, in the context of foul calls in the WNBA, where there is no problem of asymmetric information, this mechanism can be ruled out as well. 

Consistent with other sports-related studies on referee bias, implicit discrimination is the most likely explanation for overcorrection in our case \parencite{price2010,gallo2013,ppw2018,faltings2023}. This alignment is supported by the fact that referee decisions in professional sports often involve high cognitive load (e.g., time pressure) and a degree of ambiguity, primary environmental preconditions for implicit bias \parencite{bertrand2005}. Given that we find evidence of out-group favoritism, we adjust the term from implicit discrimination to implicit favoritism. Such favoritism is unintentional and outside of the individual’s awareness. It may arise due to the unconscious association of a social group with different attributes and social norms. Social norms are especially important in the WNBA because of its strong emphasis on DEI. This commitment is apparent in the league’s programs, communications, and general atmosphere and thus may affect unconscious decisions that are taken in the type of split-second, high-pressure situations that appear in basketball games.\footnote{It is worth mentioning a former player, Candice Wiggins, who claimed that the environment around the league was one of the reasons she chose to retire early. She mentioned feeling like an outsider in the WNBA as a heterosexual woman, noting that the league’s strong culture made it difficult for those who didn’t conform \parencite{wiggins2017}.}

\subsection{Limitations} \label{limits}
Despite the strengths of our approach, some limitations merit consideration. First, although we use a continuous, perceptually grounded measure of skin tone and validate racial classifications using both human raters and the FairFace algorithm, these methods depend on visual cues and are subject to noise from image quality, lighting, and subjective interpretation. While steps like HSV filtering and k-means clustering mitigate these issues, they cannot fully eliminate them. Second, our analysis relies on referee crew composition rather than individual-level foul call attribution. Thus, we cannot precisely link individual referees to the exact fouls they call, which would be ideal for isolating personal-level effects. Similarly, it would be informative to examine whether the observed favoritism arises because referees fail to call legitimate fouls or because they avoid making incorrect calls. Distinguishing between these possibilities would help determine whether overcorrection represents a genuine behavioral distortion or a more cautious, accuracy-driven adjustment in officiating, which itself warrants important future research to disentangle.

\section{Conclusion} \label{conclusion}
While increased awareness is supposed to reduce racial bias in organizations that have experienced racial discrimination, this paper aimed to investigate the effect of increased awareness in non-biased organizations regarding race and skin tone. For that, we used widespread media coverage of the racial bias of referees in the NBA to test whether it had an effect on the decision-making of referees in the WNBA, an organization that is known for its strong position on equal rights and fight against discrimination. Using state-of-the-art machine learning algorithms to define race and skin tone, our empirical analyses found that, as expected, WNBA referees were not systematically racially biased in their foul calls’ decisions before the media exposure of the NBA results. However, an increased awareness caused overcorrection, according to which players received fewer foul calls when facing more referees from the opposite race.

Despite the peculiar nature of the WNBA, our results may still have insight into the broader implementation of DEI initiatives. While we do not take a normative stance on such programs, rather, we emphasize that their impact depends heavily on context. In settings where baseline levels of bias are already low, well-intentioned interventions may unintentionally disrupt the already unbiased behaviors and create pressure to overcorrect. Moreover, while a significant body of economic research has explored out-group discrimination, our results call for a more comprehensive understanding of environments that may create overcorrection. Controlled experiments could test the generalizability of this phenomenon across other settings, while theoretical work could elucidate mechanisms linking group identity, salience, and behavior. Future research should aim to integrate these insights into models of social preferences, enriching our understanding of identity-driven decision-making in both individual and institutional contexts.

\printbibliography

\clearpage
\setcounter{page}{1}
\pagenumbering{arabic}

\appendix
\renewcommand{\thetable}{\Alph{section}.\arabic{table}}
\renewcommand{\thefigure}{\Alph{section}.\arabic{figure}}

\section*{Appendix contents}
\thispagestyle{empty}
\addcontentsline{toc}{section}{Appendix Contents}
\begin{spacing}{1.5}
\noindent
\\
\textbf{A}\quad \textbf{Referee assignments} \dotfill 1\\
\hspace{1.5em}Table \ref{table:blackstartersperteamhr} Black starters per team and refereeing crews by race \dotfill 1\\
\hspace{1.5em}Table \ref{table:tonestartersperteam} Starter skin tone per team and refereeing crews by skin tone \dotfill 1\\
\noindent
\\
\textbf{B}\quad \textbf{Methodology} \dotfill 2\\
\hspace{1.5em}B.1 Face segmentation \dotfill 2\\
\hspace{1.5em}B.2 Skin tone analysis \dotfill 3\\
\hspace{1.5em}Table \ref{table:kappa} Interrater reliability scores \dotfill 5\\
\hspace{1.5em}Table \ref{table:samplesize0406} Sample size (2004–06) \dotfill 5\\
\hspace{1.5em}Table \ref{table:samplesize0710} Sample size (2007–10) \dotfill 6\\
\hspace{1.5em}Table \ref{table:samplesize1114} Sample size (2011–14) \dotfill 7\\
\hspace{1.5em}Table \ref{table:plcharsff0406} Summary statistics: FairFace classification (2004–06) \dotfill 8\\
\hspace{1.5em}Table \ref{table:plcharsff0710} Summary statistics: FairFace classification (2007–10) \dotfill 10\\
\hspace{1.5em}Table \ref{table:plcharsff1114} Summary statistics: FairFace classification (2011–14) \dotfill 12\\
\hspace{1.5em}Table \ref{table:plcharshr} Summary statistics: Human raters \dotfill 14\\
\hspace{1.5em}Table \ref{table:plcharshr0406} Summary statistics: Human raters (2004–06) \dotfill 16\\
\hspace{1.5em}Table \ref{table:plcharshr0710} Summary statistics: Human raters (2007–10) \dotfill 18\\
\hspace{1.5em}Table \ref{table:plcharshr1114} Summary statistics: Human raters (2011–14) \dotfill 20\\
\noindent
\\
\textbf{C}\quad \textbf{Results} \dotfill 22\\
\hspace{1.5em}Table \ref{table:mlhragree} Racial bias among WNBA referees: Human-FairFace agreement \dotfill 22\\
\hspace{1.5em}Table \ref{table:app_race_subsamplesff} Racial bias among WNBA referees: Team race control using FairFace \dotfill 22\\
\hspace{1.5em}Table \ref{table:app_race_subsampleshr} Racial bias among WNBA referees: Team race control using human raters \dotfill 23\\
\hspace{1.5em}Table \ref{table:mlrace333} Racial bias among WNBA referees: FairFace classification balanced years \dotfill 23\\
\hspace{1.5em}Table \ref{table:hurace333} Racial bias among WNBA referees: Human raters balanced years \dotfill 23\\
\hspace{1.5em}Table \ref{table:tone333} Colorism among WNBA referees: Balanced years \dotfill 24\\
\hspace{1.5em}Table \ref{table:app_tone_subsamples} Colorism among WNBA referees: Team skin-tone control \dotfill 24\\
\hspace{1.5em}Figure \ref{fig:indvref0406} Distribution of colorism by referee (2004–06) \dotfill 25\\
\hspace{1.5em}Figure \ref{fig:indvref0710} Distribution of colorism by referee (2007–10) \dotfill 26\\
\hspace{1.5em}Figure \ref{fig:indvref1114} Distribution of colorism by referee (2011–14) \dotfill 26
\end{spacing}

\clearpage
\setcounter{page}{1}
\pagenumbering{arabic}

\section{Referee assignments}
\setcounter{table}{0}
\setcounter{figure}{0}
\begin{table}[H]
\centering
\captionsetup{
  justification=centering,
  singlelinecheck=false,
  labelsep=none,
  font=normalsize
}
\caption{\\Black starters per team and the distribution of refereeing crews by race}\vspace{-0.8em}
\label{table:blackstartersperteamhr}
\begin{threeparttable}
  \small
  \setlength\tabcolsep{4pt}      
  \renewcommand{\arraystretch}{0.8} 
  \begin{tabular*}{\linewidth}{@{\extracolsep{\fill}} l S S S S S @{}}
    \toprule\toprule
& \multicolumn{4}{c}{Black starters per team} \\
\cmidrule(lr){2-5}
Season & {0 non-Black refs} & {1 non-Black ref} & {2 non-Black refs} & {3 non-Black refs} & {\makecell[cb]{\(\chi^2\) test of \\ independence\(^{\alpha}\) \\ (p-value)}} \\
\midrule
    2004 & 3.83 & 3.44 & 3.51 & 3.66 & .452 \\
    2005 & 2.93 & 3.30 & 3.27 & 3.39 & .448 \\
    2006 & 3.77 & 3.40 & 3.49 & 3.26 & .190 \\
    2007 & 3.35 & 3.35 & 3.27 & 3.33 & .173 \\
    2008 & 3.93 & 3.64 & 3.78 & 3.78 & .464 \\
    2009 & 3.68 & 3.75 & 3.71 & 3.68 & .986 \\
    2010 & 3.70 & 3.77 & 3.73 & 3.63 & .246 \\
    2011 & 3.94 & 3.86 & 3.78 & 3.79 & .839 \\
    2012 & 4.19 & 4.10 & 3.87 & 4.08 & .116 \\
    2013 & 4.13 & 4.19 & 4.28 & 4.21 & .694 \\
    2014 & 3.84 & 3.86 & 3.80 & 3.85 & .272 \\
\\
\makecell[l]{Sample size (\% of \\ all player-games)} & {\makecell{406 \\ (8.61)}} & {\makecell{1,440 \\ (30.52)}} & {\makecell{1,902 \\ (40.31)}} & {\makecell{970 \\ (20.56)}} & {n = 4,718} \\
\bottomrule\bottomrule
\end{tabular*}
\begin{tablenotes}[para,flushleft]
\setstretch{0.5} 
\tiny
\setlength{\parskip}{0pt} 
\setlength{\parindent}{0pt} 
\setlength{\itemsep}{-1pt} 
\item \scriptsize \textit{Notes:} Racial classifications determined by human raters. Each observation is a team×game observation. \(^{\alpha}\)Final column tests: \(H_0\): \# non-Black referees is independent of \# Black starters.
\end{tablenotes}
\end{threeparttable}
\end{table}

\begin{table}[H]
\centering
\captionsetup{
  justification=centering,
  singlelinecheck=false,
  labelsep=none,
  font=normalsize
}
\caption{\\Starter skin tone per team and the distribution of refereeing crews by skin tone}\vspace{-0.8em}
\label{table:tonestartersperteam}
\begin{threeparttable}
  \small
  \setlength\tabcolsep{4pt}      
  \renewcommand{\arraystretch}{0.8} 
\begin{tabular*}{\linewidth}{@{\extracolsep{\fill}} l
  S[table-format=1.2] S[table-format=1.2] S[table-format=1.2] S[table-format=1.2]
  S[table-format=1.3] @{}}
\toprule\toprule
& \multicolumn{4}{c}{Player skin tone quartiles} \\
\cmidrule(lr){2-5}
Season
& \multicolumn{1}{c}{\makecell[cb]{Q1 crew\\skin tint}}
& \multicolumn{1}{c}{\makecell[cb]{Q2 crew\\skin tint}}
& \multicolumn{1}{c}{\makecell[cb]{Q3 crew\\skin tint}}
& \multicolumn{1}{c}{\makecell[cb]{Q4 crew\\skin tint}}
& \multicolumn{1}{c}{\makecell[cb]{\(\chi^2\) test of\\ independence$^{\alpha}$\\ (p-value)}} \\[-0.6ex]
\midrule

    2004 & 2.46 & 2.25 & 2.41 & 2.24 & .851 \\
    2005 & 2.48 & 2.45 & 2.27 & 2.39 & .510 \\
    2006 & 2.24 & 2.27 & 2.54 & 2.45 & .614 \\
    2007 & 2.82 & 2.81 & 2.61 & 2.83 & .858 \\
    2008 & 2.46 & 2.54 & 2.51 & 2.27 & .099 \\
    2009 & 2.36 & 2.21 & 2.28 & 2.33 & .828 \\
    2010 & 2.44 & 2.51 & 2.60 & 2.70 & .595 \\
    2011 & 2.29 & 2.46 & 2.55 & 2.47 & .690 \\
    2012 & 2.47 & 2.58 & 2.41 & 2.69 & .042 \\
    2013 & 2.48 & 2.38 & 2.71 & 2.86 & .007 \\
    2014 & 2.81 & 2.85 & 2.85 & 2.80 & .475 \\
\\
\makecell[l]{Sample size (\% of \\ all player-games)} & {\makecell{1,180 \\ (25.01)}} & {\makecell{1,180 \\ (25.01)}} & {\makecell{1,180 \\ (25.01)}} & {\makecell{1,178 \\ (24.97)}} & {n = 4,718} \\
\bottomrule\bottomrule
\end{tabular*}
\begin{tablenotes}[para,flushleft]
\setstretch{0.5} 
\tiny
\setlength{\parskip}{0pt} 
\setlength{\parindent}{0pt} 
\setlength{\itemsep}{-1pt} 
\item \scriptsize \textit{Notes:} Each observation is a team×game observation. \(^{\alpha}\)Final column tests: \(H_0\): crew skin‐tint quartile is independent of starter skin‐tone quartile.
\end{tablenotes}
\end{threeparttable}
\end{table}

\clearpage

\section{Methodology} \label{amethods}
\setcounter{table}{0}
\setcounter{figure}{0}
As mentioned in the main text, our third approach follows the ``Image-to-Data Pipeline'' in \textcite{adukia2023}, dividing the process into two main parts, i.e., face segmentation and skin tone classification. The detailed process is described here:
\label{chap:image-to-data-pipeline}
\begin{quote}
B.1 Face segmentation \\ 
The face segmentation process isolates the facial region from each headshot image. We employ a facial landmark detection technique, as detailed below:
\begin{quote}
B.1.1 \textit{Loading and converting images:} Images are loaded and converted to RGB format using the Python Imaging Library (PIL). This ensures consistency in color space for subsequent processing. \\
B.1.2 \textit{Isolating skin components\footnote{We do, however, omit the Conditional Random Field (CRF), which is used by \Textcite{adukia2023} to refine the skin mask by applying a CRF module, which predicts the labels of neighboring pixels (i.e., whether they are predicted to be skin or not skin) to produce a more fine-grained segmentation result. In our case, it is unnecessary since all our photos are human, and we will have additional skin mask refinement later (see feature processing). The added complexity a CRF module would bring is not a value add in our case scenario.}:} We use the Fully Connected Convolutional Neural Network (FC-CNN) method to extract a skin mask from faces. This involves applying a fully-convolutional neural network (FCN), a type of convolutional neural network (CNN) where the last fully connected layer is substituted with a convolutional layer that can capture the locations of the predicted labels. This enables the prediction of peripheral landmarks, such as the edges of the facial skin area, eyes, nose, and mouth. \\
B.1.3 \textit{Facial landmark detection:} Using the FaceAlignment library configured with a 2D landmark detection model, we identify 68 facial landmarks (categorized into specific facial features such as the jawline, eyebrows, nose bridge, and lips) for each image. To increase accuracy, the FaceAlignment model initializes with parameters to handle CUDA support, CNN-based face detection, and flipping input images. The library utilizes deep learning models such as Face Alignment Network (FAN) and ResNetDepth for predicting 2D and 3D landmarks, respectively. \\
B.1.4 \textit{Creating face masks:} Utilizing the predicted landmarks from the FC-CNN, we isolate the targeted facial region by extracting a skin “mask” from these predicted landmarks using the built-in Convex Hull method from SciPy’s Python library. The Convex Hull method generates the hull, and the skimage’s draw polygon function creates a binary mask isolating the face region. \\
A.1.5 \textit{Applying masks to images:} The binary mask is applied to the original image to extract the face region, resulting in the segmented skin that is then used to classify skin color. The masked face is saved as a separate image file for the skin tone analysis.
\end{quote}
\clearpage
B.2 Skin tone analysis \\ 
Following face segmentation, the skin tone of the isolated facial regions is analyzed. To conduct skin tone analysis on a dataset of facial images, we utilize their multi-step image processing pipeline, which combines color space transformations, machine learning clustering techniques, and deep learning models. This part of the pipeline comprises two main phases: (1) skin extraction and dominant color identification, as well as (2) skin tone classification. The steps involved in this process are detailed below:
\begin{quote}
B.2.1 \textit{Skin Extraction}
\begin{quote}
B.2.1.1 Each image was first converted from the RGB color space (the default for OpenCV) to the HSV color space using OpenCV's cv2.cvtColor() function. The HSV space was chosen because it provides a more intuitive representation of color, allowing for easier segmentation of skin tones. \\
B.2.1.2 A range of HSV values known to correspond to common skin tones was defined. Specifically, the lower threshold was set at [0, 48, 89] and the upper threshold at [20, 255, 255]. These values were tuned based on preliminary experiments to capture a broad range of skin tones while minimizing the inclusion of non-skin pixels. \\
B.2.1.3 A binary mask was created using the cv2.inRange() function, where pixel values falling within the specified thresholds were classified as skin, and others were discarded. \\
B.2.1.4 The generated mask was then refined using a Gaussian filter (cv2.GaussianBlur() with a kernel size of 3x3) to remove noise and smooth the edges of the detected regions. \\
B.2.1.5 The smoothed mask was applied to the original image using bitwise operations (cv2.bitwise\_and ()), resulting in an image where only the skin regions were retained. \\
B.2.1.6 Conversion to BGR Color Space: The extracted skin regions were finally converted back to the BGR color space for analysis. 
\end{quote}
B.2.2 \textit{Dominant Color Identification Using K-means}
\begin{quote}
B.2.2.1 The skin-extracted image was reshaped into a two-dimensional array where each row represented the RGB values of a pixel. \\
B.2.2.2 K-means Clustering: K-means clustering was applied to group the pixels into clusters of similar colors based on RGB pixels. The number of clusters (k) was set to 5, allowing for capturing multiple dominant skin tone shades in each image. The clustering was performed using the KMeans() function from the sci-kit-learn library. \\
B.2.2.3 The Counter() function from the collections library was used to calculate the occurrence frequency of each dominant color. This allowed us to quantify the proportion of each dominant color within the skin regions. \\
B.2.2.4 If the cluster containing black ([0, 0, 0]) pixels was detected, it was excluded from the dominant color analysis to avoid artifacts from the background or shadowed regions. This was implemented using a custom function that compared pixel values to detect black. \\
B.2.2.5 The RGB values of the five largest clusters, providing the dominant skin colors in the segmented skin, are converted to the CIELAB (Lab*) 	color space using skimage. The Lab* color space is perceptually uniform, making it more suitable for color analysis. The conversion from RGB to Lab* occurred before averaging because the Lab* color space is perceptually linear, unlike RGB color space. The Lab* values of the dominant clusters are weighted by their sizes to compute a representative skin color. The total number of pixels in the dominant clusters is calculated, and the weights are derived as the proportion of each cluster's size to the total. This weighted average provides a measure of each face’s representative skin color.
\end{quote}
\end{quote}
\end{quote}
\textbf{Output} \\
The extracted skin tone data, including the representative Lab* values, weights of the clusters, RGB values of the dominant clusters, and their sizes, are saved to CSV files. The robustness and accuracy of the skin tone classification process are validated through visual inspection and comparison with manually labeled samples using matplotlib. The entire process is automated using Python, ensuring reproducibility and efficiency.

\clearpage

\setcounter{section}{2} 
\begin{table}[H]
\centering
\captionsetup{
    justification=centering, 
    labelsep=none,           
    font=normalsize          
}
\caption{\\Kappa scores for interrater reliability}\vspace{-0.8em}
\label{table:kappa} 
\begin{threeparttable}
\renewcommand{\arraystretch}{0.8} 
\small 
\begin{tabular}{@{}lc@{}} 
\toprule
\toprule
Comparison & \multicolumn{1}{c}{Kappa Score} \\
\midrule
Human Rater 1 $\&$ Human Rater 2 & 0.892 \\
Human Rater 1 $\&$ FairFace      & 0.948 \\
Human Rater 2 $\&$ FairFace      & 0.888 \\
Human Raters  $\&$ FairFace      & 0.948 \\
\bottomrule
\bottomrule
\end{tabular}
\begin{tablenotes}[para,flushleft]
\setstretch{0.5} 
\tiny
\setlength{\parskip}{0pt} 
\setlength{\parindent}{0pt} 
\setlength{\itemsep}{-1pt} 
\item \scriptsize \textit{Notes:} Cohen's Kappa measures the agreement between two raters. The comparisons include Human Rater 1, Human Rater 2, and FairFace classifications.
\end{tablenotes}
\end{threeparttable}
\end{table}

\begin{table}[H]
\centering
\captionsetup{
  justification=centering,
  labelsep=none,
  font=normalsize
}
\caption{\\Sample size (2004-06)}\vspace{-0.8em}
\label{table:samplesize0406}
\begin{threeparttable}
  \renewcommand{\arraystretch}{0.8}
  \small
  \begin{tabular}{lccc}
    \toprule
    & Black & Non-Black & Total \\
    \midrule
    \multicolumn{4}{c}{A. FairFace} \\
    No.\ of players            & 164       & 94        & 258     \\
    \quad Mean L* value        & 42.54     & 62.45      & 49.79    \\
    \quad \quad (SD)           & (10.02)   & (6.95)     & (13.16)  \\
    \quad \quad Min–max        & 20.52–69.73 & 41.85–77.21 & 20.52–77.21 \\
    \addlinespace
    No.\ of referees           & 13        & 26         & 39      \\
    \quad Mean L* value        & 41.96     & 60.72      & 54.01    \\
    \quad \quad (SD)           & (9.53)    & (9.83)     & (12.59)  \\
    \quad \quad Min–max        & 30.06–60.45 & 41.77–76.32 & 30.06–76.32 \\
    \addlinespace
    No.\ of games              & 665     & 665      & 665    \\
    No.\ of player–games       & 8,743    & 4,153     & 12,896   \\
    No.\ of player–minutes     & 182,662   & 85,540    & 268,202  \\
    \multicolumn{4}{c}{B. Human raters} \\
    No.\ of players            & 165       & 93        & 258      \\
    \quad Mean L* value        & 42.67     & 62.42      & 49.79    \\
    \quad \quad (SD)           & (10.13)   & (6.99)     & (13.16)  \\
    \quad \quad Min–max        & 20.52–69.73 & 41.85–77.21 & 20.52–77.21 \\
    \addlinespace
    No.\ of referees           & 14        & 25         & 39      \\
    \quad Mean L* value        & 41.95     & 61.48      & 54.47    \\
    \quad \quad (SD)           & (9.16)    & (9.23)     & (13.14)  \\
    \quad \quad Min–max        & 30.06–60.45 & 43.89–76.32 & 30.06–76.32 \\
    \addlinespace
    No.\ of games              & 665     & 663      & 665    \\
    No.\ of player–games       & 8,831    & 4,065     & 12,896   \\
    No.\ of player–minutes     & 184,253   & 83,949    & 268,202  \\
    \multicolumn{4}{c}{C. Referee crew compositions} \\
    No.\ of unique crews       &           &            & 594    \\
    \quad Mean abs.\ L* diff.  &           &            & 8.80     \\
    \quad \quad (SD)           &           &            & (4.29)   \\
    \quad \quad Min–max        &           &            & 0.481–27.39 \\
    \bottomrule
  \end{tabular}
  \begin{tablenotes}[para,flushleft]
    \setstretch{0.5}\tiny
    \setlength{\parskip}{0pt}
    \setlength{\parindent}{0pt}
    \setlength{\itemsep}{-1pt}
    \item \scriptsize \textit{Notes:} L* values represent skin tone measurements using the CIELAB color space. The conditional descriptives are based on the racial classifications made by FairFace predictions and human raters as discussed in Section 3.1.
  \end{tablenotes}
\end{threeparttable}
\end{table}

\begin{table}[H]
\centering
\captionsetup{
  justification=centering,
  labelsep=none,
  font=normalsize
}
\caption{\\Sample size (2007-10)}\vspace{-0.8em}
\label{table:samplesize0710}
\begin{threeparttable}
  \renewcommand{\arraystretch}{0.8}
  \small
  \begin{tabular}{lccc}
    \toprule
    & Black & Non-Black & Total \\
    \midrule
    \multicolumn{4}{c}{A. FairFace} \\
    No.\ of players            & 191       & 92        & 283     \\
    \quad Mean L* value        & 44.80     & 64.00      & 49.79    \\
    \quad \quad (SD)           & (10.21)   & (6.63)     & (13.16)  \\
    \quad \quad Min–max        & 20.52–69.73 & 45.13–77.21 & 20.52–77.21 \\
    \addlinespace
    No.\ of referees           & 16        & 29         & 45      \\
    \quad Mean L* value        & 42.24     & 59.90      & 54.01    \\
    \quad \quad (SD)           & (10.94)   & (7.94)     & (12.59)  \\
    \quad \quad Min–max        & 20.56–60.75 & 49.14–76.32 & 20.56–76.32 \\
    \addlinespace
    No.\ of games              & 878     & 878      & 878    \\
    No.\ of player–games       & 12,486    & 4,787     & 17,273   \\
    No.\ of player–minutes     & 259,401   & 95,504    & 354,905  \\
    \multicolumn{4}{c}{B. Human raters} \\
    No.\ of players            & 190       & 93        & 283      \\
    \quad Mean L* value        & 44.78     & 63.84      & 51.04    \\
    \quad \quad (SD)           & (10.24)   & (6.74)     & (12.87)  \\
    \quad \quad Min–max        & 20.52–69.73 & 45.13–77.21 & 20.52–77.21 \\
    \addlinespace
    No.\ of referees           & 18        & 27         & 45      \\
    \quad Mean L* value        & 43.34     & 60.47      & 53.62    \\
    \quad \quad (SD)           & (10.78)   & (7.93)     & (12.41)  \\
    \quad \quad Min–max        & 20.56–60.75 & 49.14–76.32 & 20.56–76.32 \\
    \addlinespace
    No.\ of games              & 878     & 878      & 878    \\
    No.\ of player–games       & 12,429    & 4,844     & 17,273   \\
    No.\ of player–minutes     & 257,752   & 97,153    & 354,905  \\
    \multicolumn{4}{c}{C. Referee crew compositions} \\
    No.\ of unique crews       &           &            & 749    \\
    \quad Mean abs.\ L* diff.  &           &            & 8.55     \\
    \quad \quad (SD)           &           &            & (4.31)   \\
    \quad \quad Min–max        &           &            & 0.548–33.46 \\
    \bottomrule
  \end{tabular}
  \begin{tablenotes}[para,flushleft]
    \setstretch{0.5}\tiny
    \setlength{\parskip}{0pt}
    \setlength{\parindent}{0pt}
    \setlength{\itemsep}{-1pt}
    \item \scriptsize \textit{Notes:} L* values represent skin tone measurements using the CIELAB color space. The conditional descriptives are based on the racial classifications made by FairFace predictions and human raters as discussed in Section 3.1.
  \end{tablenotes}
\end{threeparttable}
\end{table}

\begin{table}[H]
\centering
\captionsetup{
  justification=centering,
  labelsep=none,
  font=normalsize
}
\caption{\\Sample size (2011-14)}\vspace{-0.8em}
\label{table:samplesize1114}
\begin{threeparttable}
  \renewcommand{\arraystretch}{0.8}
  \small
  \begin{tabular}{lccc}
    \toprule
    & Black & Non-Black & Total \\
    \midrule
    \multicolumn{4}{c}{A. FairFace} \\
    No.\ of players            & 178       & 69        & 247     \\
    \quad Mean L* value        & 47.45     & 64.92      & 52.33    \\
    \quad \quad (SD)           & (10.59)   & (7.62)     & (12.59)  \\
    \quad \quad Min–max        & 20.52–71.26 & 41.85–79.4813 & 20.52–79.4813 \\
    \addlinespace
    No.\ of referees           & 17        & 24         & 41      \\
    \quad Mean L* value        & 43.01     & 57.88      & 51.71    \\
    \quad \quad (SD)           & (9.40)    & (7.47)     & (11.07)  \\
    \quad \quad Min–max        & 30.06–60.75 & 43.21–72.92 & 30.06–72.92 \\
    \addlinespace
    No.\ of games              & 816     & 813      & 816    \\
    No.\ of player–games       & 12,377    & 3,332     & 15,709   \\
    No.\ of player–minutes     & 265,311   & 64,516    & 329,827  \\
    \multicolumn{4}{c}{B. Human raters} \\
    No.\ of players            & 178       & 69        & 247      \\
    \quad Mean L* value        & 47.50     & 64.79      & 52.33    \\
    \quad \quad (SD)           & (10.64)   & (7.77)     & (12.59)  \\
    \quad \quad Min–max        & 20.52–71.26 & 41.85–79.4813 & 20.52–79.4813 \\
    \addlinespace
    No.\ of referees           & 20        & 21         & 41      \\
    \quad Mean L* value        & 44.94     & 58.17      & 51.71    \\
    \quad \quad (SD)           & (10.06)   & (7.69)     & (11.07)  \\
    \quad \quad Min–max        & 30.06–63.32 & 43.21–72.92 & 30.06–72.92 \\
    \addlinespace
    No.\ of games              & 816     & 813      & 816    \\
    No.\ of player–games       & 12,267    & 3,442     & 15,709   \\
    No.\ of player–minutes     & 262,005   & 67,822    & 329,827  \\
    \multicolumn{4}{c}{C. Referee crew compositions} \\
    No.\ of unique crews       &           &            & 666    \\
    \quad Mean abs.\ L* diff.  &           &            & 7.87     \\
    \quad \quad (SD)           &           &            & (3.91)   \\
    \quad \quad Min–max        &           &            & 0.414–25.06 \\
    \bottomrule
  \end{tabular}
  \begin{tablenotes}[para,flushleft]
    \setstretch{0.5}\tiny
    \setlength{\parskip}{0pt}
    \setlength{\parindent}{0pt}
    \setlength{\itemsep}{-1pt}
    \item \scriptsize \textit{Notes:} L* values represent skin tone measurements using the CIELAB color space. The conditional descriptives are based on the racial classifications made by FairFace predictions and human raters as discussed in Section 3.1.
  \end{tablenotes}
\end{threeparttable}
\end{table}

\begin{table}[H]
\centering
\captionsetup{
  justification=centering,
  labelsep=none,
  font=normalsize
}
\caption{\\Summary statistics: FairFace classification (2004-2006)}\vspace{-0.8em}
\label{table:plcharsff0406}
\begin{threeparttable}
\renewcommand{\arraystretch}{0.8}
\small
\begin{tabular}{lccc}
\toprule\toprule
& \multicolumn{1}{c}{Black players} 
& \multicolumn{1}{c}{Non-Black players} 
& \multicolumn{1}{c}{} \\
\cmidrule(lr){2-2} \cmidrule(lr){3-3}
& \multicolumn{1}{c}{\makecell{Mean \\ (SD)}} 
& \multicolumn{1}{c}{\makecell{Mean \\ (SD)}} 
& \multicolumn{1}{c}{Difference} \\
\midrule
\multicolumn{4}{c}{Raw player statistics} \\
Minutes played                    & 26.30  & 26.32  & -0.03 \\
                                  & (8.93) & (9.03) &       \\
Fouls                             & 2.42   & 2.24   & 0.17\tnote{***} \\
                                  & (1.53) & (1.51) &                \\
Points                            & 9.69   & 9.52   & 0.17            \\
                                  & (6.54) & (6.84) &                \\
\multicolumn{4}{c}{Player productivity: stats × 40/minutes played} \\
Fouls                             & 3.95   & 3.68   & 0.27\tnote{***} \\
                                  & (2.99) & (2.99) &                \\
Points                            & 13.99  & 13.66  & 0.33\tnote{**} \\
                                  & (7.94) & (8.35) &                \\
Free throws made                  & 2.93   & 2.73   & 0.20\tnote{***} \\
                                  & (3.21) & (3.21) &                \\
Free throws missed                & 1.11   & 0.77   & 0.34\tnote{***} \\
                                  & (1.76) & (1.45) &                \\
2 point goals made                & 4.45   & 3.61   & 0.84\tnote{***} \\
                                  & (3.16) & (2.84) &                \\
2 point goals missed              & 5.59   & 4.31   & 1.28\tnote{***} \\
                                  & (3.60) & (3.16) &                \\
3 point goals made                & 0.72   & 1.24   & -0.52\tnote{***} \\
                                  & (1.33) & (1.75) &                \\
3 point goals missed              & 1.47   & 2.26   & -0.79\tnote{***} \\
                                  & (2.09) & (2.52) &                \\
Offensive rebounds                & 2.06   & 1.62   & 0.44\tnote{***} \\
                                  & (2.30) & (2.04) &                \\
Defensive rebounds                & 4.37   & 4.20   & 0.17\tnote{**} \\
                                  & (3.22) & (3.27) &                \\
Assists                           & 2.89   & 3.45   & -0.56\tnote{***} \\
                                  & (2.71) & (2.88) &                \\
Steals                            & 1.56   & 1.48   & 0.07\tnote{**} \\
                                  & (1.79) & (1.73) &                \\
Blocks                            & 0.69   & 0.76   & -0.07\tnote{**} \\
                                  & (1.32) & (1.50) &                \\
Turnovers                         & 2.87   & 2.82   & 0.05            \\
                                  & (2.50) & (2.45) &                \\
\multicolumn{4}{c}{Game information} \\
Attendance (1,000s)               & 8.08   & 7.92   & 0.16\tnote{**} \\
                                  & (2.57) & (2.46) &                \\
Out of contention                 & 0.04   & 0.03   & 0.01\tnote{***} \\
                                  & (0.20) & (0.16) &                \\
\bottomrule
\end{tabular}
\end{threeparttable}
\end{table}

\begin{table}[H]
\ContinuedFloat
\centering
\captionsetup{
labelformat=empty,
labelsep=none,
justification=centering,
font=normalsize
}
\caption{\\(Continued)}\vspace{-0.8em}
\label{table:plcharsff0406}
\begin{threeparttable}
\renewcommand{\arraystretch}{0.8}
\small
\begin{tabular}{lccc}
\toprule\toprule
& \multicolumn{1}{c}{Black players} 
& \multicolumn{1}{c}{Non-Black players} 
& \multicolumn{1}{c}{} \\
\cmidrule(lr){2-2} \cmidrule(lr){3-3}
& \multicolumn{1}{c}{\makecell{Mean \\ (SD)}} 
& \multicolumn{1}{c}{\makecell{Mean \\ (SD)}} 
& \multicolumn{1}{c}{Difference} \\
\midrule
\multicolumn{4}{c}{Player characteristics} \\
Age                               & 27.66  & 26.44  & 1.23\tnote{***} \\
                                  & (4.14) & (3.42) &                \\
WNBA experience (yrs)             & 5.01   & 4.48   & 0.53\tnote{*}  \\
                                  & (2.63) & (2.38) &                \\
All-WNBA this year                & 0.12   & 0.12   & -0.01         \\
                                  & (0.32) & (0.33) &                \\
Center                            & 0.19   & 0.13   & 0.06          \\
                                  & (0.39) & (0.34) &                \\
Forward                           & 0.42   & 0.35   & 0.07          \\
                                  & (0.49) & (0.48) &                \\
Guard                             & 0.39   & 0.52   & -0.13         \\
                                  & (0.49) & (0.50) &                \\
Starter                           & 0.70   & 0.71   & -0.01         \\
                                  & (0.46) & (0.45) &                \\
Height (cm.)                      & 183.57 & 184.57 & -1.00         \\
                                  & (7.92) & (10.48)&                \\
Weight (kg.)                      & 76.30  & 76.02  & 0.28          \\
                                  & (9.90) & (10.29)&                \\

\multicolumn{4}{c}{Referees} \\
0 non-Black referees              & 0.01   & 0.01   & 0.00          \\
                                  & (0.11) & (0.10) &                \\
1 non-Black referees              & 0.21   & 0.22   & -0.01         \\
                                  & (0.41) & (0.42) &                \\
2 non-Black referees              & 0.47   & 0.47   & 0.00          \\
                                  & (0.50) & (0.50) &                \\
3 non-Black referees              & 0.30   & 0.29   & 0.01          \\
                                  & (0.46) & (0.46) &                \\
\text{\#} non-Black referees      & 2.06   & 2.05   & 0.01          \\
                                  & (0.75) & (0.75) &                \\
\bottomrule\bottomrule
\end{tabular}

\begin{tablenotes}[para,flushleft]
\scriptsize
\item All observations are weighted by minutes played. \\
\(***\), \(**\), and \(*\) Differences statistically significant at 1\%, 5\%, and 10\%, respectively.
\end{tablenotes}
\end{threeparttable}
\end{table}

\begin{table}[H]
\centering
\captionsetup{
  justification=centering,
  labelsep=none,
  font=normalsize
}
\caption{\\Summary statistics: FairFace classification (2007-2010)}\vspace{-0.8em}
\label{table:plcharsff0710}
\begin{threeparttable}
\renewcommand{\arraystretch}{0.8}
\small
\begin{tabular}{lccc}
\toprule\toprule
& \multicolumn{1}{c}{Black players} 
& \multicolumn{1}{c}{Non-Black players} 
& \multicolumn{1}{c}{} \\
\cmidrule(lr){2-2} \cmidrule(lr){3-3}
& \multicolumn{1}{c}{\makecell{Mean \\ (SD)}} 
& \multicolumn{1}{c}{\makecell{Mean \\ (SD)}} 
& \multicolumn{1}{c}{Difference} \\
\midrule
\multicolumn{4}{c}{Raw player statistics} \\
Minutes played                    & 25.88  & 25.81  & 0.07          \\
                                  & (8.76) & (9.13) &               \\
Fouls                             & 2.41   & 2.16   & 0.26\tnote{***} \\
                                  & (1.49) & (1.52) &                \\
Points                            & 10.56  & 10.19  & 0.37\tnote{***} \\
                                  & (6.94) & (7.68) &                \\
\multicolumn{4}{c}{Player productivity: stats × 40/minutes played} \\
Fouls                             & 4.05   & 3.61   & 0.44\tnote{***} \\
                                  & (3.11) & (3.09) &                \\
Points                            & 15.61  & 14.80  & 0.81\tnote{***} \\
                                  & (8.66) & (9.34) &                \\
Free throws made                  & 3.31   & 2.78   & 0.53\tnote{***} \\
                                  & (3.53) & (3.44) &                \\
Free throws missed                & 1.08   & 0.61   & 0.47\tnote{***} \\
                                  & (1.78) & (1.31) &                \\
2 point goals made                & 4.79   & 3.59   & 1.20\tnote{***} \\
                                  & (3.39) & (3.06) &                \\
2 point goals missed              & 5.81   & 4.23   & 1.57\tnote{***} \\
                                  & (3.71) & (3.22) &                \\
3 point goals made                & 0.91   & 1.61   & -0.71\tnote{***} \\
                                  & (1.53) & (2.02) &                \\
3 point goals missed              & 1.78   & 2.94   & -1.16\tnote{***} \\
                                  & (2.30) & (2.87) &                \\
Offensive rebounds                & 2.20   & 1.51   & 0.69\tnote{***} \\
                                  & (2.44) & (2.04) &                \\
Defensive rebounds                & 4.84   & 4.34   & 0.51\tnote{***} \\
                                  & (3.43) & (3.40) &                \\
Assists                           & 3.08   & 3.80   & -0.73\tnote{***} \\
                                  & (2.86) & (3.28) &                \\
Steals                            & 1.64   & 1.53   & 0.11\tnote{***}  \\
                                  & (1.87) & (1.80) &                \\
Blocks                            & 0.77   & 0.68   & 0.09\tnote{***} \\
                                  & (1.42) & (1.44) &                \\
Turnovers                         & 3.06   & 2.89   & 0.18\tnote{***} \\
                                  & (2.62) & (2.62) &                \\
\multicolumn{4}{c}{Game information} \\
Attendance (1,000s)               & 7.88   & 7.76   & 0.12\tnote{**} \\
                                  & (2.56) & (2.45) &                \\
Out of contention                 & 0.03   & 0.03   & 0.00          \\
                                  & (0.16) & (0.16) &                \\
\bottomrule
\end{tabular}
\end{threeparttable}
\end{table}

\begin{table}[H]
\ContinuedFloat
\centering
\captionsetup{
labelformat=empty,
labelsep=none,
justification=centering,
font=normalsize
}
\caption{\\(Continued)}\vspace{-0.8em}
\label{table:plcharsff0710}
\begin{threeparttable}
\renewcommand{\arraystretch}{0.8}
\small
\begin{tabular}{lccc}
\toprule\toprule
& \multicolumn{1}{c}{Black players} 
& \multicolumn{1}{c}{Non-Black players} 
& \multicolumn{1}{c}{} \\
\cmidrule(lr){2-2} \cmidrule(lr){3-3}
& \multicolumn{1}{c}{\makecell{Mean \\ (SD)}} 
& \multicolumn{1}{c}{\makecell{Mean \\ (SD)}} 
& \multicolumn{1}{c}{Difference} \\
\midrule
\multicolumn{4}{c}{Player characteristics} \\
Age                               & 27.24  & 28.22  & -0.98\tnote{***} \\
                                  & (4.03) & (3.50) &                \\
WNBA experience (yrs)             & 5.42   & 6.47   & -1.05\tnote{***} \\
                                  & (3.41) & (3.21) &                \\
All-WNBA this year                & 0.08   & 0.17   & -0.09\tnote{**} \\
                                  & (0.28) & (0.38) &                \\
Center                            & 0.14   & 0.09   & 0.05          \\
                                  & (0.34) & (0.28) &                \\
Forward                           & 0.44   & 0.30   & 0.13          \\
                                  & (0.50) & (0.46) &                \\
Guard                             & 0.42   & 0.61   & -0.18\tnote{**} \\
                                  & (0.49) & (0.49) &                \\
Starter                           & 0.69   & 0.69   & -0.00         \\
                                  & (0.46) & (0.46) &                \\
Height (cm.)                      & 183.37 & 182.95 & 0.42          \\
                                  & (8.36) & (9.21) &                \\
Weight (kg.)                      & 77.01  & 76.04  & 0.97          \\
                                  & (9.77) & (10.74)&                \\

\multicolumn{4}{c}{Referees} \\
0 non-Black referees              & 0.07   & 0.07   & 0.00          \\
                                  & (0.26) & (0.25) &                \\
1 non-Black referees              & 0.27   & 0.26   & 0.01          \\
                                  & (0.45) & (0.44) &                \\
2 non-Black referees              & 0.42   & 0.42   & 0.00          \\
                                  & (0.49) & (0.49) &                \\
3 non-Black referees              & 0.23   & 0.25   & -0.02\tnote{*}  \\
                                  & (0.42) & (0.43) &                \\
\text{\#} non-Black referees      & 1.82   & 1.85   & -0.03\tnote{**} \\
                                  & (0.87) & (0.87) &                \\
\bottomrule\bottomrule
\end{tabular}

\begin{tablenotes}[para,flushleft]
\scriptsize
\item All observations are weighted by minutes played. \\
\(***\), \(**\), and \(*\) Differences statistically significant at 1\%, 5\%, and 10\%, respectively.
\end{tablenotes}
\end{threeparttable}
\end{table}

\begin{table}[H]
\centering
\captionsetup{
  justification=centering,
  labelsep=none,
  font=normalsize
}
\caption{\\Summary statistics: FairFace classification (2011-2014)}\vspace{-0.8em}
\label{table:plcharsff1114}
\begin{threeparttable}
\renewcommand{\arraystretch}{0.8}
\small
\begin{tabular}{lccc}
\toprule\toprule
& \multicolumn{1}{c}{Black players} 
& \multicolumn{1}{c}{Non-Black players} 
& \multicolumn{1}{c}{} \\
\cmidrule(lr){2-2} \cmidrule(lr){3-3}
& \multicolumn{1}{c}{\makecell{Mean \\ (SD)}} 
& \multicolumn{1}{c}{\makecell{Mean \\ (SD)}} 
& \multicolumn{1}{c}{Difference} \\
\midrule
\multicolumn{4}{c}{Raw player statistics} \\
Minutes played                    & 26.14  & 24.62  & 1.52\tnote{***} \\
                                  & (8.37) & (8.53) &                \\
Fouls                             & 2.21   & 1.89   & 0.32\tnote{***} \\
                                  & (1.45) & (1.41) &                \\
Points                            & 10.66  & 9.29   & 1.37\tnote{***} \\
                                  & (7.00) & (6.95) &                \\
\multicolumn{4}{c}{Player productivity: stats × 40/minutes played} \\
Fouls                             & 3.63   & 3.32   & 0.31\tnote{***} \\
                                  & (2.83) & (2.99) &                \\
Points                            & 15.50  & 14.08  & 1.42\tnote{***} \\
                                  & (8.51) & (8.96) &                \\
Free throws made                  & 2.95   & 2.50   & 0.46\tnote{***} \\
                                  & (3.26) & (3.21) &                \\
Free throws missed                & 0.89   & 0.63   & 0.26\tnote{***} \\
                                  & (1.51) & (1.28) &                \\
2 point goals made                & 4.84   & 3.67   & 1.17\tnote{***} \\
                                  & (3.35) & (3.02) &                \\
2 point goals missed              & 5.71   & 4.31   & 1.40\tnote{***} \\
                                  & (3.62) & (3.24) &                \\
3 point goals made                & 0.95   & 1.42   & -0.46\tnote{***} \\
                                  & (1.57) & (1.92) &                \\
3 point goals missed              & 1.92   & 2.42   & -0.50\tnote{***} \\
                                  & (2.39) & (2.59) &                \\
Offensive rebounds                & 2.04   & 1.40   & 0.64\tnote{***} \\
                                  & (2.37) & (2.06) &                \\
Defensive rebounds                & 4.88   & 4.40   & 0.48\tnote{***} \\
                                  & (3.55) & (3.40) &                \\
Assists                           & 3.10   & 4.17   & -1.07\tnote{***} \\
                                  & (2.78) & (3.46) &                \\
Steals                            & 1.61   & 1.37   & 0.23\tnote{***} \\
                                  & (1.78) & (1.69) &                \\
Blocks                            & 0.79   & 0.70   & 0.09\tnote{***} \\
                                  & (1.43) & (1.42) &                \\
Turnovers                         & 2.76   & 2.78   & -0.02          \\
                                  & (2.39) & (2.57) &                \\
\multicolumn{4}{c}{Game information} \\
Attendance (1,000s)               & 7.61   & 7.69   & -0.08\tnote{*} \\
                                  & (2.30) & (2.28) &                \\
Out of contention                 & 0.04   & 0.02   & 0.02\tnote{***} \\
                                  & (0.19) & (0.15) &                \\
\bottomrule
\end{tabular}
\end{threeparttable}
\end{table}

\begin{table}[H]
\ContinuedFloat
\centering
\captionsetup{
labelformat=empty,
labelsep=none,
justification=centering,
font=normalsize
}
\caption{\\(Continued)}\vspace{-0.8em}
\label{table:plcharsff1114}
\begin{threeparttable}
\renewcommand{\arraystretch}{0.8}
\small
\begin{tabular}{lccc}
\toprule\toprule
& \multicolumn{1}{c}{Black players} 
& \multicolumn{1}{c}{Non-Black players} 
& \multicolumn{1}{c}{} \\
\cmidrule(lr){2-2} \cmidrule(lr){3-3}
& \multicolumn{1}{c}{\makecell{Mean \\ (SD)}} 
& \multicolumn{1}{c}{\makecell{Mean \\ (SD)}} 
& \multicolumn{1}{c}{Difference} \\
\midrule
\multicolumn{4}{c}{Player characteristics} \\
Age                               & 27.30  & 28.80  & -1.50\tnote{**} \\
                                  & (3.82) & (4.80) &                \\
WNBA experience (yrs)             & 5.74   & 7.04   & -1.30\tnote{**} \\
                                  & (3.53) & (4.77) &                \\
All-WNBA this year                & 0.12   & 0.15   & -0.03           \\
                                  & (0.32) & (0.36) &                \\
Center                            & 0.13   & 0.12   & 0.00           \\
                                  & (0.33) & (0.33) &                \\
Forward                           & 0.43   & 0.30   & 0.13           \\
                                  & (0.50) & (0.46) &                \\
Guard                             & 0.44   & 0.61   & -0.18\tnote{**} \\
                                  & (0.50) & (0.49) &                \\
Starter                           & 0.70   & 0.65   & 0.05\tnote{***} \\
                                  & (0.46) & (0.48) &                \\
Height (cm.)                      & 183.31 & 182.51 & 0.80           \\
                                  & (8.81) & (8.44) &                \\
Weight (kg.)                      & 78.01  & 76.88  & 1.13           \\
                                  & (11.34)& (10.86)&                \\

\multicolumn{4}{c}{Referees} \\
0 non-Black referees              & 0.10   & 0.09   & 0.01           \\
                                  & (0.30) & (0.29) &                \\
1 non-Black referees              & 0.33   & 0.33   & 0.00           \\
                                  & (0.47) & (0.47) &                \\
2 non-Black referees              & 0.39   & 0.39   & -0.01          \\
                                  & (0.49) & (0.49) &                \\
3 non-Black referees              & 0.18   & 0.18   & -0.00          \\
                                  & (0.39) & (0.39) &                \\
\text{\#} non-Black referees      & 1.65   & 1.67   & -0.1           \\
                                  & (0.89) & (0.88) &                \\
\bottomrule\bottomrule
\end{tabular}

\begin{tablenotes}[para,flushleft]
\scriptsize
\item All observations are weighted by minutes played. \\
\(***\), \(**\), and \(*\) Differences statistically significant at 1\%, 5\%, and 10\%, respectively.
\end{tablenotes}
\end{threeparttable}
\end{table}

\begin{table}[H]
\centering
\captionsetup{
  justification=centering,
  labelsep=none,
  font=normalsize
}
\caption{\\Summary statistics: Human raters}\vspace{-0.8em}
\label{table:plcharshr}
\begin{threeparttable}
\renewcommand{\arraystretch}{0.8}
\small
\begin{tabular}{lccc}
\toprule\toprule
& \multicolumn{1}{c}{Black players} & \multicolumn{1}{c}{Non-Black players} & \multicolumn{1}{c}{} \\
\cmidrule(lr){2-2} \cmidrule(lr){3-3}
& \multicolumn{1}{c}{\makecell{Mean \\ (SD)}} & \multicolumn{1}{c}{\makecell{Mean \\ (SD)}} & \multicolumn{1}{c}{Difference} \\
\midrule
\multicolumn{4}{c}{Raw player statistics} \\
Minutes played                    & 26.04  & 25.80  & 0.24\tnote{***} \\
                                  & (8.67) & (8.93) &                \\
Fouls                             & 2.34   & 2.13   & 0.21\tnote{***} \\
                                  & (1.49) & (1.50) &                \\
Points                            & 10.34  & 9.82   & 0.53\tnote{***} \\
                                  & (6.88) & (7.20) &                \\
\multicolumn{4}{c}{Player productivity: stats × 40/minutes played} \\
Fouls                             & 3.88   & 3.55   & 0.33\tnote{***} \\
                                  & (2.99) & (3.00) &                \\
Points                            & 15.12  & 14.31  & 0.81\tnote{***} \\
                                  & (8.46) & (8.90) &                \\
Free throws made                  & 3.08   & 2.68   & 0.40\tnote{***} \\
                                  & (3.36) & (3.29) &                \\
Free throws missed                & 1.01   & 0.68   & 0.33\tnote{***} \\
                                  & (1.68) & (1.36) &                \\
2 point goals made                & 4.70   & 3.70   & 1.00\tnote{***} \\
                                  & (3.31) & (3.03) &                \\
2 point goals missed              & 5.70   & 4.33   & 1.38\tnote{***} \\
                                  & (3.65) & (3.22) &                \\
3 point goals made                & 0.88   & 1.41   & -0.53\tnote{***} \\
                                  & (1.51) & (1.91) &                \\
3 point goals missed              & 1.76   & 2.52   & -0.76\tnote{***} \\
                                  & (2.29) & (2.70) &                \\
Offensive rebounds                & 2.09   & 1.57   & 0.53\tnote{***} \\
                                  & (2.37) & (2.09) &                \\
Defensive rebounds                & 4.72   & 4.37   & 0.35\tnote{***} \\
                                  & (3.42) & (3.40) &                \\
Assists                           & 3.04   & 3.75   & -0.70\tnote{***} \\
                                  & (2.80) & (3.20) &                \\
Steals                            & 1.61   & 1.48   & 0.13\tnote{***} \\
                                  & (1.82) & (1.74) &                \\
Blocks                            & 0.75   & 0.74   & 0.00           \\
                                  & (1.39) & (1.48) &                \\
Turnovers                         & 2.90   & 2.84   & 0.06\tnote{**} \\
                                  & (2.51) & (2.54) &                \\
\multicolumn{4}{c}{Game information} \\
Attendance (1,000s)               & 7.84   & 7.78   & 0.06\tnote{*}  \\
                                  & (2.48) & (2.41) &                \\
Out of contention                 & 0.03   & 0.03   & 0.01\tnote{***} \\
                                  & (0.18) & (0.16) &                \\
\bottomrule
\end{tabular}
\end{threeparttable}
\end{table}

\begin{table}[H]
\ContinuedFloat
\centering
\captionsetup{
labelformat=empty,
labelsep=none,
justification=centering,
font=normalsize
}
\caption{\\(Continued)}\vspace{-0.8em}
\label{table:plcharshr}
\begin{threeparttable}
\renewcommand{\arraystretch}{0.8}
\small
\begin{tabular}{lccc}
\toprule\toprule
& \multicolumn{1}{c}{Black players} & \multicolumn{1}{c}{Non-Black players} & \multicolumn{1}{c}{} \\
\cmidrule(lr){2-2} \cmidrule(lr){3-3}
& \multicolumn{1}{c}{\makecell{Mean \\ (SD)}} & \multicolumn{1}{c}{\makecell{Mean \\ (SD)}} & \multicolumn{1}{c}{Difference} \\
\midrule
\multicolumn{4}{c}{Player characteristics} \\
Age                               & 27.34  & 27.83  & -0.49\tnote{*}  \\
                                  & (3.99) & (3.95) &                \\
WNBA experience (yrs)             & 5.39   & 6.05   & -0.67\tnote{***} \\
                                  & (3.26) & (3.65) &                \\
All-WNBA this year                & 0.11   & 0.15   & -0.04\tnote{*}  \\
                                  & (0.31) & (0.36) &                \\
Center                            & 0.15   & 0.11   & 0.04           \\
                                  & (0.35) & (0.31) &                \\
Forward                           & 0.43   & 0.33   & 0.10           \\
                                  & (0.49) & (0.47) &                \\
Guard                             & 0.42   & 0.56   & -0.14\tnote{*} \\
                                  & (0.49) & (0.50) &                \\
Starter                           & 0.69   & 0.70   & -0.00          \\
                                  & (0.46) & (0.46) &                \\
Height (cm.)                      & 183.33 & 183.60 & -0.27          \\
                                  & (8.38) & (9.60) &                \\
Weight (kg.)                      & 77.17  & 76.34  & 0.83           \\
                                  & (10.46)& (10.57)&                \\

\multicolumn{4}{c}{Referees} \\
0 non-Black referees              & 0.09   & 0.08   & 0.01\tnote{***} \\
                                  & (0.28) & (0.27) &                \\
1 non-Black referees              & 0.31   & 0.29   & 0.01\tnote{**}  \\
                                  & (0.46) & (0.46) &                \\
2 non-Black referees              & 0.40   & 0.41   & -0.01\tnote{*} \\
                                  & (0.49) & (0.49) &                \\
3 non-Black referees              & 0.20   & 0.22   & -0.01\tnote{**} \\
                                  & (0.40) & (0.41) &                \\
\text{\#} non-Black referees      & 1.71   & 1.77   & -0.05\tnote{***} \\
                                  & (0.89) & (0.87) &                \\
\bottomrule\bottomrule
\end{tabular}
\begin{tablenotes}[para,flushleft]
\scriptsize
\item All observations are weighted by minutes played. \\
\(***\), \(**\), and \(*\) Differences statistically significant at 1\%, 5\%, and 10\%, respectively.
\end{tablenotes}
\end{threeparttable}
\end{table}

\begin{table}[H]
\centering
\captionsetup{
  justification=centering,
  labelsep=none,
  font=normalsize
}
\caption{\\Summary statistics: Human raters (2004-2006)}\vspace{-0.8em}
\label{table:plcharshr0406}
\begin{threeparttable}
\renewcommand{\arraystretch}{0.8}
\small
\begin{tabular}{lccc}
\toprule\toprule
& \multicolumn{1}{c}{Black players} 
& \multicolumn{1}{c}{Non-Black players} 
& \multicolumn{1}{c}{} \\
\cmidrule(lr){2-2} \cmidrule(lr){3-3}
& \multicolumn{1}{c}{\makecell{Mean \\ (SD)}} 
& \multicolumn{1}{c}{\makecell{Mean \\ (SD)}} 
& \multicolumn{1}{c}{Difference} \\
\midrule
\multicolumn{4}{c}{Raw player statistics} \\
Minutes played                    & 26.25  & 26.42  & -0.17          \\
                                  & (8.92) & (9.03) &                \\
Fouls                             & 2.43   & 2.22   & 0.20\tnote{***} \\
                                  & (1.53) & (1.51) &                \\
Points                            & 9.66   & 9.59   & 0.08           \\
                                  & (6.53) & (6.87) &                \\
\multicolumn{4}{c}{Player productivity: stats × 40/minutes played} \\
Fouls                             & 3.97   & 3.63   & 0.35\tnote{***} \\
                                  & (3.00) & (2.96) &                \\
Points                            & 13.96  & 13.71  & 0.26\tnote{*}   \\
                                  & (7.94) & (8.37) &                \\
Free throws made                  & 2.92   & 2.74   & 0.18\tnote{***} \\
                                  & (3.21) & (3.22) &                \\
Free throws missed                & 1.10   & 0.76   & 0.34\tnote{***} \\
                                  & (1.76) & (1.45) &                \\
2 point goals made                & 4.45   & 3.60   & 0.85\tnote{***} \\
                                  & (3.16) & (2.84) &                \\
2 point goals missed              & 5.58   & 4.32   & 1.26\tnote{***} \\
                                  & (3.59) & (3.17) &                \\
3 point goals made                & 0.71   & 1.25   & -0.54\tnote{***}\\
                                  & (1.32) & (1.76) &                \\
3 point goals missed              & 1.46   & 2.28   & -0.82\tnote{***}\\
                                  & (2.08) & (2.52) &                \\
Offensive rebounds                & 2.07   & 1.60   & 0.47\tnote{***} \\
                                  & (2.30) & (2.01) &                \\
Defensive rebounds                & 4.39   & 4.16   & 0.23\tnote{***} \\
                                  & (3.23) & (3.25) &                \\
Assists                           & 2.88   & 3.48   & -0.60\tnote{***}\\
                                  & (2.71) & (2.89) &                \\
Steals                            & 1.55   & 1.49   & 0.06\tnote{*}   \\
                                  & (1.79) & (1.74) &                \\
Blocks                            & 0.69   & 0.75   & -0.06\tnote{**}\\
                                  & (1.32) & (1.49) &                \\
Turnovers                         & 2.87   & 2.83   & 0.04           \\
                                  & (2.50) & (2.45) &                \\
\multicolumn{4}{c}{Game information} \\
Attendance (1,000s)               & 8.09   & 7.90   & 0.19\tnote{***}  \\
                                  & (2.57) & (2.45) &                \\
Out of contention                 & 0.04   & 0.03   & 0.01\tnote{***}\\
                                  & (0.19) & (0.16) &                \\
\bottomrule
\end{tabular}
\end{threeparttable}
\end{table}

\begin{table}[H]
\ContinuedFloat
\centering
\captionsetup{
labelformat=empty,
labelsep=none,
justification=centering,
font=normalsize
}
\caption{\\(Continued)}\vspace{-0.8em}
\label{table:plcharshr0406}
\begin{threeparttable}
\renewcommand{\arraystretch}{0.8}
\small
\begin{tabular}{lccc}
\toprule\toprule
& \multicolumn{1}{c}{Black players} 
& \multicolumn{1}{c}{Non-Black players} 
& \multicolumn{1}{c}{} \\
\cmidrule(lr){2-2} \cmidrule(lr){3-3}
& \multicolumn{1}{c}{\makecell{Mean \\ (SD)}} 
& \multicolumn{1}{c}{\makecell{Mean \\ (SD)}} 
& \multicolumn{1}{c}{Difference} \\
\midrule
\multicolumn{4}{c}{Player characteristics} \\
Age                               & 27.62  & 26.50  & 1.12\tnote{***} \\
                                  & (4.14) & (3.41) &                \\
WNBA experience (yrs)             & 4.98   & 4.53   & 0.46\tnote{*}  \\
                                  & (2.63) & (2.38) &                \\
All-WNBA this year                & 0.11   & 0.13   & -0.01         \\
                                  & (0.32) & (0.33) &                \\
Center                            & 0.19   & 0.13   & 0.06          \\
                                  & (0.39) & (0.34) &                \\
Forward                           & 0.43   & 0.34   & 0.09          \\
                                  & (0.49) & (0.47) &                \\
Guard                             & 0.38   & 0.53   & -0.15         \\
                                  & (0.49) & (0.50) &                \\
Starter                           & 0.70   & 0.72   & -0.02\tnote{**}\\
                                  & (0.46) & (0.45) &                \\
Height (cm.)                      & 183.65 & 184.41 & -0.76         \\
                                  & (7.93) & (10.51)&                \\
Weight (kg.)                      & 76.42  & 75.76  & 0.66          \\
                                  & (9.94) & (10.20)&                \\

\multicolumn{4}{c}{Referees} \\
0 non-Black referees              & 0.03   & 0.03   & 0.00          \\
                                  & (0.18) & (0.18) &                \\
1 non-Black referees              & 0.24   & 0.25   & -0.01         \\
                                  & (0.43) & (0.43) &                \\
2 non-Black referees              & 0.46   & 0.46   & 0.00          \\
                                  & (0.50) & (0.50) &                \\
3 non-Black referees              & 0.27   & 0.26   & 0.01          \\
                                  & (0.44) & (0.44) &                \\
\# non-Black referees             & 1.96   & 1.94   & 0.02          \\
                                  & (0.80) & (0.80) &                \\
\bottomrule\bottomrule
\end{tabular}

\begin{tablenotes}[para,flushleft]
\scriptsize
\item All observations are weighted by minutes played. \\
\(***\), \(**\), and \(*\) Differences statistically significant at 1\%, 5\%, and 10\%, respectively.
\end{tablenotes}
\end{threeparttable}
\end{table}

\begin{table}[H]
\centering
\captionsetup{
  justification=centering,
  labelsep=none,
  font=normalsize
}
\caption{\\Summary statistics: Human raters (2007-2010)}\vspace{-0.8em}
\label{table:plcharshr0710}
\begin{threeparttable}
\renewcommand{\arraystretch}{0.8}
\small
\begin{tabular}{lccc}
\toprule\toprule
& \multicolumn{1}{c}{Black players} 
& \multicolumn{1}{c}{Non-Black players} 
& \multicolumn{1}{c}{Difference} \\
\cmidrule(lr){2-2} \cmidrule(lr){3-3}
& \multicolumn{1}{c}{\makecell{Mean \\ (SD)}} 
& \multicolumn{1}{c}{\makecell{Mean \\ (SD)}} 
& \\
\midrule
\multicolumn{4}{c}{Raw player statistics} \\
Minutes played                    & 25.85  & 25.87  & -0.02          \\
                                  & (8.77) & (9.07) &                \\
Fouls                             & 2.41   & 2.17   & 0.24\tnote{***} \\
                                  & (1.49) & (1.53) &                \\
Points                            & 10.54  & 10.25  & 0.29\tnote{**} \\
                                  & (6.95) & (7.65) &                \\
\multicolumn{4}{c}{Player productivity: stats × 40/minutes played} \\
Fouls                             & 4.06   & 3.61   & 0.44\tnote{***} \\
                                  & (3.12) & (3.05) &                \\
Points                            & 15.58  & 14.87  & 0.71\tnote{***} \\
                                  & (8.67) & (9.30) &                \\
Free throws made                  & 3.31   & 2.76   & 0.55\tnote{***} \\
                                  & (3.53) & (3.43) &                \\
Free throws missed                & 1.08   & 0.63   & 0.44\tnote{***} \\
                                  & (1.78) & (1.34) &                \\
2 point goals made                & 4.77   & 3.68   & 1.09\tnote{***} \\
                                  & (3.38) & (3.12) &                \\
2 point goals missed              & 5.80   & 4.27   & 1.54\tnote{***} \\
                                  & (3.72) & (3.22) &                \\
3 point goals made                & 0.91   & 1.58   & -0.67\tnote{***}\\
                                  & (1.54) & (2.02) &                \\
3 point goals missed              & 1.79   & 2.88   & -1.09\tnote{***}\\
                                  & (2.31) & (2.87) &                \\
Offensive rebounds                & 2.18   & 1.57   & 0.61\tnote{***} \\
                                  & (2.42) & (2.12) &                \\
Defensive rebounds                & 4.82   & 4.40   & 0.43\tnote{***} \\
                                  & (3.42) & (3.45) &                \\
Assists                           & 3.09   & 3.77   & -0.68\tnote{***}\\
                                  & (2.86) & (3.28) &                \\
Steals                            & 1.64   & 1.53   & 0.12\tnote{***}\\
                                  & (1.87) & (1.79) &                \\
Blocks                            & 0.76   & 0.72   & 0.04\tnote{*}  \\
                                  & (1.41) & (1.47) &                \\
Turnovers                         & 3.06   & 2.89   & 0.17\tnote{***}\\
                                  & (2.61) & (2.63) &                \\
\multicolumn{4}{c}{Game information} \\
Attendance (1,000s)               & 7.88   & 7.76   & 0.12\tnote{**} \\
                                  & (2.56) & (2.45) &                \\
Out of contention                 & 0.03   & 0.03   & 0.00          \\
                                  & (0.16) & (0.16) &                \\
\bottomrule
\end{tabular}
\end{threeparttable}
\end{table}

\begin{table}[H]
\ContinuedFloat
\centering
\captionsetup{
labelformat=empty,
labelsep=none,
justification=centering,
font=normalsize
}
\caption{\\(Continued)}\vspace{-0.8em}
\label{table:plcharshr0710}
\begin{threeparttable}
\renewcommand{\arraystretch}{0.8}
\small
\begin{tabular}{lccc}
\toprule\toprule
& \multicolumn{1}{c}{Black players} 
& \multicolumn{1}{c}{Non-Black players} 
& \multicolumn{1}{c}{Difference} \\
\cmidrule(lr){2-2} \cmidrule(lr){3-3}
& \multicolumn{1}{c}{\makecell{Mean \\ (SD)}} 
& \multicolumn{1}{c}{\makecell{Mean \\ (SD)}} 
& \\
\midrule
\multicolumn{4}{c}{Player characteristics} \\
Age                               & 27.23  & 28.23  & -0.99\tnote{***} \\
                                  & (4.04) & (3.46) &                \\
WNBA experience (yrs)             & 5.39   & 6.53   & -1.13\tnote{***}\\
                                  & (3.41) & (3.18) &                \\
All-WNBA this year                & 0.08   & 0.17   & -0.09\tnote{**} \\
                                  & (0.28) & (0.38) &                \\
Center                            & 0.14   & 0.09   & 0.05          \\
                                  & (0.34) & (0.28) &                \\
Forward                           & 0.43   & 0.32   & 0.12          \\
                                  & (0.50) & (0.46) &                \\
Guard                             & 0.43   & 0.60   & -0.17\tnote{*} \\
                                  & (0.49) & (0.49) &                \\
Starter                           & 0.69   & 0.70   & -0.01         \\
                                  & (0.46) & (0.46) &                \\
Height (cm.)                      & 183.29 & 183.19 & 0.10          \\
                                  & (8.31) & (9.31) &                \\
Weight (kg.)                      & 76.97  & 76.17  & 0.81          \\
                                  & (9.79) & (10.69)&                \\

\multicolumn{4}{c}{Referees} \\
0 non-Black referees              & 0.07   & 0.07   & 0.00          \\
                                  & (0.26) & (0.25) &                \\
1 non-Black referees              & 0.29   & 0.28   & 0.01          \\
                                  & (0.45) & (0.45) &                \\
2 non-Black referees              & 0.42   & 0.42   & 0.00          \\
                                  & (0.49) & (0.49) &                \\
3 non-Black referees              & 0.22   & 0.23   & -0.01         \\
                                  & (0.42) & (0.42) &                \\
\text{\#} non-Black referees      & 1.79   & 1.82   & -0.03         \\
                                  & (0.87) & (0.87) &                \\
\bottomrule\bottomrule
\end{tabular}
\begin{tablenotes}[para,flushleft]
\scriptsize
\item All observations are weighted by minutes played. \\
\(***\), \(**\), and \(*\) Differences statistically significant at 1\%, 5\%, and 10\%, respectively.
\end{tablenotes}
\end{threeparttable}
\end{table}

\begin{table}[H]
\centering
\captionsetup{
  justification=centering,
  labelsep=none,
  font=normalsize
}
\caption{\\Summary statistics: Human raters (2011–2014)}\vspace{-0.8em}
\label{table:plcharshr1114}
\begin{threeparttable}
\renewcommand{\arraystretch}{0.8}
\small
\begin{tabular}{lccc}
\toprule\toprule
& \multicolumn{1}{c}{Black players} 
& \multicolumn{1}{c}{Non-Black players} 
& \multicolumn{1}{c}{Difference} \\
\cmidrule(lr){2-2} \cmidrule(lr){3-3}
& \multicolumn{1}{c}{\makecell{Mean \\ (SD)}} 
& \multicolumn{1}{c}{\makecell{Mean \\ (SD)}} 
& \\
\midrule
\multicolumn{4}{c}{Raw player statistics} \\
Minutes played                    & 26.08  & 24.92  & 1.16\tnote{***} \\
                                  & (8.38) & (8.50) &                \\
Fouls                             & 2.20   & 1.94   & 0.26\tnote{***} \\
                                  & (1.45) & (1.42) &                \\
Points                            & 10.63  & 9.49   & 1.14\tnote{***} \\
                                  & (7.01) & (6.93) &                \\
\multicolumn{4}{c}{Player productivity: stats × 40/minutes played} \\
Fouls                             & 3.63   & 3.35   & 0.28\tnote{***} \\
                                  & (2.84) & (2.95) &                \\
Points                            & 15.47  & 14.25  & 1.23\tnote{***} \\
                                  & (8.54) & (8.88) &                \\
Free throws made                  & 2.96   & 2.49   & 0.48\tnote{***} \\
                                  & (3.27) & (3.18) &                \\
Free throws missed                & 0.89   & 0.65   & 0.24\tnote{***} \\
                                  & (1.52) & (1.27) &                \\
2 point goals made                & 4.80   & 3.86   & 0.94\tnote{***} \\
                                  & (3.34) & (3.12) &                \\
2 point goals missed              & 5.70   & 4.42   & 1.28\tnote{***} \\
                                  & (3.62) & (3.28) &                \\
3 point goals made                & 0.97   & 1.35   & –0.38\tnote{***} \\
                                  & (1.58) & (1.90) &                \\
3 point goals missed              & 1.95   & 2.30   & –0.36\tnote{***} \\
                                  & (2.39) & (2.58) &                \\
Offensive rebounds                & 2.02   & 1.52   & 0.50\tnote{***} \\
                                  & (2.36) & (2.13) &                \\
Defensive rebounds                & 4.84   & 4.58   & 0.26\tnote{***} \\
                                  & (3.53) & (3.49) &                \\
Assists                           & 3.12   & 4.04   & –0.93\tnote{***} \\
                                  & (2.79) & (3.43) &                \\
Steals                            & 1.61   & 1.39   & 0.22\tnote{***} \\
                                  & (1.79) & (1.68) &                \\
Blocks                            & 0.77   & 0.77   & 0.00           \\
                                  & (1.42) & (1.48) &                \\
Turnovers                         & 2.76   & 2.76   & 0.00           \\
                                  & (2.39) & (2.54) &                \\
\multicolumn{4}{c}{Game information} \\
Attendance (1,000s)               & 7.62   & 7.65   & –0.03           \\
                                  & (2.29) & (2.29) &                \\
Out of contention                 & 0.04   & 0.02   & 0.02\tnote{***} \\
                                  & (0.19) & (0.15) &                \\
\bottomrule
\end{tabular}
\end{threeparttable}
\end{table}

\begin{table}[H]
\ContinuedFloat
\centering
\captionsetup{
labelformat=empty,
labelsep=none,
justification=centering,
font=normalsize
}
\caption{\\(Continued)}\vspace{-0.8em}
\label{table:plcharshr1114_cont}
\begin{threeparttable}
\renewcommand{\arraystretch}{0.8}
\small
\begin{tabular}{lccc}
\toprule\toprule
& \multicolumn{1}{c}{Black players} 
& \multicolumn{1}{c}{Non-Black players} 
& \multicolumn{1}{c}{Difference} \\
\cmidrule(lr){2-2} \cmidrule(lr){3-3}
& \multicolumn{1}{c}{\makecell{Mean \\ (SD)}} 
& \multicolumn{1}{c}{\makecell{Mean \\ (SD)}} 
& \\
\midrule
\multicolumn{4}{c}{Player characteristics} \\
Age                               & 27.25  & 28.91  & –1.66\tnote{***} \\
                                  & (3.82) & (4.71) &                \\
WNBA experience (yrs)             & 5.67   & 7.27   & –1.60\tnote{***} \\
                                  & (3.48) & (4.77) &                \\
All-WNBA this year                & 0.12   & 0.15   & –0.03           \\
                                  & (0.33) & (0.35) &                \\
Center                            & 0.13   & 0.12   & 0.01           \\
                                  & (0.33) & (0.32) &                \\
Forward                           & 0.42   & 0.33   & 0.09           \\
                                  & (0.49) & (0.47) &                \\
Guard                             & 0.45   & 0.55   & –0.10          \\
                                  & (0.50) & (0.50) &                \\
Starter                           & 0.69   & 0.67   & 0.02\tnote{***} \\
                                  & (0.46) & (0.47) &                \\
Height (cm.)                      & 183.15 & 183.17 & –0.02          \\
                                  & (8.75) & (8.73) &                \\
Weight (kg.)                      & 77.90  & 77.33  & 0.57           \\
                                  & (11.38)& (10.78)&                \\
\multicolumn{4}{c}{Referees} \\
0 non-Black referees              & 0.14   & 0.14   & 0.00           \\
                                  & (0.35) & (0.35) &                \\
1 non-Black referees              & 0.38   & 0.37   & 0.01           \\
                                  & (0.48) & (0.48) &                \\
2 non-Black referees              & 0.34   & 0.35   & –0.01          \\
                                  & (0.48) & (0.48) &                \\
3 non-Black referees              & 0.14   & 0.14   & 0.00           \\
                                  & (0.34) & (0.34) &                \\
\# non-Black referees             & 1.47   & 1.48   & 0.01           \\
                                  & (0.90) & (0.90) &                \\
\bottomrule\bottomrule
\end{tabular}

\begin{tablenotes}[para,flushleft]
\scriptsize
\item All observations are weighted by minutes played. \\
\(***\), \(**\), and \(*\) Differences statistically significant at 1\%, 5\%, and 10\%, respectively.
\end{tablenotes}
\end{threeparttable}
\end{table}

\clearpage

\section{Results}
\setcounter{table}{0}
\setcounter{figure}{0}

\begin{table}[H]
\centering
\captionsetup{
    justification=centering, 
    labelsep=none,           
    font=normalsize          
}
\caption{\\Racial bias among WNBA referees subsample where FairFace and human raters agree}\vspace{-0.8em}
\label{table:mlhragree}
\begin{threeparttable}
\renewcommand{\arraystretch}{0.8} 
\footnotesize 
\begin{tabular}{lccccc} 
\toprule
\toprule
 & Pre-awareness & \multicolumn{2}{c}{Post-awareness} & \multicolumn{2}{c}{Change in coefficient} \\
\cmidrule(r){2-2} \cmidrule(r){3-4} \cmidrule(r){5-6}
 & 2004--2006 & 2007--2010 & 2011--2014 & \makecell{From 2004--2006\\to 2007--2010} & \makecell{From 2007--2010\\to 2011--2014} \\
\midrule
Black \(\times\) & 0.236 & -0.264\textsuperscript{*} & 0.334\textsuperscript{*} & 0.237 & -0.262\textsuperscript{*} \\
fraction non-Black referees & \small{(0.204)} & \small{(0.141)} & \small{(0.174)} & \small{(0.204)} & \small{(0.141)} \\
\addlinespace
Post Black \(\times\) &  &  &  & -0.499\textsuperscript{*} & 0.593\textsuperscript{***} \\
fraction non-Black referees &  &  &  & \small{(0.257)} & \small{(0.211)} \\
\addlinespace
N & 11,425 & 16,508 & 12,861 & 27,933 & 29,369 \\
Sample mean & 4.38 & 4.52 & 4.10 & 4.46 & 4.34 \\
\bottomrule
\bottomrule
\end{tabular}
\begin{tablenotes}[para,flushleft]
\setstretch{0.5} 
\tiny
\setlength{\parskip}{0pt} 
\setlength{\parindent}{0pt} 
\setlength{\itemsep}{-1pt} 
\item \scriptsize \textit{Notes:} Estimated for a sub-sample restricted to cases where FairFace and human raters agree on racial classification. The dependent variable is defined as 40×fouls/minutes. Each regression includes player-year fixed effects, game fixed effects, whether a player is a starter, whether a player is playing in her home arena, and the skin color of her coach. The last two columns report the results of a pooled regression, including both a relevant primary period and a subsequent period to measure the change in our variable of interest between the two periods. Each observation is weighted by the number of minutes played. Standard errors are clustered at the player and game levels and appear in parentheses.\\
\textsuperscript{***}p \textless 0.01; \textsuperscript{**}p \textless 0.05; \textsuperscript{*}p \textless 0.1.
\end{tablenotes}
\end{threeparttable}
\end{table}

\begin{table}[H]
\centering
\captionsetup{
    justification=centering,
    labelsep=none,
    font=normalsize
}
\caption{\\Racial bias among WNBA referees using FairFace when controlling for team race}
\vspace{-0.8em}
\label{table:app_race_subsamplesff}
\begin{threeparttable}
\renewcommand{\arraystretch}{0.82}
\footnotesize
\begin{tabular}{lccccc}
\toprule
\toprule
 & Pre-awareness & \multicolumn{2}{c}{Post-awareness} & \multicolumn{2}{c}{Change in coefficient} \\
\cmidrule(r){2-2} \cmidrule(r){3-4} \cmidrule(r){5-6}
 & 2004--2006 & 2007--2010 & 2011--2014 &
 \makecell{From 2004--2006\\to 2007--2010} &
 \makecell{From 2007--2010\\to 2011--2014} \\
\midrule

Black $\times$
& 0.077 & -0.182 & 0.212 & 0.076 & -0.181 \\
fraction non-Black referees & (0.206) & (0.144) & (0.165) & (0.206) & (0.144) \\
\addlinespace
Post Black $\times$
&  &  &  & -0.258 & 0.392\textsuperscript{*} \\
fraction non-Black referees &  &  &  & (0.262) & (0.206) \\
\addlinespace
N & 12,893 & 17,266 & 15,705 & 30,159 & 32,971 \\

\bottomrule
\bottomrule
\end{tabular}

\begin{tablenotes}[para,flushleft]
\setstretch{0.5}
\tiny
\setlength{\parskip}{0pt}
\setlength{\parindent}{0pt}
\setlength{\itemsep}{-1pt}
\item \scriptsize \textit{Notes:} Dependent variable is defined as 40×fouls/minutes. Each regression includes player-year fixed effects, game fixed effects, an indicator for whether the player is a starter, an indicator for whether the player is playing in her home arena, and the skin color of her coach. Also, a control for the player’s team’s racial composition (relative to the opponent) to account for differences in the on-court racial mix that could be correlated with officiating. The last two columns report pooled specifications that combine the primary period and subsequent period, allowing the coefficient on the variable of interest to differ across periods; the reported “post” interaction captures the change in the coefficient between the two periods. Each observation is weighted by minutes played. Standard errors are clustered at the player and game levels and reported in parentheses.
\\
\textsuperscript{***}p \textless 0.01; \textsuperscript{**}p \textless 0.05; \textsuperscript{*}p \textless 0.1.
\end{tablenotes}
\end{threeparttable}
\end{table}

\begin{table}[H]
\centering
\captionsetup{
    justification=centering,
    labelsep=none,
    font=normalsize
}
\caption{\\Racial bias among WNBA referees using human raters when controlling for team race}
\vspace{-0.8em}
\label{table:app_race_subsampleshr}
\begin{threeparttable}
\renewcommand{\arraystretch}{0.82}
\footnotesize
\begin{tabular}{lccccc}
\toprule
\toprule
 & Pre-awareness & \multicolumn{2}{c}{Post-awareness} & \multicolumn{2}{c}{Change in coefficient} \\
\cmidrule(r){2-2} \cmidrule(r){3-4} \cmidrule(r){5-6}
 & 2004--2006 & 2007--2010 & 2011--2014 &
 \makecell{From 2004--2006\\to 2007--2010} &
 \makecell{From 2007--2010\\to 2011--2014} \\
\midrule

Black $\times$
& 0.105 & -0.258\textsuperscript{**} & 0.217 & 0.108 & -0.257\textsuperscript{**} \\
fraction non-Black referees & (0.172) & (0.129) & (0.146) & (0.172) & (0.129) \\
\addlinespace
Post Black $\times$
&  &  &  & -0.364\textsuperscript{*} & 0.472\textsuperscript{**} \\
fraction non-Black referees &  &  &  & (0.220) & (0.185) \\
\addlinespace
N & 12,893 & 17,266 & 15,705 & 30,159 & 32,971 \\

\bottomrule
\bottomrule
\end{tabular}

\begin{tablenotes}[para,flushleft]
\setstretch{0.5}
\tiny
\setlength{\parskip}{0pt}
\setlength{\parindent}{0pt}
\setlength{\itemsep}{-1pt}
\item \scriptsize \textit{Notes:} See comments for Table~\ref{table:app_race_subsamplesff}.
\textsuperscript{***}p \textless 0.01; \textsuperscript{**}p \textless 0.05; \textsuperscript{*}p \textless 0.1.
\end{tablenotes}
\end{threeparttable}
\end{table}

\begin{table}[H]
\centering
\captionsetup{
    justification=centering, 
    labelsep=none,           
    font=normalsize          
}
\caption{\\Racial bias among WNBA referees using FairFace and even 3-year intervals}\vspace{-0.8em}
\label{table:mlrace333}
\begin{threeparttable}
\renewcommand{\arraystretch}{0.8} 
\footnotesize 
\begin{tabular}{lccccc} 
\toprule
\toprule
 & Pre-awareness & \multicolumn{2}{c}{Post-awareness} & \multicolumn{2}{c}{Change in coefficient} \\
\cmidrule(r){2-2} \cmidrule(r){3-4} \cmidrule(r){5-6}
 & 2004--2006 & 2007--2009 & 2010--2012 & \makecell{From 2004--2006\\to 2007--2009} & \makecell{From 2007--2009\\to 2010--2012} \\
\midrule
Black \(\times\) & 0.076 & -0.288 & 0.207 & 0.078 & -0.291\textsuperscript{*} \\
fraction non-Black referees & \small{(0.206)} & \small{(0.176)} & \small{(0.162)} & \small{(0.206)} & \small{(0.176)} \\
\addlinespace
Post Black \(\times\) &  &  &  & -0.366 & 0.499\textsuperscript{**} \\
fraction non-Black referees &  &  &  & \small{(0.275)} & \small{(0.231)} \\
\addlinespace
N & 12,893 & 13,372 & 11,735 & 26,265 & 25,107 \\
Sample mean & 4.38 & 4.64 & 4.12 & 4.51 & 4.40 \\
\bottomrule
\bottomrule
\end{tabular}
\begin{tablenotes}[para,flushleft]
\setstretch{0.5} 
\tiny
\setlength{\parskip}{0pt} 
\setlength{\parindent}{0pt} 
\setlength{\itemsep}{-1pt} 
\item \scriptsize \textit{Notes:} Estimated for even three-year intervals. The dependent variable is defined as 40×fouls/minutes. Crew distance is the mean of the absolute distance in skin tone values between the player and each referee in the officiating crew. Each regression includes player-year fixed effects, game fixed effects, whether a player is a starter, whether a player is playing in her home arena, and her coach's skin tone. The last two columns report the results of a pooled regression, including both a primary period and a subsequent period to measure the changes throughout our main sample years. Each observation is weighted by the number of minutes played. Standard errors are clustered at the player and game levels and appear in parentheses.\\
\textsuperscript{***}p \textless 0.01; \textsuperscript{**}p \textless 0.05; \textsuperscript{*}p \textless 0.1.  
\end{tablenotes}
\end{threeparttable}
\end{table}

\begin{table}[H]
\centering
\captionsetup{
    justification=centering, 
    labelsep=none,           
    font=normalsize          
}
\caption{\\Racial bias among WNBA referees using human raters and even 3-year intervals}\vspace{-0.8em}
\label{table:hurace333}
\begin{threeparttable}
\renewcommand{\arraystretch}{0.8} 
\footnotesize 
\begin{tabular}{lccccc} 
\toprule
\toprule
 & Pre-awareness & \multicolumn{2}{c}{Post-awareness} & \multicolumn{2}{c}{Change in coefficient} \\
\cmidrule(r){2-2} \cmidrule(r){3-4} \cmidrule(r){5-6}
 & 2004--2006 & 2007--2009 & 2010--2012 & \makecell{From 2004--2006\\to 2007--2009} & \makecell{From 2007--2009\\to 2010--2012} \\
\midrule
Black \(\times\) & 0.106 & -0.358\textsuperscript{**} & 0.244 & 0.108 & -0.362\textsuperscript{**} \\
fraction non-Black referees & \small{(0.172)} & \small{(0.168)} & \small{(0.150)} & \small{(0.172)} & \small{(0.168)} \\
\addlinespace
Post Black \(\times\) &  &  &  & -0.466 & 0.608\textsuperscript{**} \\
fraction non-Black referees &  &  &  & \small{(0.238)} & \small{(0.240)} \\
\addlinespace
N & 12,893 & 13,372 & 11,735 & 26,265 & 25,107 \\
Sample mean & 4.38 & 4.64 & 4.12 & 4.51 & 4.40 \\
\bottomrule
\bottomrule
\end{tabular}
\begin{tablenotes}[para,flushleft]
\setstretch{0.5} 
\tiny
\setlength{\parskip}{0pt} 
\setlength{\parindent}{0pt} 
\setlength{\itemsep}{-1pt} 
\item \scriptsize \textit{Notes:} See comments for Table~\ref{table:mlrace333}. \textsuperscript{***}p \textless 0.01; \textsuperscript{**}p \textless 0.05; \textsuperscript{*}p \textless 0.1.  
\end{tablenotes}
\end{threeparttable}
\end{table}

\begin{table}[H]
\centering
\captionsetup{
    justification=centering, 
    labelsep=none, 
    font=normalsize 
}
\caption{\\Colorism among WNBA referees using even 3-year intervals}\vspace{-0.8em}
\label{table:tone333}
\begin{threeparttable}
\renewcommand{\arraystretch}{0.8} 
\footnotesize 
\begin{tabular}{lccccc}        
\toprule
\toprule
 & Pre-awareness & \multicolumn{2}{c}{Post-awareness} & \multicolumn{2}{c}{Change in coefficient} \\
\cmidrule(r){2-2} \cmidrule(r){3-4} \cmidrule(r){5-6}
 & 2004--2006 & 2007--2009 & 2010--2012 & \makecell{From 2004--2006\\to 2007--2009} & \makecell{From 2007--2009\\to 2010--2012} \\
\midrule
Crew distance & -0.000 & -0.021\textsuperscript{***} & 0.002 & -0.000 & -0.021\textsuperscript{***} \\
 & \small{(0.006)} & \small{(0.008)} & \small{(0.008)} & \small{(0.006)} & \small{(0.008)} \\
\addlinespace
Post crew distance &  &  &  & -0.021\textsuperscript{**} & 0.023\textsuperscript{**} \\
 &  &  &  & \small{(0.010)} & \small{(0.012)} \\
\addlinespace
N & 12,893 & 13,372 & 11,735 & 26,265 & 25,107 \\
Sample mean & 4.38 & 4.56 & 4.09 & 4.48 & 4.64 \\
\bottomrule
\bottomrule
\end{tabular}
\begin{tablenotes}[para,flushleft]
\setstretch{0.5} 
\tiny
\setlength{\parskip}{0pt} 
\setlength{\parindent}{0pt} 
\setlength{\itemsep}{-1pt} 
\item \scriptsize \textit{Notes:} See comments for Table~\ref{table:mlrace333}. \textsuperscript{***}p \textless 0.01; \textsuperscript{**}p \textless 0.05; \textsuperscript{*}p \textless 0.1.
\end{tablenotes}
\end{threeparttable}
\end{table}

\begin{table}[H]
\centering
\captionsetup{
    justification=centering,
    labelsep=none,
    font=normalsize
}
\caption{\\Colorism among WNBA referees when controlling for team skin-tone}
\vspace{-0.8em}
\label{table:app_tone_subsamples}
\begin{threeparttable}
\renewcommand{\arraystretch}{0.82}
\footnotesize
\begin{tabular}{lccccc}
\toprule
\toprule
 & Pre-awareness & \multicolumn{2}{c}{Post-awareness} & \multicolumn{2}{c}{Change in coefficient} \\
\cmidrule(r){2-2} \cmidrule(r){3-4} \cmidrule(r){5-6}
 & 2004--2006 & 2007--2010 & 2011--2014 &
 \makecell{From 2004--2006\\to 2007--2010} &
 \makecell{From 2007--2010\\to 2011--2014} \\
\midrule

Crew distance         & -0.000     & -0.020$^{***}$ & 0.006      & -0.000     & -0.020$^{***}$ \\
                      & (0.006)    & (0.007)        & (0.007)    & (0.006)    & (0.007)        \\
\addlinespace
Post crew distance    &            &                &            & -0.020$^{**}$ & 0.026$^{***}$ \\
                      &            &                &            & (0.009)      & (0.010)       \\
\addlinespace
N                     & 30,159     & 30,159         & 32,971     & 30,159      & 32,971        \\

\bottomrule
\bottomrule
\end{tabular}

\begin{tablenotes}[para,flushleft]
\setstretch{0.5}
\tiny
\setlength{\parskip}{0pt}
\setlength{\parindent}{0pt}
\setlength{\itemsep}{-1pt}
\item \scriptsize \textit{Notes:} See comments for Table~\ref{table:app_race_subsamplesff}. \textsuperscript{***}p \textless 0.01; \textsuperscript{**}p \textless 0.05; \textsuperscript{*}p \textless 0.1.
\end{tablenotes}
\end{threeparttable}
\end{table}

\begin{figure} [H]
\centering
\captionsetup{labelfont=normalfont,labelsep=none,justification=centering}
\caption{\\Distribution of colorism by referee in the pre-awareness period (2004-2006)}\vspace{-0.8em}
\captionsetup{justification=raggedright, singlelinecheck=false}
\adjustbox{max totalheight=\textheight,keepaspectratio,center,valign=c}{%
\begin{minipage}{\textwidth}
\centering
\subfloat{\includegraphics[width=1.0\linewidth]{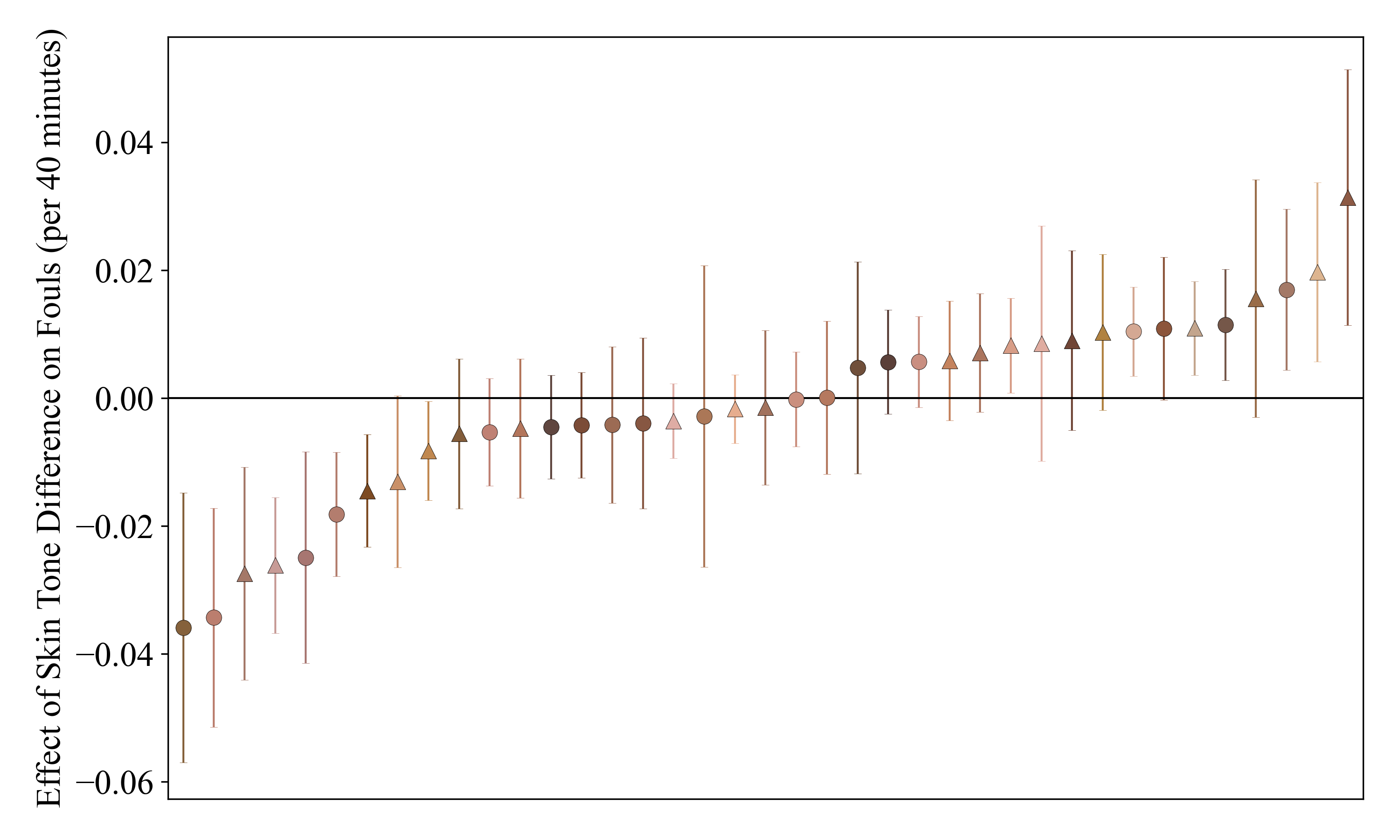}}\\[-0.05em]
\caption*{\justifying\scriptsize 
\textit{Notes:} Each triangle (woman) or circle (man) represents an estimate of the number of extra fouls per 40 minutes an individual WNBA referee calls based on the skin tone of players relative to the referee, with the bars representing the 95\% confidence interval around these estimates. We run separate regressions for each referee using equation (\ref{eq:2}), where instead of the average absolute difference in skin tone between a player and each member of the three-person referee crew, we use the absolute distance of each referee to a specific player. The color of the point estimates and error bars corresponds to the referee’s representative skin tone.
}
\label{fig:indvref0406}
\end{minipage}%
}
\end{figure}

\begin{figure}[h]
\centering
\captionsetup{labelfont=normalfont,labelsep=none,justification=centering}
\caption{\\Distribution of colorism by referee in the \textit{immediate} post-awareness period (2007-2010)}\vspace{-0.8em}
\captionsetup{justification=raggedright, singlelinecheck=false}
\adjustbox{max totalheight=\textheight,keepaspectratio,center,valign=c}{%
\begin{minipage}{\textwidth}
\centering
\subfloat{\includegraphics[width=1.0\linewidth]{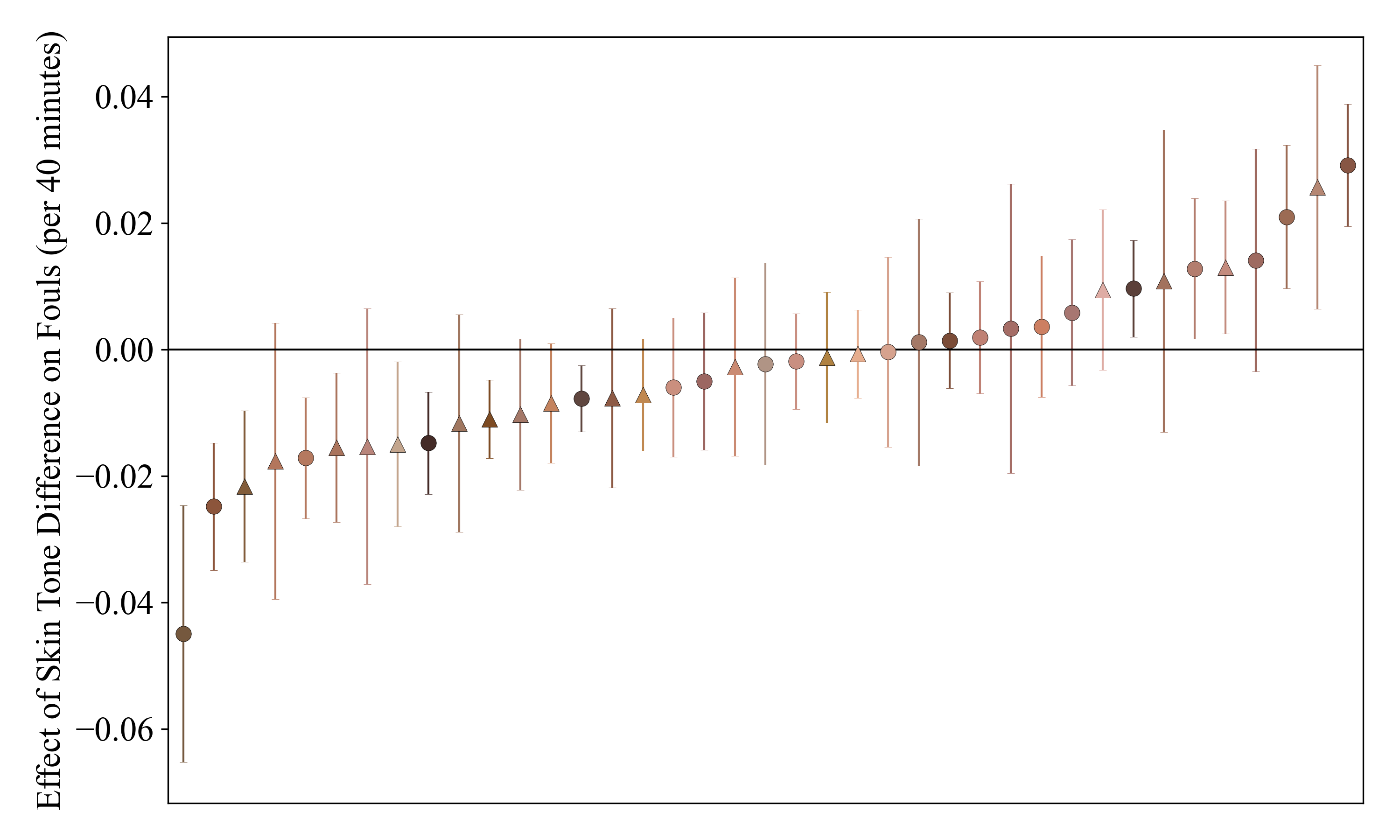}}\\[-0.05em]
\caption*{\justifying\scriptsize 
\textit{Notes:} See comments for Figure~\ref{fig:indvref0406}.
}
\label{fig:indvref0710}
\end{minipage}%
}
\end{figure}

\begin{figure} [H]
\centering
\captionsetup{labelfont=normalfont,labelsep=none,justification=centering}
\caption{\\Distribution of colorism by referee in the \textit{later} post-awareness period (2011-2014)}\vspace{-0.8em}
\captionsetup{justification=raggedright, singlelinecheck=false}
\adjustbox{max totalheight=\textheight,keepaspectratio,center,valign=c}{%
\begin{minipage}{\textwidth}
\centering
\subfloat{\includegraphics[width=1.0\linewidth]{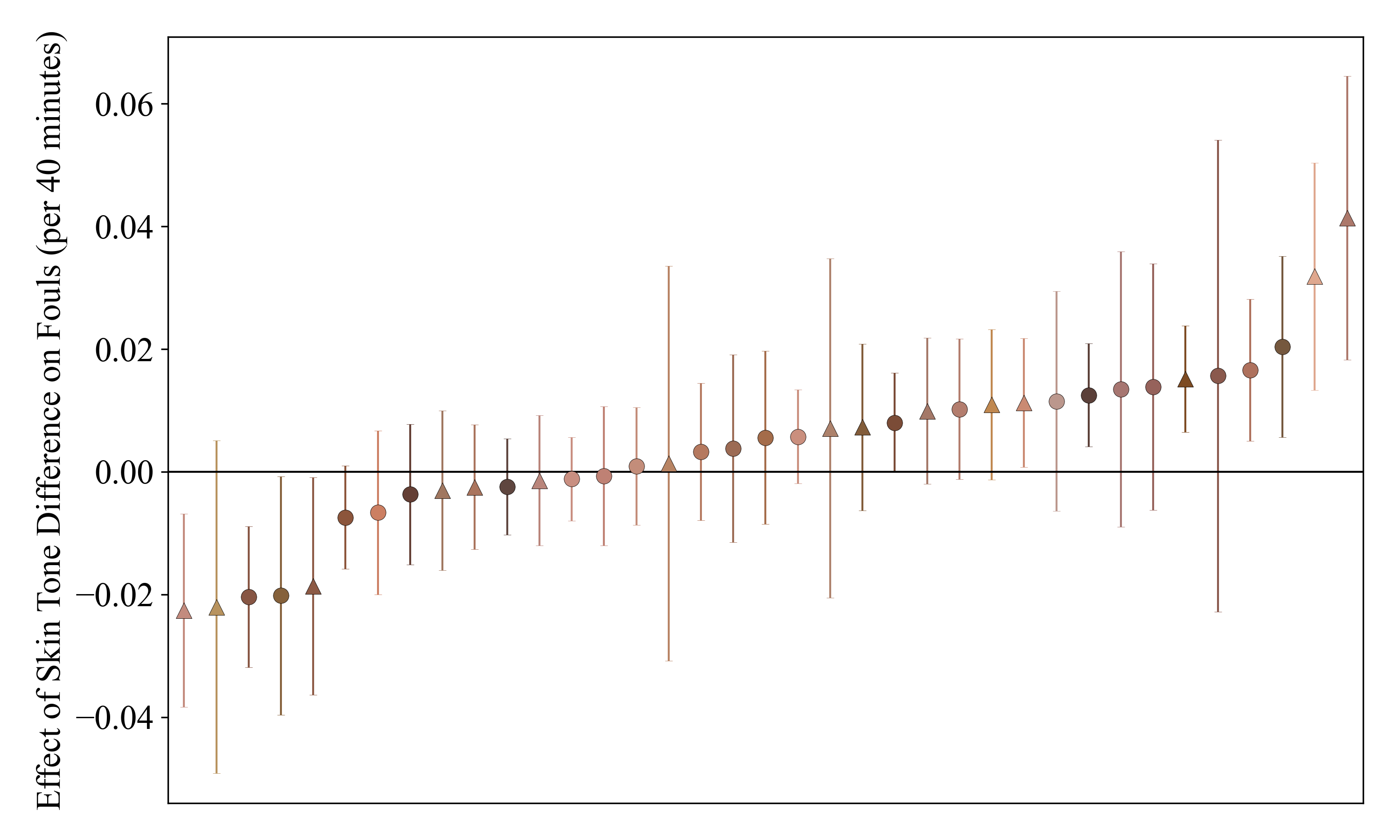}}\\[-0.05em]
\caption*{\justifying\scriptsize 
\textit{Notes:} See comments for Figure~\ref{fig:indvref0406}.
}
\label{fig:indvref1114}
\end{minipage}%
}
\end{figure}

\end{document}